\documentclass{iopjournal}

\usepackage{xcolor}
\usepackage{graphicx}
\usepackage{url}
\usepackage{siunitx}

\usepackage{amsmath,amssymb}

\newcommand{\rhot}{\rho_{\mathrm{tor}}}
\newcommand{\fsavg}[1]{\left\langle #1\right\rangle_{\mathrm{FS}}}

\begin{document}

\title{Edge transport mechanisms in small-ELM and type-I ELM regimes at JET using global gyrokinetic simulations}
\author{Mattia Dicorato$^{1,*}$\orcid{0000-0001-7335-5062}, Jeronimo Garcia$^{2}$\orcid{0000-0003-0900-5564}, Tobias G\"{o}rler$^{1}$\orcid{0000-0002-0851-6699}, Elena de la Luna$^{3}$, Gabriele Merlo$^{1}$, Emiliano Fable$^{1}$, Frank Jenko$^{1,4}$, \v{Z}iga \v{S}tancar$^{5}$, Luca Garzotti$^{5}$, Vito K. Zotta$^{6}$, Fernanda Rimini$^{5}$, Domenico Frigione$^{7}$, JET Contributors$^{\dagger}$, and the EUROfusion Tokamak Exploitation Team$^{\ddagger}$}

\affil{$^1$Max Planck Institute for Plasma Physics, Boltzmannstr. 2, 85748 Garching, Germany}

\affil{$^2$CEA, IRFM, Saint-Paul-lez-Durance, F-13108, FR}

\affil{$^3$Laboratorio Nacional de Fusión, CIEMAT, Madrid, ES}

\affil{$^4$Institute for Fusion Studies, UT at Austin, Austin, USA}

\affil{$^5$United Kingdom Atomic Energy Authority, Culham Science Centre, Abingdon, UK}

\affil{$^6$Department of Astronautical, Electrical and Energy Engineering, Sapienza University, Rome, IT}

\affil{$^7$University of Rome Tor Vergata, Rome, IT}

\affil{$^\dagger$See the author list of C. F. Maggi et al. \textit{Nucl.~Fusion} 64 (2024) 112012}

\affil{$^\ddagger$ See the author list of N. Vianello et al. \textit{Nucl.~Fusion} 66 (2026) 116010}

\affil{$^*$Author to whom any correspondence should be addressed.}

\email{mattia.dicorato@ipp.mpg.de}


\begin{abstract}

To explore tokamak operating regimes that combine high energy confinement with edge dynamics relevant for ITER, new regimes with small edge-localized modes (ELMs), referred to as baseline small-ELM (BSE), have been obtained at JET with the Be wall and W divertor. In this work, the edge transport mechanisms in a selected BSE discharge are compared with those in a standard H-mode with type-I ELMs using radially global gyrokinetic simulations with the GENE code. The simulations use experimental input profiles and include kinetic electrons and deuterium ions, electromagnetic (EM) effects, and the radially varying equilibrium $E\times B$ flow. At low toroidal mode number $n$, the type-I ELM pedestal is found to be unstable to kinetic-ballooning modes (KBMs), with radial mode structures covering the full pedestal. In the BSE regime, no KBM branch is found, and the same spectral range is dominated by hybrid ITG--TEM modes. The modifications of the linear modes by finite$-\beta$ effects are discussed in details. In nonlinear simulations of the type-I ELM regime, KBM-driven turbulence produces high transport levels that are resilient to the equilibrium $E\times B$ flow. By contrast, in the BSE regime, EM effects moderately increase transport, which is strongly regulated by the equilibrium $E\times B$ flow. The two regimes also differ in their radial turbulence organization and in the amplitude and persistence of the zonal shearing rate. These results identify the interplay between EM effects and equilibrium $E\times B$ flow as a key element of edge transport regulation.

\end{abstract}


\section{Introduction}
\label{sec:introduction}

The development of fusion energy based on magnetically confined plasmas relies on the integration of plasma regimes combining high core pressure and energy-confinement time with edge dynamics compatible with the heat- and particle-load tolerances of plasma-facing components (PFCs). High confinement is achieved in tokamak experiments operating in H-mode through the formation of an edge transport barrier and the consequent steepening of the edge pressure profiles, known as the pedestal. Because the pedestal provides the boundary condition for the stiff core profiles, its properties strongly affect the achievable core pressure and fusion performance~\cite{urano_pedestal_2014,groebner_elements_2023}. The edge dynamics of conventional H-mode plasmas is typically characterized by edge-localized modes (ELMs), which transiently relax the pedestal and expel heat and particles. However, the compatibility of the most common ELMs, the so-called type-I ELMs, with the PFC heat- and particle-load limits is not guaranteed when extrapolated to reactor-scale fusion devices~\cite{snyder_pedestal_2009,fenstermacher_progress_2025}.

This issue has motivated the development of ELM-free and small-ELM regimes that preserve high confinement while avoiding large transient pedestal losses. At JET with the Be wall and W divertor, experiments aimed at developing the ITER baseline scenario have led to a new regime with small-ELMs, referred to as the baseline small-ELM (BSE)~\cite{garcia_new_2022,de_la_luna_exploring_2024,garzotti_2025,garzotti_2026}. This regime retains high confinement, with an enhancement factor relative to the IPB98(y,2) scaling law~\cite{confinement_chapter_1999} of $H_{98}\simeq1$, while reducing the energy expelled per ELM. It also reaches normalized plasma parameters relevant to the ITER baseline scenario, including $\beta_N\simeq1.8$ and pedestal electron collisionality up to $\nu^*_{e,\mathrm{ped}}\sim 0.1$~\cite{garcia_new_2022,de_la_luna_exploring_2024}. Because these ITER-relevant parameters are rarely achieved in regimes without large type-I ELMs, it is important to understand the physical mechanisms that regulate the BSE edge structure.

Because the BSE regime has only recently been identified, its regulating mechanisms have been less extensively explored than those of the type-I ELM pedestal. The latter is commonly interpreted within a theoretical framework in which coupled peeling--ballooning (P--B) MHD instabilities, driven by the edge pressure gradient and current density, constrain the maximum pedestal pressure gradient and provide a model for type-I ELM onset~\cite{snyder_edge_2002,snyder_pedestal_2009}, while kinetic-ballooning-mode (KBM)-driven transport is proposed to regulate the pedestal width~\cite{snyder_development_2009,snyder_first-principles_2011}. Although this framework has successfully described many conventional type-I ELM pedestals, additional physics may be required for regimes whose pedestal structure is not limited by proximity to the P--B boundary~\cite{groebner_elements_2023,fenstermacher_progress_2025}.

This is particularly relevant to the BSE regime, whose atypical pedestal structure may reflect different regulating mechanisms. Ideal-MHD analyses have shown that the BSE pedestal remains stable to P--B modes and far from the local ideal-ballooning boundary, leaving the triggering mechanism of the small-ELMs unexplained~\cite{garcia_new_2022,de_la_luna_exploring_2024}. The present work investigates the turbulent transport mechanisms in the BSE edge through a comparative analysis with an ITER-baseline discharge with type-I ELMs. Gyrokinetic theory provides the first-principles framework used to study the low-frequency microinstabilities and turbulent transport driven by the steep pedestal gradients~\cite{brizard_foundations_2007,garbet_gyrokinetic_2010}. Previous gyrokinetic studies of JET with the Be/W wall pedestals have identified several candidate transport mechanisms, including KBMs, microtearing modes (MTM), ion-temperature-gradient (ITG) and trapped-electron modes (TEM), and electron-temperature-gradient (ETG) instabilities~\cite{hatch_gyrokinetic_2017,kotschenreuther_gyrokinetic_2019,chapman-oplopoiou_role_2022}.

Previous local gyrokinetic analyses around the pedestal top of the discharges considered here indicated different instability and turbulent transport regimes in the type-I ELM and BSE cases and highlighted the importance of electromagnetic effects and equilibrium $E\times B$ shear~\cite{dicorato_gyrokinetic_2022,dicorato_turbulent_2024}. Local simulations, however, cannot retain the strong radial variation of the equilibrium profiles, instability drive, magnetic geometry, and equilibrium flow across the full edge, nor determine the radially global localization and coupling of the unstable modes. These limitations become particularly relevant when the equilibrium scale lengths approach the turbulence radial scales and global effects can modify mode structures and turbulent transport~\cite{mcmillan_system_2010}. The present analysis therefore employs state-of-the-art radially global simulations with the GENE code~\cite{gorler_global_2011}, allowing the transport mechanisms to be characterized across the edge region inside the last closed flux surface.

The aim of this work is to characterize and compare the global turbulent transport mechanisms of a pre-type-I ELM pedestal and a time-averaged BSE edge. A comparison of electrostatic-limit (ES$_l$) and electromagnetic (EM) simulations, with and without the equilibrium $E\times B$ flow, is used to assess the impact of finite$-\beta$ effects and equilibrium-flow regulation in the two discharges. The analysis addresses the global linear stability, radial mode localization, nonlinear effective diffusivities, and radial and zonal organization of the turbulence.

The remainder of the paper is organized as follows. Section~\ref{sec:exp_prof} describes the experimental conditions and edge profiles of the two JET discharges. Section~\ref{sec:gene} introduces the global gyrokinetic model and simulation matrix. Sections~\ref{sec:linear} and~\ref{sec:nonlinear} present the linear stability and nonlinear turbulence analyses, respectively. The implications of the results and conclusions are discussed in section~\ref{sec:conclusion}.


\section{JET edge profiles in type-I ELM and small-ELM regimes}
\label{sec:exp_prof}

\begin{figure}
\centering
\includegraphics[width=0.33\textwidth]{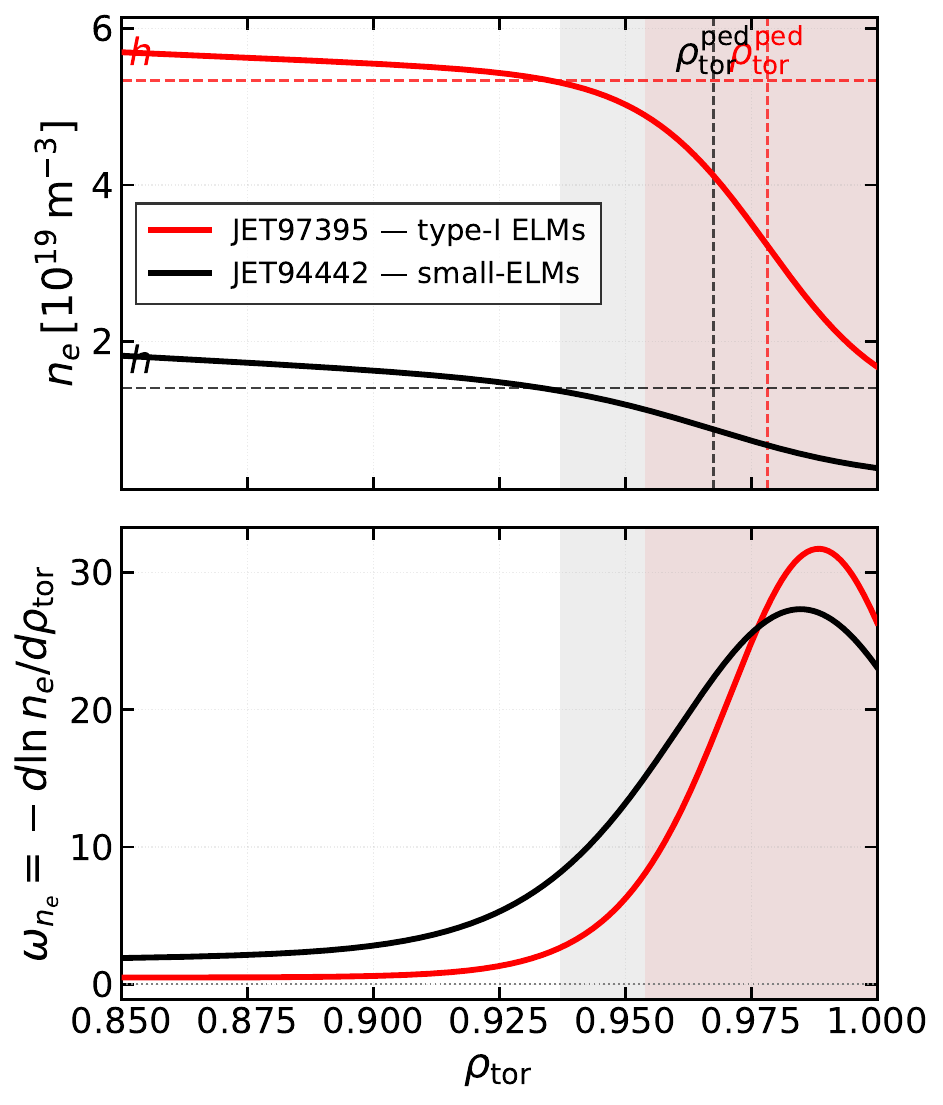}\hfill
\includegraphics[width=0.33\textwidth]{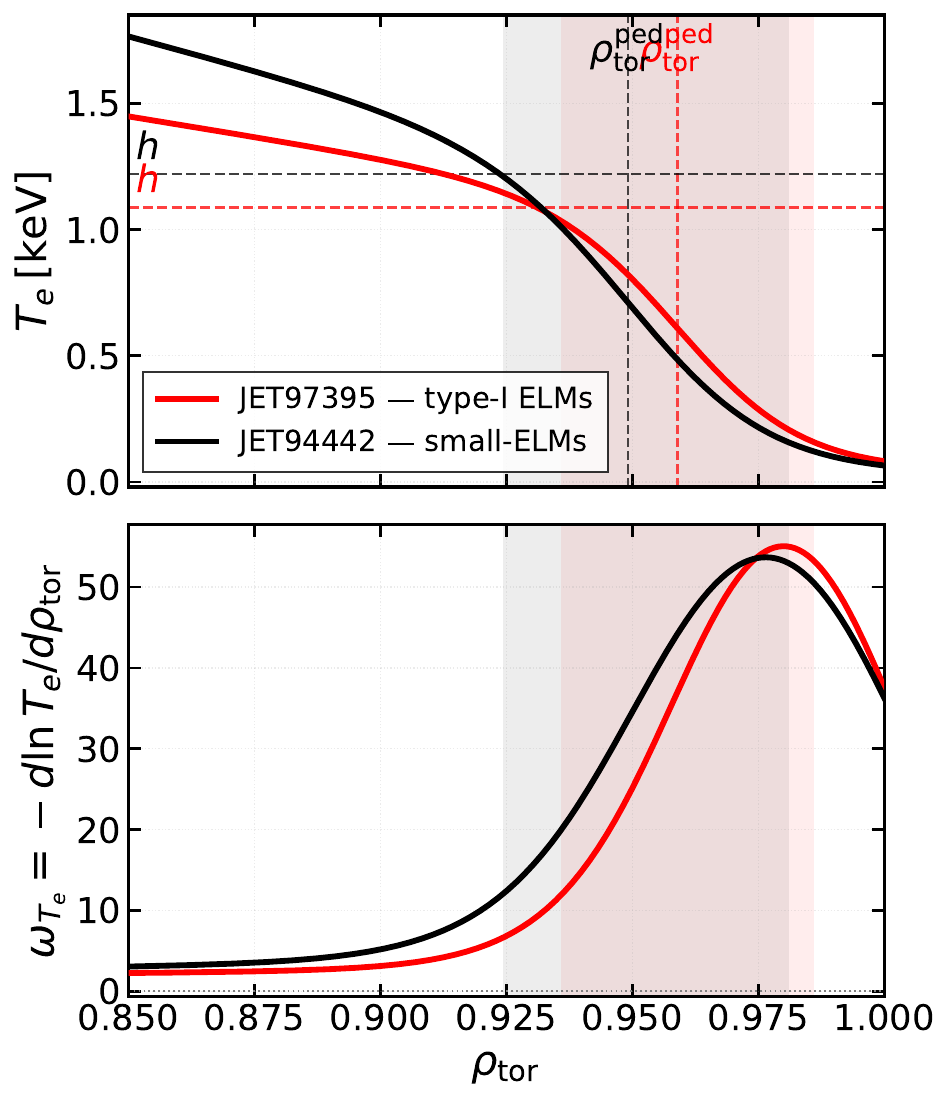}\hfill
\includegraphics[width=0.33\textwidth]{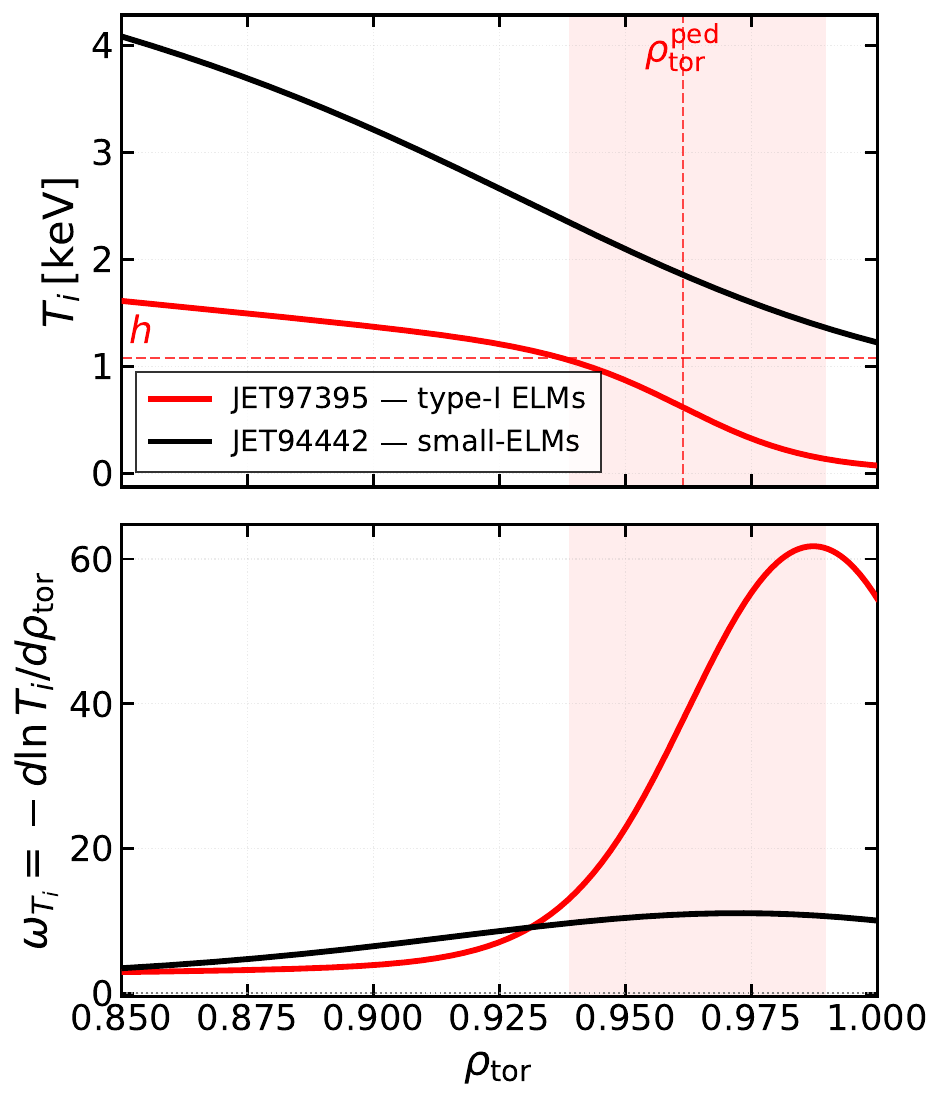}\hfill\\
\makebox[0.25\textwidth][c]{\small (a)}\hfill
\makebox[0.25\textwidth][c]{\small (b)}\hfill
\makebox[0.25\textwidth][c]{\small (c)}\hfill\\
\caption{Comparison of the edge profiles of (a) electron density $n_e$, (b) electron temperature $T_e$, and (c) ion temperature $T_i$, as functions of $\rhot$, for the type-I ELMy discharge \#97395 (red) and the BSE discharge \#94442 (black). The upper and lower panels of (a)--(c) show the fitted profiles and the corresponding normalized logarithmic gradients, respectively. The shaded regions indicate the fitted pedestal widths, while $h$ and $\rhot^{\mathrm{ped}}$ denote the pedestal height and the position of maximum absolute fitted gradient, respectively.}
\label{fig:input_profiles_basic}
\end{figure}

\begin{table*}[t]
\caption{Modified-hyperbolic-tangent characterization of the density and temperature pedestals. Density heights $h$ are expressed in $10^{19}\,\mathrm{m}^{-3}$, temperature heights in keV, widths in $\Delta\rhot$, and radial positions in $\rhot$. The pedestal position $\rhot^{\mathrm{ped}}$ is the location of the maximum absolute gradient of the fitted profile. The pedestal-top position $\rhot^{\mathrm{top}}$ is the radial position corresponding to the fitted pedestal height. No pedestal parameters are reported for the \#94442 ion-temperature profile because it does not exhibit a sufficiently well-defined pedestal.}
\label{tab:pedestal_structure}
\centering
\begin{tabular}{llcccc}
\hline
Quantity & Discharge & Height $h$ & Width $w$ &
$\rhot^{\mathrm{ped}}$ & $\rhot^{\mathrm{top}}$ \\
\hline
$n_e$ & \#97395 & 5.34 & 0.049 & 0.978 & 0.935 \\
      & \#94442 & 1.40 & 0.067 & 0.968 & 0.933 \\
\hline
$T_e$ & \#97395 & 1.09 & 0.050 & 0.959 & 0.931 \\
      & \#94442 & 1.22 & 0.057 & 0.949 & 0.924 \\
\hline
$T_i$ & \#97395 & 1.08 & 0.051 & 0.962 & 0.937 \\
      & \#94442 & N.A. & N.A. & N.A. & N.A. \\
\hline
\end{tabular}
\end{table*}

To identify the dominant transport mechanisms in the BSE edge, a comparative analysis is performed against a standard gas-fuelled H-mode discharge with type-I ELMs. The nominal profiles are taken from the type-I ELM discharge \#97395 and the BSE discharge \#94442, which was obtained without an external edge particle source. The experimental conditions have been presented in previous works~\cite{garcia_new_2022,de_la_luna_exploring_2024,dicorato_turbulent_2024} and are summarized here for the present analysis. The two discharges have the same plasma current, $I_p=\qty{3}{\mega\ampere}$, toroidal magnetic field on axis, $B_{\mathrm{tor}}=\qty{2.8}{\tesla}$, safety factor at the flux surface containing $95\%$ of the poloidal magnetic flux, $q_{95}=3.2$, and average triangularity, $\delta=0.23$. These parameters produce similar plasma shapes and reduce plasma shaping as a possible source of the differences identified below. Both discharges are heated predominantly by neutral beam injection (NBI), with an additional contribution from ion cyclotron resonance heating (ICRH): discharge \#97395 has $P_{\mathrm{NBI}}=\qty{28}{\mega\watt}$ and $P_{\mathrm{ICRH}}=\qty{4}{\mega\watt}$, whereas discharge \#94442 has $P_{\mathrm{NBI}}=\qty{17.5}{\mega\watt}$ and $P_{\mathrm{ICRH}}=\qty{3}{\mega\watt}$.

The two discharges exhibit different edge dynamics. Discharge \#97395 has regular type-I ELMs with frequency $f_{\mathrm{ELM}}=\qty{45}{\hertz}$ and relative plasma-energy loss per ELM $\Delta W_{\mathrm{ELM}}/W_{\mathrm{MHD}}=2.2\%$. By contrast, discharge \#94442 develops small-ELMs with frequencies in the range $f_{\mathrm{ELM}}=\qtyrange[range-units=single,range-phrase=-]{100}{600}{\hertz}$ and an estimated relative energy loss $\Delta W_{\mathrm{ELM}}/W_{\mathrm{MHD}}=0.6\%$. Because the small-ELMs are frequent and have low amplitude, they produce density and temperature perturbations that remain within the measurement uncertainties~\cite{de_la_luna_exploring_2024}. Different time intervals are therefore selected to characterize the edge profiles in the two regimes. For the type-I ELM discharge, a narrow time interval immediately preceding an ELM, $t=\qtyrange[range-units=single,range-phrase=-]{11.95}{11.98}{\second}$, was selected, representing the maximum recovery of the pedestal before the ELM crash. For the BSE discharge \#94442, the profiles were instead averaged over several small-ELM cycles, using $t=\qtyrange[range-units=single,range-phrase=-]{11.35}{11.45}{\second}$. The following analysis should therefore be interpreted as a comparison between a pre-type-I ELM pedestal at the highest gradient during the ELM cycle, and a time-averaged BSE edge, for which averaging over small-ELM cycles may smooth transiently steeper gradients.

The resulting time-averaged and fitted edge profiles of electron density $n_e$, electron and ion temperatures $T_e$ and $T_i$ are shown in figure~\ref{fig:input_profiles_basic} as functions of the normalized toroidal-flux coordinate $\rhot=\sqrt{\Phi_{\mathrm{tor}}/\Phi_{\mathrm{edge}}}$, where $\Phi_{\mathrm{tor}}$ is the toroidal magnetic flux and $\Phi_{\mathrm{edge}}$ is its value at the last closed flux surface. The edge profiles were fitted over the radial domain $0.85\leq\rhot\leq1$ using a five-parameter modified-hyperbolic-tangent analytical function (see appendix~\ref{app:profiles} for definitions). This procedure allows the pedestal structure to be characterized quantitatively through its height $h$ and width $w$. The pedestal position, $\rhot^{\mathrm{ped}}$, is defined as the location of the maximum absolute gradient of the fitted profile. These quantities are indicated in figure~\ref{fig:input_profiles_basic} and summarized in table~\ref{tab:pedestal_structure}. In addition, the lower panels of figures~\ref{fig:input_profiles_basic}(a)--(c) show the normalized logarithmic gradients $\omega_X=-\mathrm{d}(\ln X)/\mathrm{d}\rhot$ with $X\in\{n_e,T_e,T_i\}$ which constitute the driving parameters for turbulent transport.

\begin{figure}[t]
\centering
\includegraphics[width=0.45\textwidth]{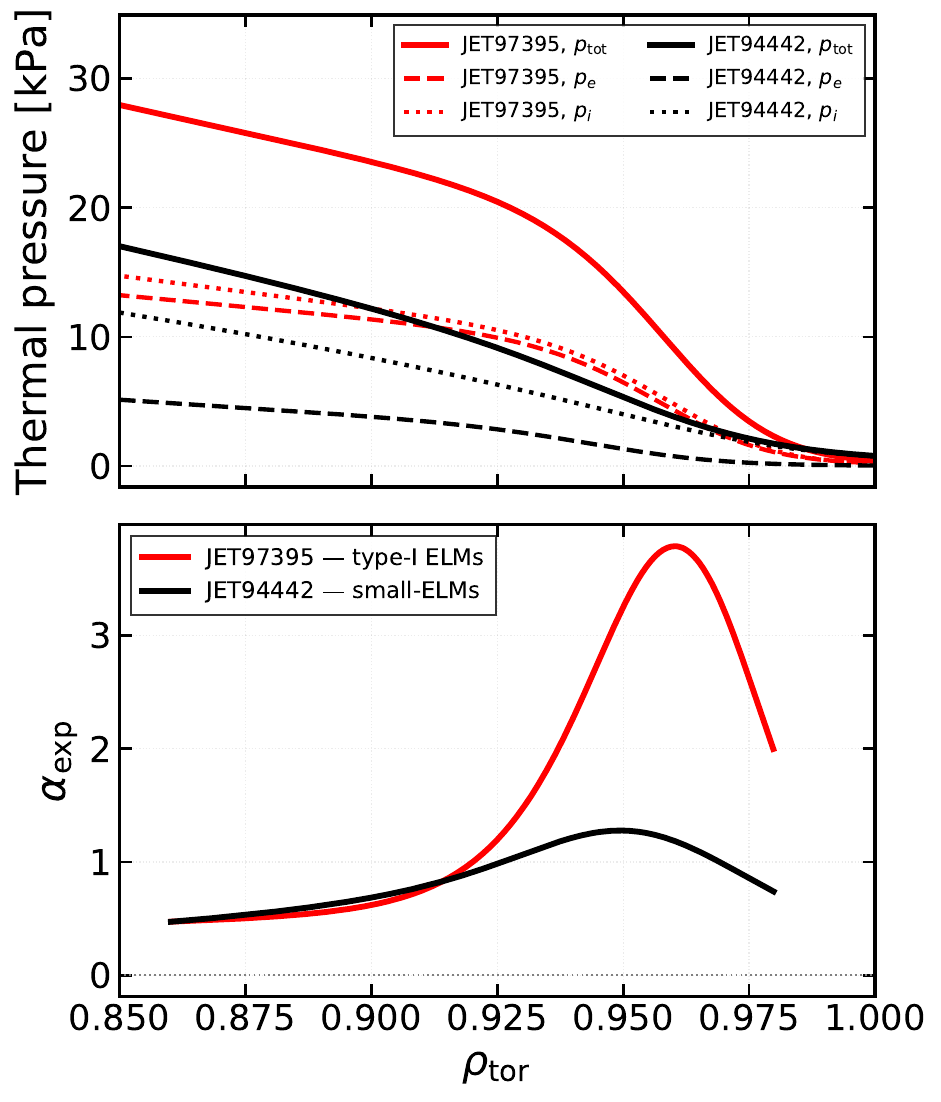}
\caption{Comparison of the thermal-pressure profiles and normalized pressure-gradient parameter for the type-I ELMy discharge \#97395 (red) and the BSE discharge \#94442 (black). The upper panel shows the electron pressure $p_e$ (dashed), ion pressure $p_i$ (dotted), and total thermal pressure $p_{\mathrm{tot}}$ (solid). The lower panel shows the normalized pressure-gradient parameter relevant to ballooning stability $\alpha_{\mathrm{exp}}$.}
\label{fig:pressure_alpha}
\end{figure}

As shown in figure~\ref{fig:input_profiles_basic} and table~\ref{tab:pedestal_structure}, operation without gas puffing reduces the $n_e$ pedestal height $h$ by approximately a factor of $4$ in the BSE discharge. It also increases the pedestal width $w$ and shifts $\rhot^{\mathrm{ped}}$ inward. Because $n_e$ is lower, the normalized density-gradient drive $\omega_{n_e}$ is up to approximately $3$ times larger in the BSE case over the inner part of the edge region. Towards the outer pedestal, $\omega_{n_e}$ becomes larger in the type-I ELM case. The $T_e$ pedestal height $h$ is moderately larger in the BSE discharge, while its width $w$ increases and $\rhot^{\mathrm{ped}}$ shifts slightly inward. The resulting $\omega_{T_e}$ is up to approximately $80\%$ larger over a substantial part of the BSE pedestal. In addition, $T_i$ is significantly higher in the BSE discharge but does not develop a steep pedestal region. Consequently, $\omega_{T_i}$ is smaller by up to approximately a factor of $4$ than in the type-I ELM case. Notice, however, that edge $T_i$ measurements have substantial uncertainty within the pedestal, whereas the pedestal-top value is better constrained. The present analysis uses the nominal fitted profiles; the propagation of the uncertainty in $T_i$ into $\omega_{T_i}$, the ion-pressure gradient, and
the diamagnetic contribution to the reconstructed $E_r$ are therefore neglected. The ratio $T_i/T_e$ is therefore substantially larger in the BSE regime than in the type-I ELM H-mode, for which $T_i/T_e\simeq1$. The resulting total thermal pressure $p_{\mathrm{tot}}$, together with the electron and ion contributions $p_e$ and $p_i$, are shown in figure~\ref{fig:pressure_alpha}. The normalized pressure-gradient parameter relevant to ballooning stability is defined as
\begin{equation}
\alpha_{\mathrm{exp}} = -q^2R_0\frac{\mathrm{d}\beta_{\mathrm{tot}}}{\mathrm{d}r_{\mathrm{tor}}},
\qquad 
\beta_{\mathrm{tot}} = \frac{2\mu_0p_{\mathrm{tot}}}{B_0^2},
\qquad
r_{\mathrm{tor}}=a_{\mathrm{tor}}\rhot,
\label{eq:alpha_exp}
\end{equation}
where $R_0$ and $B_0$ are the reference major radius and magnetic field, respectively. Despite the strong reduction in density, the total pressure in the BSE pedestal is only approximately $50\%$ lower than in the type-I ELM discharge because of the larger ion-pressure contribution, $p_i\propto n_iT_i$. The broader and inward-shifted pedestal profiles of the BSE regime are clearly reflected in the corresponding $\alpha_{\mathrm{exp}}$ profile, whose maximum is approximately $3$ times smaller than in the type-I ELM discharge. Indeed, previous ideal-MHD analyses found the type-I ELM pedestal to be close to the P--B boundary, whereas the BSE pedestal remains P--B stable and far from the local ideal-ballooning boundary across the pedestal~\cite{garcia_new_2022,de_la_luna_exploring_2024}.

\begin{figure}[t]
\centering
\includegraphics[width=0.35\textwidth]{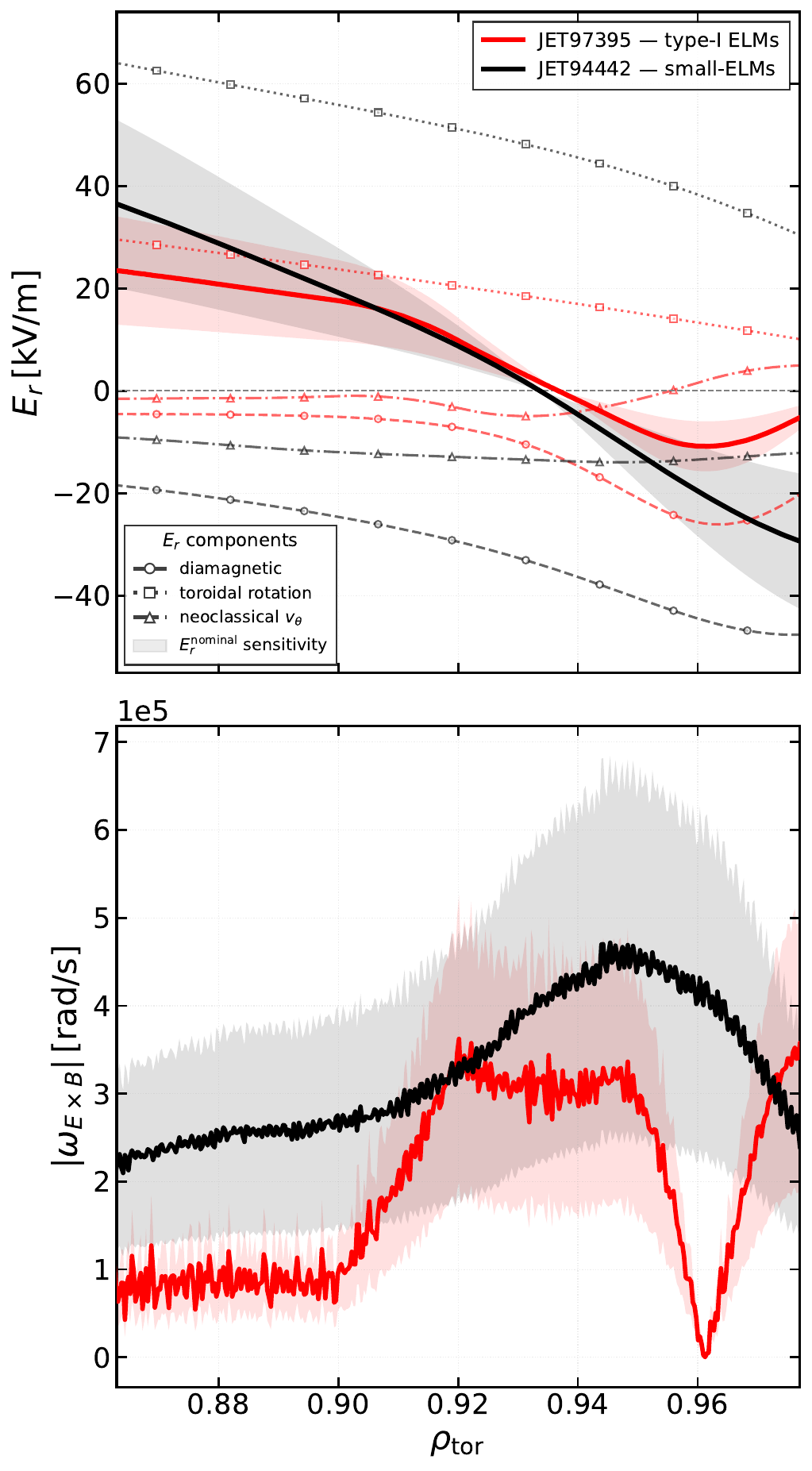}
\caption{Main-ion force-balance radial electric field and corresponding Hahm--Burrell $E\times B$ shearing rate for the type-I ELMy discharge \#97395 (red) and the BSE discharge \#94442 (black). The upper panel shows the complete reconstructed $E_r$ and its diamagnetic, toroidal-rotation, and neoclassical poloidal-flow contributions. The lower panel shows $|\omega_{E\times B}^{\mathrm{HB}}|$. The shaded regions indicate the range obtained by varying the reconstructed $E_r$ amplitude between $0.5E_r$ and $1.5E_r$ to account for its uncertainty.}
\label{fig:equilibrium_Er_shear}
\end{figure}

Finally, figure~\ref{fig:equilibrium_Er_shear} compares the equilibrium radial electric field $E_r$ reconstructed from the main-ion radial force balance, including the diamagnetic, neoclassical poloidal-flow calculated with NCLASS~\cite{houlberg_bootstrap_1997}, and toroidal-rotation contributions, together with the corresponding Hahm--Burrell shearing rate $\omega_{E\times B}^{\mathrm{HB}}$, which provides a measure of the effectiveness of the equilibrium flow in shearing turbulent structures, given by~\cite{hahm_flow_1995},
\begin{equation}
\omega_{E\times B}^{\mathrm{HB}} = \frac{R^2B_\theta^2}{B\,\Delta\psi} \frac{\partial}{\partial\psi_N} \left( \frac{E_r}{RB_\theta} \right), \qquad B=\sqrt{B_\varphi^2+B_\theta^2},
\label{eq:hahm_burrell}
\end{equation}
where $\psi$ is the poloidal flux, $\Delta\psi=\left|\psi_{\mathrm{bnd}}-\psi_{\mathrm{axis}}\right|$ and $\psi_N=\frac{\psi-\psi_{\mathrm{axis}}}{\Delta\psi}$. The resulting $|\omega_{E\times B}^{\mathrm{HB}}|$ profiles are shown in the lower panel of figure~\ref{fig:equilibrium_Er_shear}. The reconstructed profiles exhibit substantial differences in their radial shape. In the type-I ELM discharge, $E_r$ develops a well around $\rhot\simeq0.96$, close to the region of maximum $\alpha_{\mathrm{exp}}$, while the corresponding $|\omega_{E\times B}^{\mathrm{HB}}|$ exhibits a local minimum in this strongly driven region. In the BSE discharge, $E_r$, and consequently $|\omega_{E\times B}^{\mathrm{HB}}|$, is larger in magnitude over most of the radial domain, does not develop the same clear well structure, and exhibits a shearing-rate maximum close to the steepest-gradient region. Because the reconstruction contains substantial uncertainties associated with the individual force-balance contributions and their radial derivatives, the shaded regions show the range obtained by varying the reconstructed $E_r$ amplitude by $\pm50\%$. This amplitude dependence is considered explicitly in the nonlinear simulation matrix introduced in section~\ref{sec:gene}.


\section{Global electromagnetic gyrokinetic simulations}
\label{sec:gene}
\begin{figure}[t]
\centering
\includegraphics[width=0.47\textwidth]{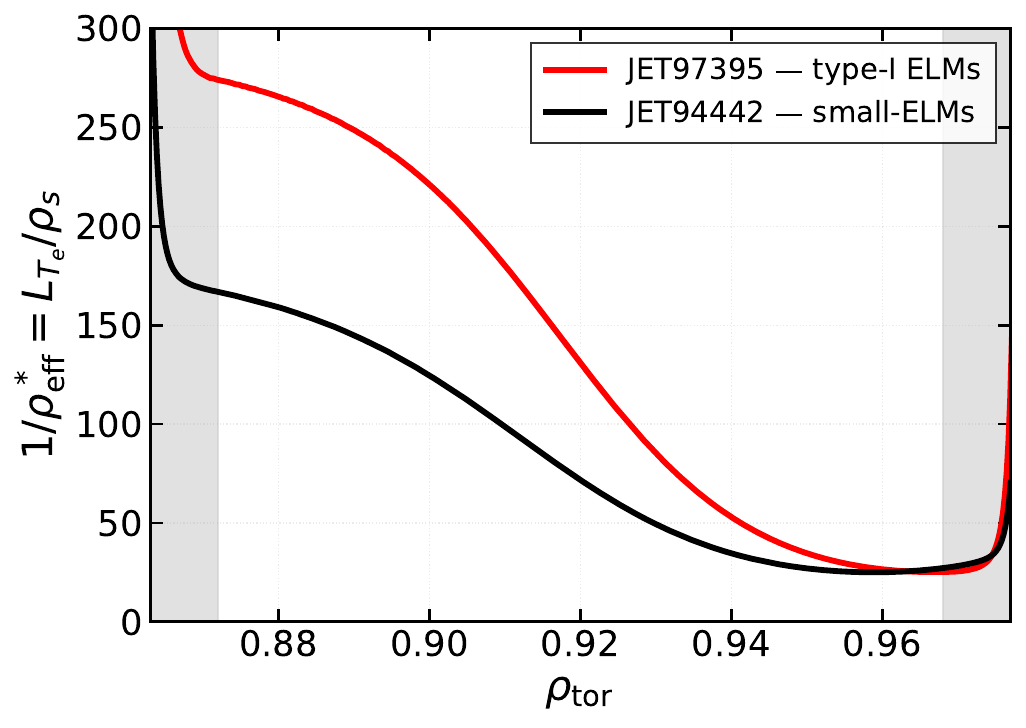}\qquad
\includegraphics[width=0.47\textwidth]{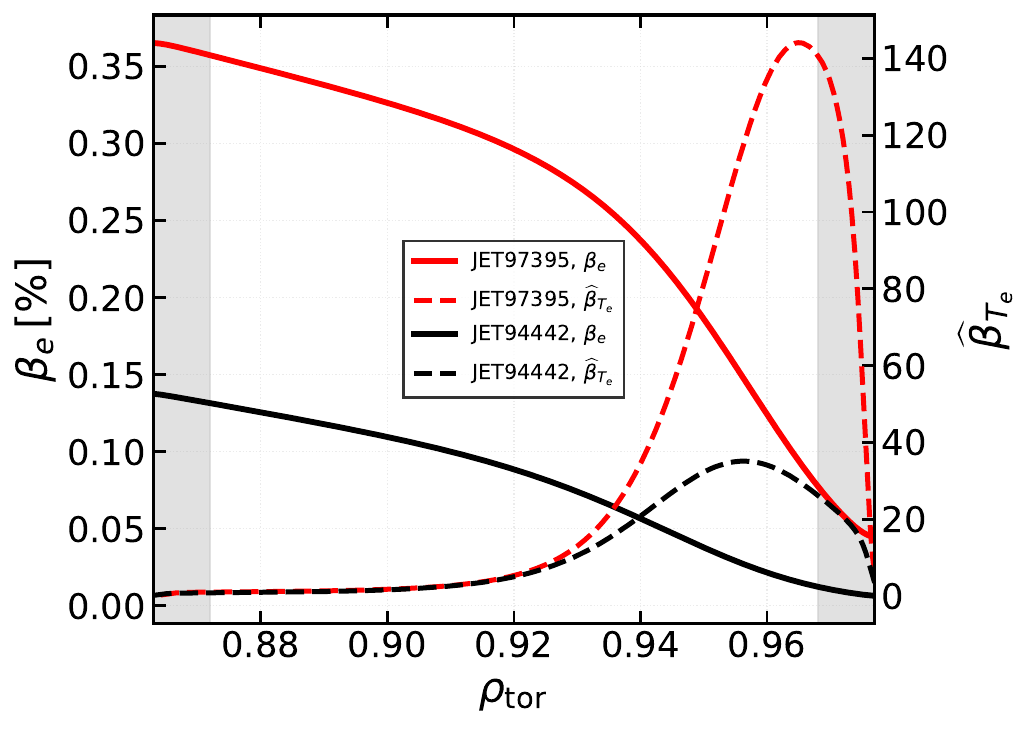}
\makebox[0.47\textwidth][c]{\small (a)}\qquad
\makebox[0.47\textwidth][c]{\small (b)}
\caption{Radial profiles of the quantities characterizing the global gyrokinetic edge domains: (a) the ratio $L_{T_e}/\rho_s$ and (b) the electron beta $\beta_e$ and normalized drift--Alfv\'en parameter $\widehat{\beta}_{T_e}$. The reduction of $L_{T_e}/\rho_s$ in the steep-gradient region indicates weakened scale separation between equilibrium and turbulent radial scales. Grey regions denote the numerical radial buffers.}
\label{fig:input_profiles_beta_globality}
\end{figure}

In this work, the edge transport mechanisms are investigated by means of gyrokinetic simulations performed with the GENE code~\cite{jenko_electron_2000,gorler_global_2011,germaschewski2021}. GENE is an Eulerian code that numerically solves the gyrokinetic Vlasov--Maxwell system for the time evolution of the perturbed distribution function $F_{1\sigma}$ of each plasma species $\sigma$ in the five-dimensional phase space $(x,y,z,v_\parallel,\mu)$. It employs a field-aligned coordinate system~\cite{beer_field-aligned_1995}, where $x$, $y$, and $z$ are the radial, binormal, and parallel along the magnetic field oriented coordinates, respectively, while $v_\parallel$ is the parallel velocity and $\mu$ the magnetic moment. The simulations include kinetic electrons and deuterium ions with the realistic electron-to-ion mass ratio, linearized Landau--Boltzmann collisional operator, electromagnetic fluctuations through $\hat{A}_{1\parallel}$, and the experimental magnetic equilibrium via field-line tracing.

In the edge, since the density and temperature profiles vary strongly over a narrow radial region, the scale separation between equilibrium quantities and the typical turbulence correlation length, as measured by the effective parameter $\rho_{\mathrm{eff}}^*=\rho_s/L_{T_e}$, is weakened. This is shown in figure~\ref{fig:input_profiles_beta_globality}(a), where $L_{T_e}/\rho_s$ decreases to values of approximately $30$ in the steep-gradient region. Previous studies have shown that global effects can become relevant at these values~\cite{mcmillan_system_2010}. This motivates the use of radially global simulations covering the domain $\rhot\in[0.86,0.98]$. This setup retains self-consistently the radial variation of the experimental profiles and magnetic geometry and allows for non-local radial coupling between turbulent structures, potentially affecting saturation mechanisms~\cite{ida_non-local_2022}. The simulations are gradient-driven, and the experimental profiles are prescribed in order to characterize the resulting turbulent transport. The simulations presented in the following use the nominal fitted profiles and do not consider their experimental uncertainties. A systematic profile-uncertainty or flux-matching analysis is beyond the scope of the present analysis. Zero-Dirichlet boundary conditions (B.C.) are imposed at the radial boundaries. Because the non-periodic radial B.C. allow the net radial transport to relax the input density and temperature profiles, Krook-type particle and heat sources are employed to maintain their time averages close to the initial values. In addition, buffer regions covering $10\%$ of the radial domain at each boundary are used to reduce the fluctuation amplitudes towards the boundary and hence avoid numerical inconsistencies with the $F_1|_{\rm boundary}=0$ assumption. Recent numerical developments required for high-fidelity global edge-turbulence simulations are employed, including the block-structured discretization of the velocity-space grid $(v_\parallel,\mu)$~\cite{jarema_block-structured_2016} and a revised $\hat{A}_{1\parallel}$~\cite{crandall_collisional_2019,leppin_turbulence_nodate} solver that ensures numerical stability also at large $\beta$ values. The numerical setup is reported in appendix~\ref{app:numerics}.

Because a $\delta f$ approach is employed, all equilibrium quantities must be prescribed, including the equilibrium $E_r\times B$ flow associated with the radial electric field $E_r$. A new interface has been implemented in GENE to reconstruct $E_r$ from the main-ion radial force-balance equation. The complete expressions, including the GENE geometrical coefficients, sign conventions, and normalization, are given in appendix~\ref{app:Er}. The resulting $E_r$ profiles and their three contributions are shown in the upper panel of figure~\ref{fig:equilibrium_Er_shear}. When the equilibrium potential $\Phi_0(x)$ is retained, the following contribution is included explicitly in the gyrokinetic Vlasov equation, which can be written schematically as,
\begin{equation}
\frac{\partial F_{1\sigma}}{\partial t} = \mathrm{RHS} - \frac{c}{\mathcal C(x)}\frac{B_0}{B_\parallel^*}\partial_x\Phi_0(x) \left[ \partial_yF_{1\sigma} - \frac{q_\sigma}{cm_\sigma} \partial_y\overline{A}_{1\parallel} \partial_{v_\parallel}F_{0\sigma}
\right].
\label{eq:equilibrium_ExB_term_main}
\end{equation}
The first term in brackets arises from the $E\times B$ advection of the perturbed distribution function, while the second involving $\hat{A}_{1\parallel}$ arises from the parallel acceleration, and $F_{0\sigma}$ is the local background Maxwellian. The global implementation retains the radial dependence of $\partial_x\Phi_0$ without linearizing the equilibrium-flow profile, as is commonly done in local flux-tube simulations. The complete derivation and normalization are reported in appendix~\ref{app:Er}.

In the following sections, different linear and nonlinear numerical experiments are used to assess the impact of two key mechanisms in edge turbulence: electromagnetic effects associated with finite$-\beta$~\cite{scott_drift_2005,scott_nonlinear_2010} and the equilibrium $E_r\times B$ flow~\cite{burrell_role_2020}. In figure~\ref{fig:input_profiles_beta_globality}(b), the profiles of $\beta_e$ and the normalized drift--Alfv\'en parameter $\widehat{\beta}_{T_e} = \beta_e \left(qR_0/L_{T_e}\right)^2$, which provides a measure of the coupling between drift-wave and shear-Alfv\'en dynamics and, consequently, of the potential relevance of electromagnetic effects~\cite{scott_drift_2005,scott_nonlinear_2010,scott_turbulence_2021,scott_turbulence_2021-1}, are shown for the two discharges. Despite the relatively low electron beta, $\beta_e<1\%$ in both regimes, $\widehat{\beta}_{T_e} > 30$ in the steep-gradient region. Since electromagnetic effects are expected to become relevant for $\widehat{\beta}_{T_e}\gtrsim1$~\cite{scott_drift_2005,scott_nonlinear_2010}, this comparison indicates that they may play an important role in both discharges. In the following, simulations in the electrostatic limit ($\mathrm{ES}_l$), performed by artificially reducing $\beta_e$ at the centre of the radial domain to $\beta_e=10^{-5}$ are compared with electromagnetic (EM) simulations performed at the nominal experimental $\beta_e$ to assess their impact. In addition, nonlinear simulations are performed with and without the equilibrium radial electric field $E_r$ in order to assess its impact. The equilibrium $E_r$ is written as
\begin{equation}
E_r^{\mathrm{ext}}(\rhot) = f_EE_r^{\mathrm{FB}}(\rhot),
\label{eq:Er_scaling}
\end{equation}
where $E_r^{\mathrm{FB}}$ is the complete force-balance reconstruction and $f_E$ is the imposed amplitude factor. For the type-I ELM discharge, the nominal reconstructed profile is used, corresponding to $f_E=1$. For the BSE discharge, the principal analysis focuses on the reduced profile $f_E=0.5$, which lies within the shaded range shown in figure~\ref{fig:equilibrium_Er_shear}. An additional EM simulation with $f_E=0.8$ is used to test the sensitivity to the $E_r$ amplitude; at this larger value, the turbulent transport is almost completely suppressed (see section~\ref{sec:nonlinear}). The simulation matrix is summarized in table~\ref{tab:simulation_matrix}.

\begin{table}[t]
\caption{Global nonlinear simulation matrix. Here, $\mathrm{ES}_l$ and EM denote the electrostatic-limit and electromagnetic models, respectively. For simulations with equilibrium $E\times B$ flow, $E_r^{\mathrm{ext}}=f_EE_r^{\mathrm{FB}}$.}
\label{tab:simulation_matrix}
\centering
\begin{tabular}{llll}
\hline
Edge profile & Field model & Equilibrium flow & $f_E$ \\
\hline
Type-I ELM & $\mathrm{ES}_l$ & without & 0 \\
Type-I ELM & $\mathrm{ES}_l$ & with & 1.0 \\
Type-I ELM & EM & without & 0 \\
Type-I ELM & EM & with & 1.0 \\
\hline
BSE & $\mathrm{ES}_l$ & without & 0 \\
BSE & $\mathrm{ES}_l$ & with & 0.5 \\
BSE & EM & without & 0 \\
BSE & EM & with & 0.5 \\
BSE & EM & with & 0.8 \\
\hline
\end{tabular}
\end{table}


\section{Global linear stability and mode identification}
\label{sec:linear}

In this section, a global gyrokinetic stability analysis is performed to characterize the modes responsible for driving transport in the two discharges. The analysis focuses on the impact of EM effects on the most unstable modes through a detailed characterization of their transport fingerprints, parity, and global radial structures.

\subsection{Global spectra and electromagnetic response}

\begin{figure}[t]
\centering
\includegraphics[width=0.45\textwidth]{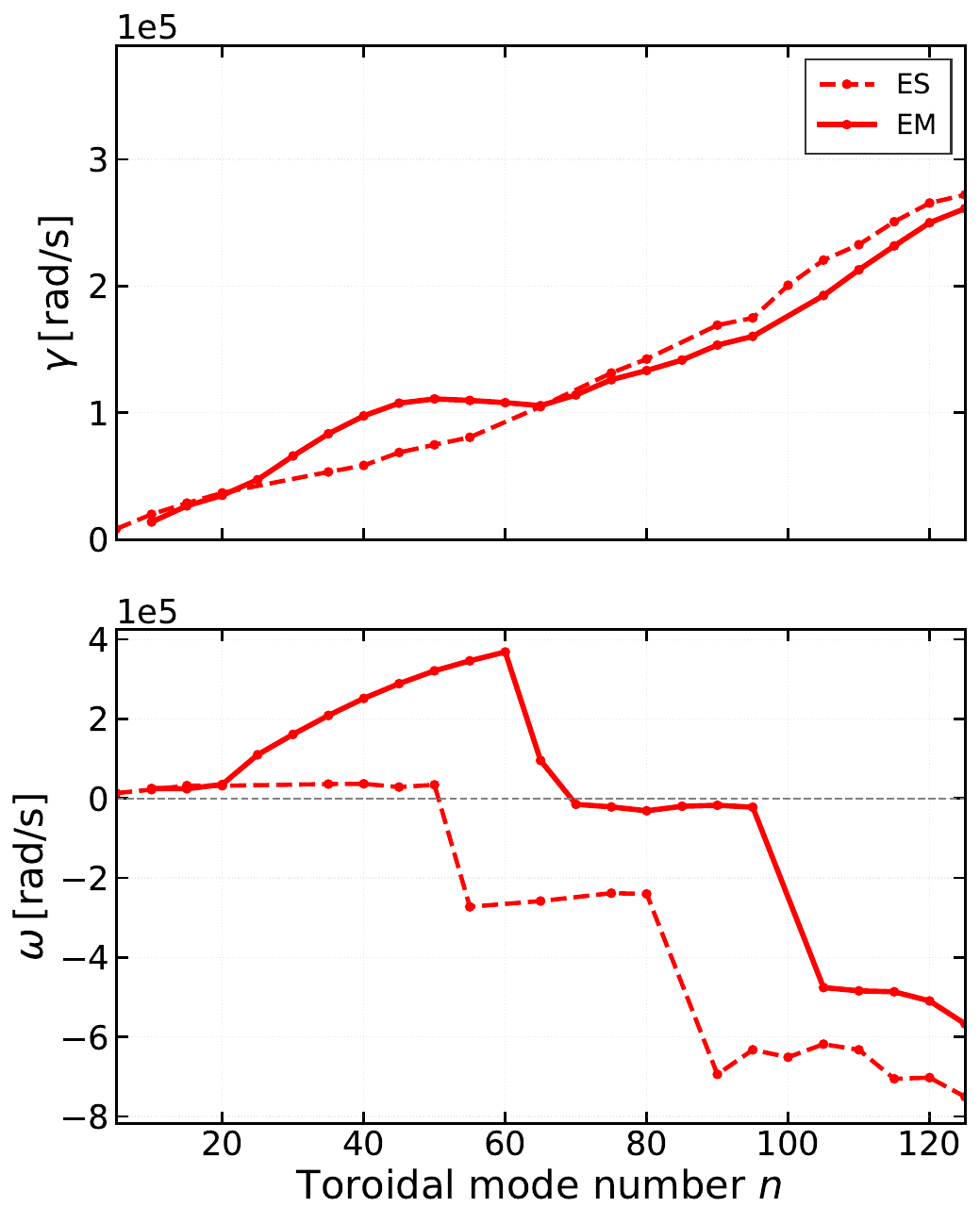}\qquad
\includegraphics[width=0.45\textwidth]{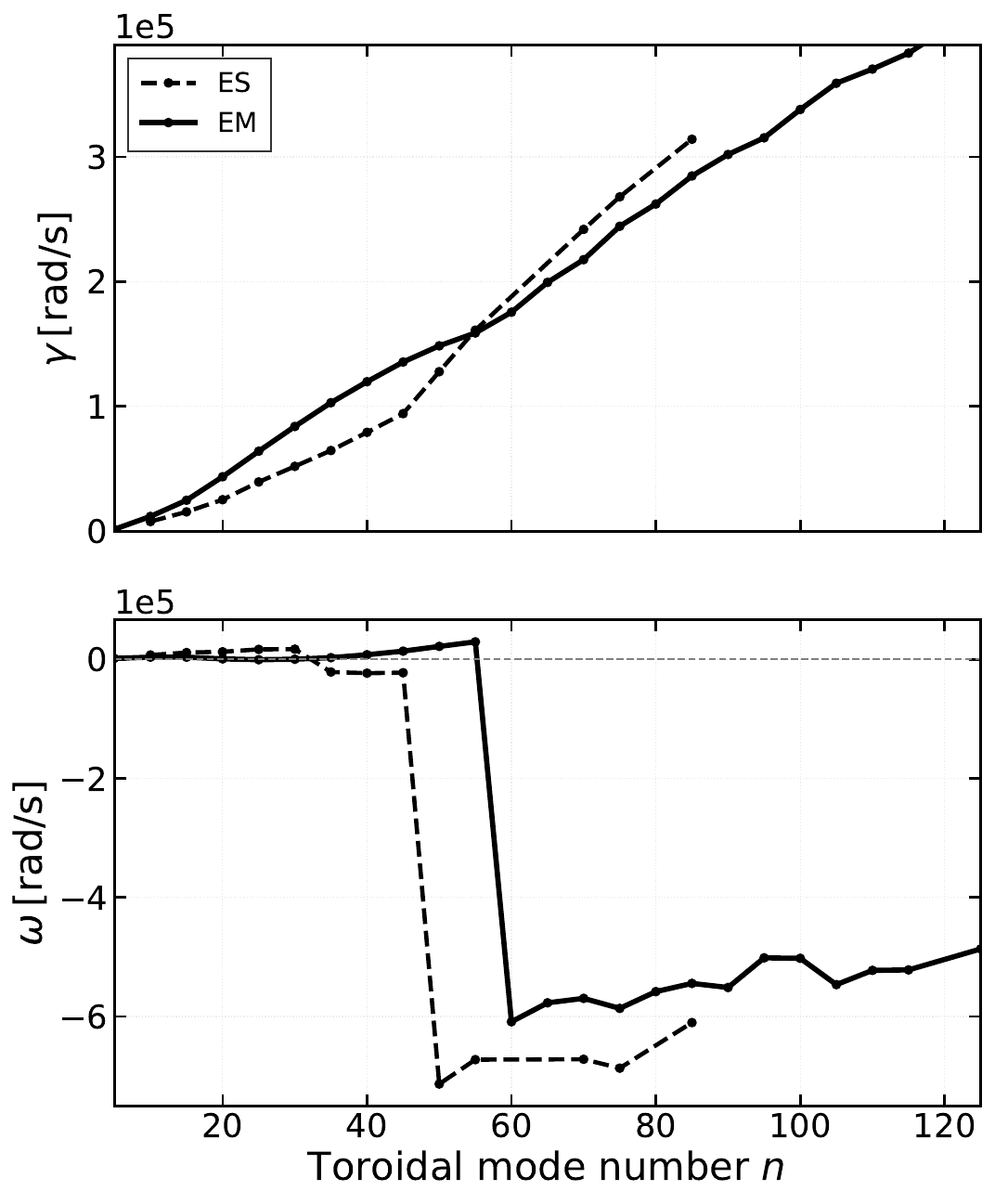}
\makebox[0.45\textwidth][c]{\small (a) Type-I ELM}\qquad
\makebox[0.45\textwidth][c]{\small (b) BSE}
\caption{Global linear growth rates $\gamma$ and real frequencies $\omega$ as functions of the toroidal mode number $n$ for (a) the type-I ELM discharge \#97395 and (b) the BSE discharge \#94442. Electrostatic-limit, $\mathrm{ES}_l$, and electromagnetic, EM, simulations are compared. Positive and negative real frequencies denote propagation in the ion- and electron-diamagnetic directions, respectively, according to the GENE sign convention.}
\label{fig:linear_spectra}
\end{figure}

\begin{figure}[t]
\centering
\includegraphics[width=0.47\textwidth]{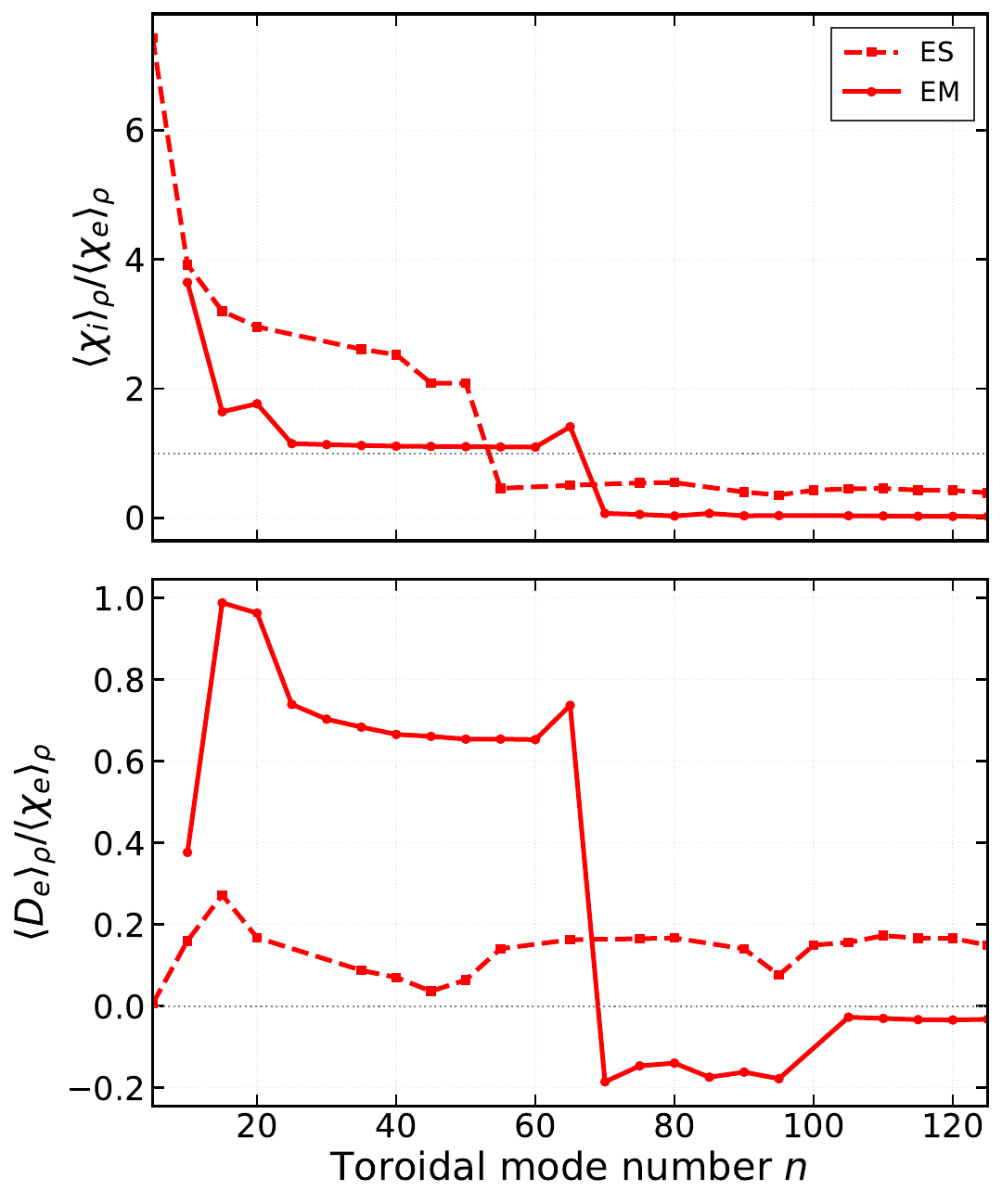}\qquad
\includegraphics[width=0.47\textwidth]{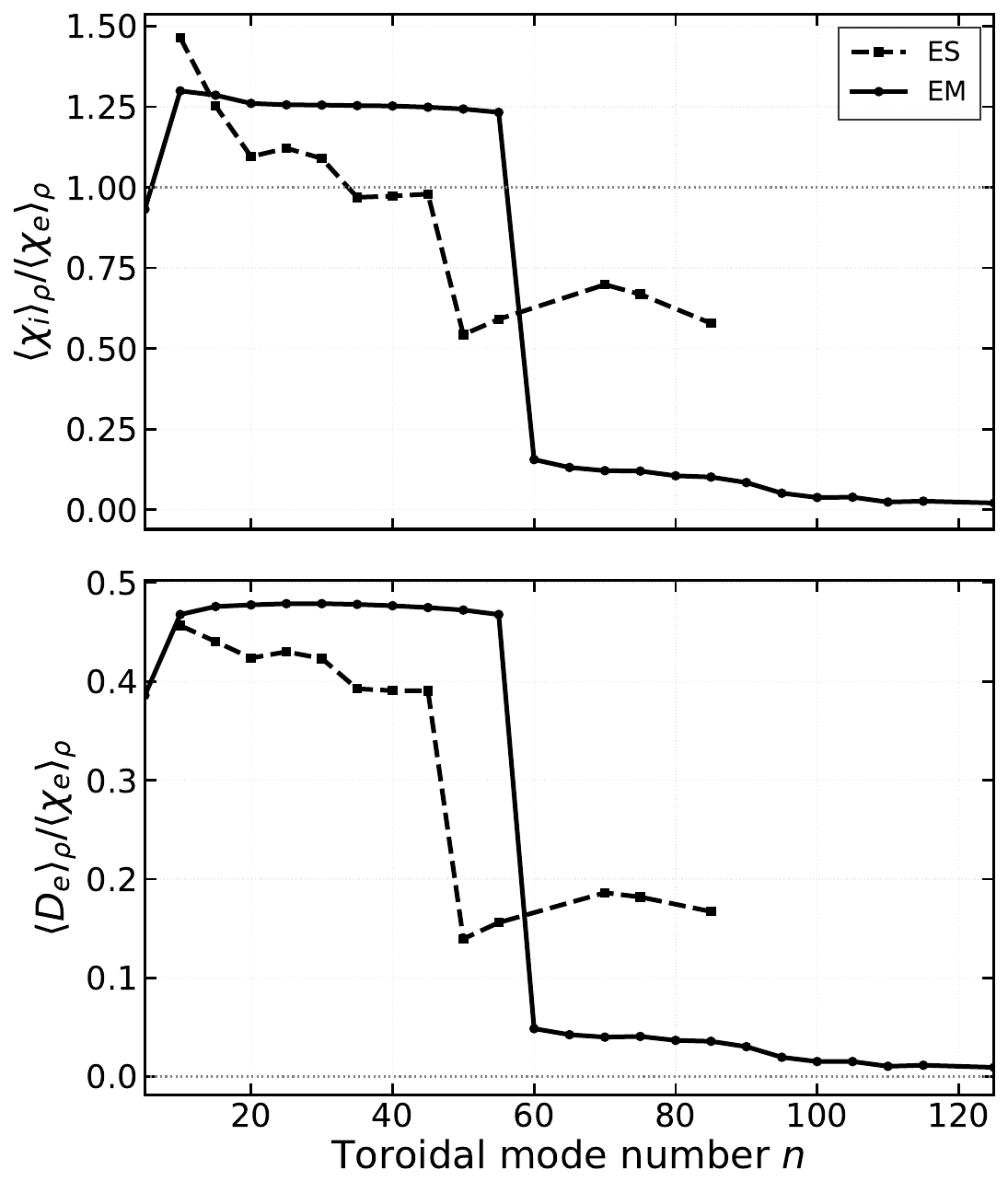}
\makebox[0.47\textwidth][c]{\small (a) Type-I ELM}\qquad
\makebox[0.47\textwidth][c]{\small (b) BSE}
\caption{Reduced linear transport fingerprints for (a) the type-I ELM discharge \#97395 and (b) the BSE discharge \#94442. The buffer-excluded ratios $\langle\chi_i\rangle_\rho/\langle\chi_e\rangle_\rho$ and $\langle D_e\rangle_\rho/\langle\chi_e\rangle_\rho$ are formed from separately radially averaged effective transport coefficients.}
\label{fig:linear_reduced_identification}
\end{figure}

\begin{figure}[t]
\centering
\includegraphics[width=0.47\textwidth]{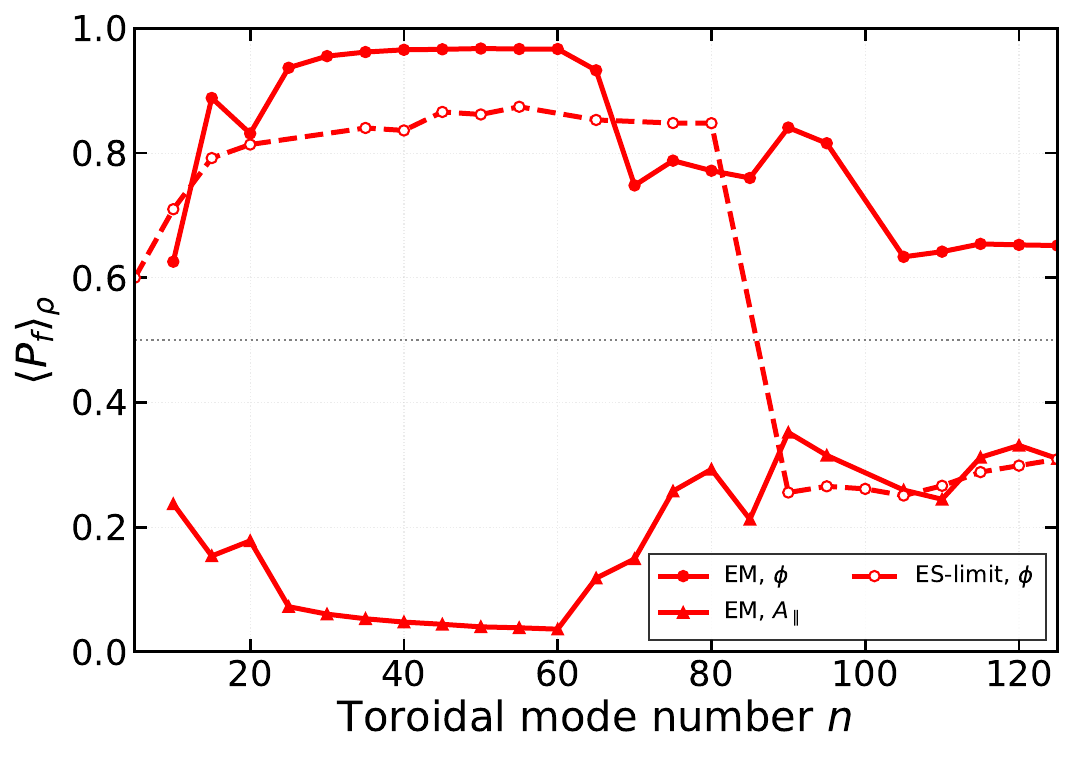}\qquad
\includegraphics[width=0.47\textwidth]{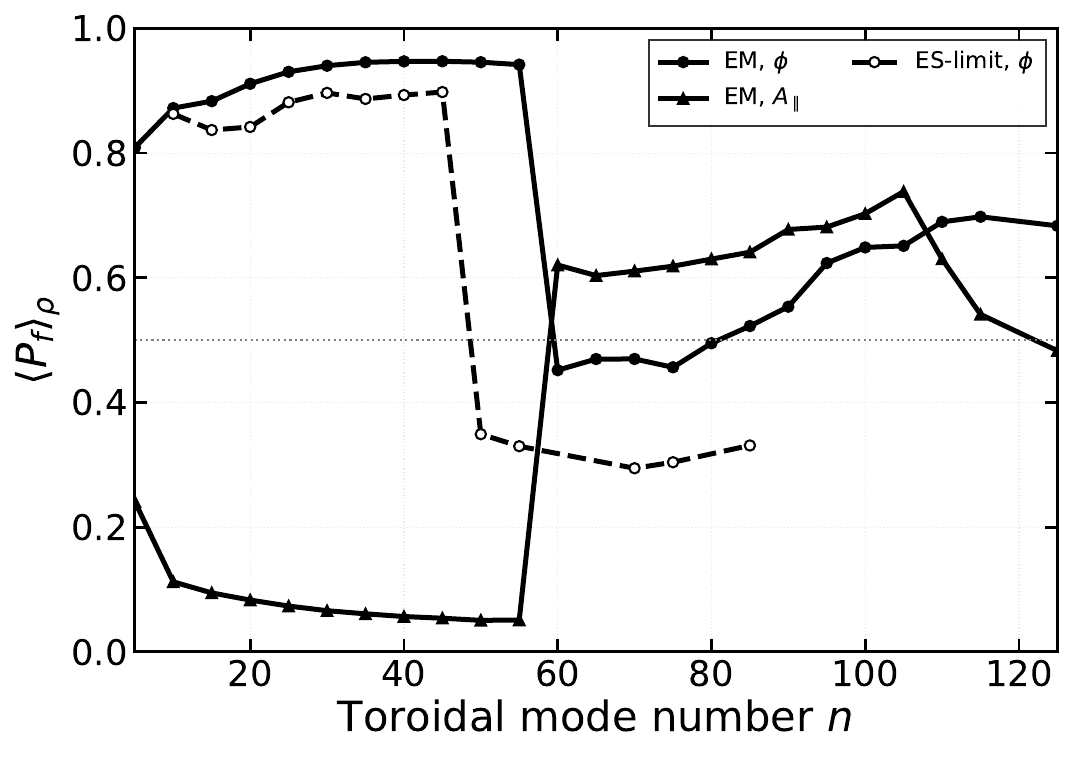}
\makebox[0.47\textwidth][c]{\small (a) Type-I ELM}\qquad
\makebox[0.47\textwidth][c]{\small (b) BSE}
\caption{Field-envelope-weighted parity of $\hat{\phi})_1$ and $\hat{A}_{1\parallel}$ for (a) the type-I ELM discharge \#97395 and (b) the BSE discharge \#94442. Values approaching unity and zero indicate even and odd parity, respectively.}
\label{fig:linear_parity}
\end{figure}

The growth rate $\gamma$ and real frequency $\omega$ of the most unstable modes are shown in figure~\ref{fig:linear_spectra} as functions of the toroidal mode number $n$ for the two discharges, for the $\mathrm{ES}_l$ and EM simulations. Positive and negative frequencies correspond to propagation in the ion- and electron-diamagnetic directions, respectively. To complete the mode characterization and compare their relative contributions to the different transport channels, the corresponding transport fingerprints~\cite{kotschenreuther_gyrokinetic_2019} are shown in figure~\ref{fig:linear_reduced_identification} (see appendix~\ref{app:linear_fingerprints} for the definition). In addition, the radially averaged parity of $\hat{\phi}_1$ and $\hat{A}_{1\parallel}$ is shown in figure~\ref{fig:linear_parity} (see appendix~\ref{app:linear_fingerprints} for the definition).
The dominant modes are thus classified based on their frequency, transport fingerprints, and parity together with their finite$-\beta$ response and radial localization.

For the type-I ELM discharge, retaining EM effects destabilizes the largest-scale modes in the range $20\leq n\leq60$, accompanied by an increase of the positive real frequency. The corresponding transport fingerprints have $\chi_i/\chi_e\simeq1$ and $D_e/\chi_e\simeq2/3$, indicating that these modes are similarly effective in transporting ion and electron heat and also produce non-negligible particle transport. In addition, the parity exhibits the characteristic ballooning structure, with even $\hat{\phi}_1$ and odd $\hat{A}_{1\parallel}$. These properties are compatible with KBMs, in agreement with the current understanding of the transport mechanisms that regulate type-I ELMy pedestals~\cite{snyder_development_2009} and with the previous local GK analyses of the same discharge~\cite{dicorato_gyrokinetic_2022,dicorato_turbulent_2024}.

At smaller scales, $n>60$, two branches with negative frequencies appear. Their transport fingerprints have $\chi_i/\chi_e\ll1$ and $D_e/\chi_e<0$, indicating transport predominantly in the electron heat channel and an inward effective particle flux. These branches are compatible with TEM and intermediate-scale ETG modes occurring at the upper end of the ion-scale spectral range retained, rather than at the full electron-scale range conventionally associated with ETG turbulence~\cite{jenko_electron_2000,dicorato_turbulent_2024}.

A different picture would emerge from the $\mathrm{ES}_l$ simulations. At the largest scales, $5<n<50$, the modes have significantly lower positive frequencies, $\chi_i/\chi_e\simeq3$ (with larger values for individual $n$), and $D_e/\chi_e<0$, which are compatible with an ITG-dominated branch. At smaller scales, $n>50$, modes with negative frequencies and $\chi_i/\chi_e<1$ are compatible with TEM and intermediate-scale ETG modes. These branches are also present in the EM simulations, although their frequencies and transport fingerprints are modified by finite$-\beta$ effects.

For the BSE discharge, EM effects again destabilize the largest-scale modes, approximately over $5<n<50$, and modify their real frequencies, which remain relatively small in magnitude and change continuously from the electron- to the ion-diamagnetic direction. As inferred from the fingerprints, this corresponds to a single hybrid branch characterized by $\chi_i/\chi_e\simeq5/4$ and $D_e/\chi_e\simeq1/2$, compatible with hybrid ITG--TEM modes modified by finite$-\beta$ effects. In contrast to the type-I ELM discharge, no KBM destabilization is observed across the edge region considered here. The global domain includes both the maximum of $\alpha_{\mathrm{exp}}$ at $\rhot\simeq0.95$ and the maximum of $\widehat{\beta}_{T_e}$ at $\rhot\simeq0.96$, showing that the absence of a KBM branch does not result from excluding the most strongly driven part of the pedestal. This result confirms and strengthens the previous local analyses around the pedestal top~\cite{dicorato_gyrokinetic_2022,dicorato_turbulent_2024}, indicating that KBMs are not the dominant transport mechanism at the nominal profiles of this BSE discharge across the pedestal region.

At smaller scales, $n>60$, the BSE discharge exhibits a negative-frequency branch that is slightly stabilized by EM effects. These modes have $\chi_i/\chi_e\ll1$ and $D_e/\chi_e\ll1$ and are therefore compatible with ETG modes. Also in this case, they occur at intermediate scales between ion and electron scales, with a binormal wavenumber at the radial position of maximum mode amplitude of approximately $k_y\rho_i\simeq0.5$--$0.6$. As shown in figure~\ref{fig:linear_parity}, this branch also develops tearing-like features. It therefore has characteristics of tearing-parity ETG modes affected by finite$-\beta$ effects, resembling the modes reported in references~\cite{pueschel_microinstabilities_2019,geng_physics_2020}. More generally, recent experimental measurements at DIII-D have also highlighted the role of ETG-driven drift-wave turbulence and its nonlinear interaction with larger-scale edge modes during ELM suppression~\cite{li_multi-scale_2025}.

\subsection{Radial localization}

\begin{figure}[t]
\centering
\includegraphics[width=0.49\textwidth]{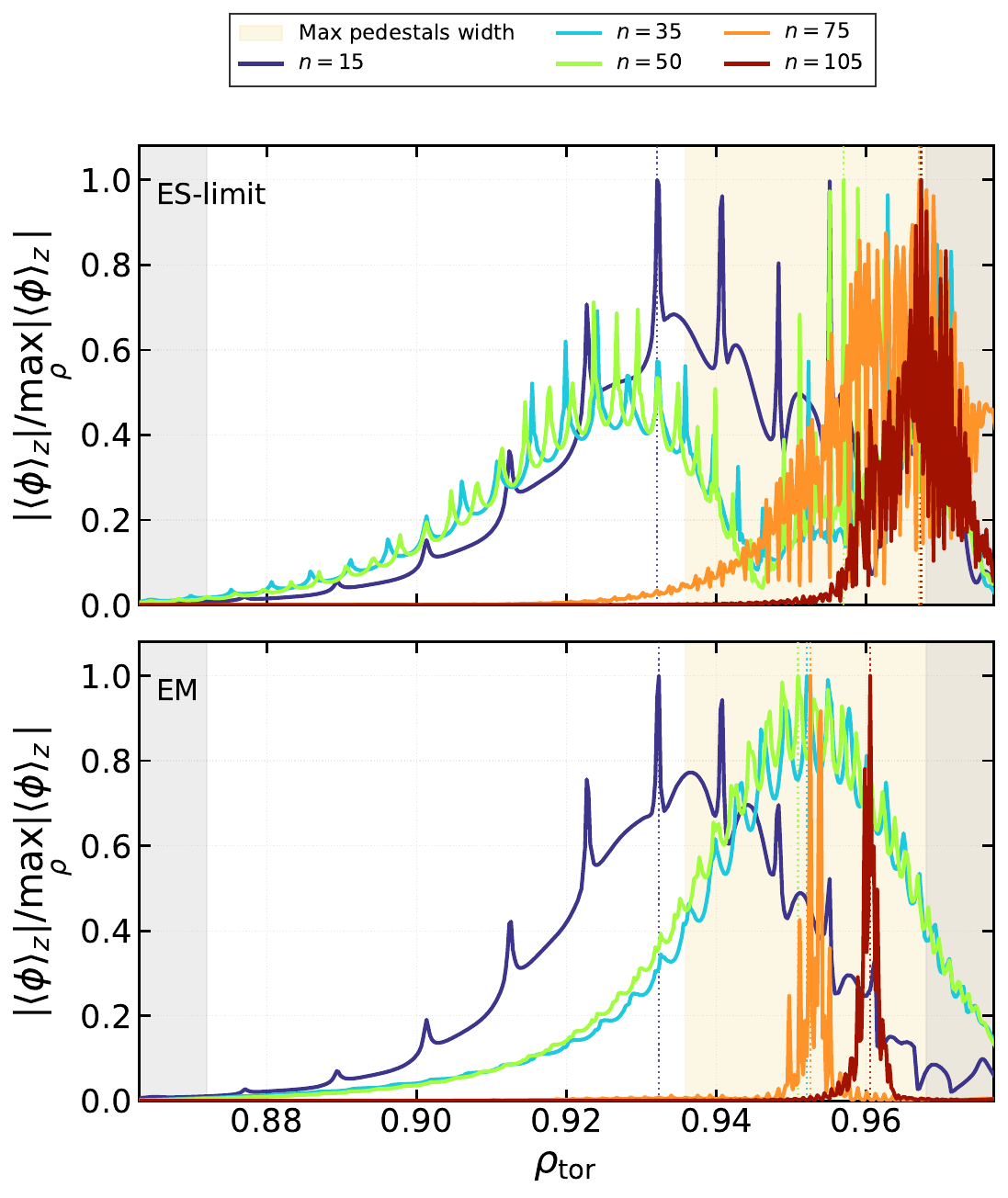}\hfill
\includegraphics[width=0.49\textwidth]{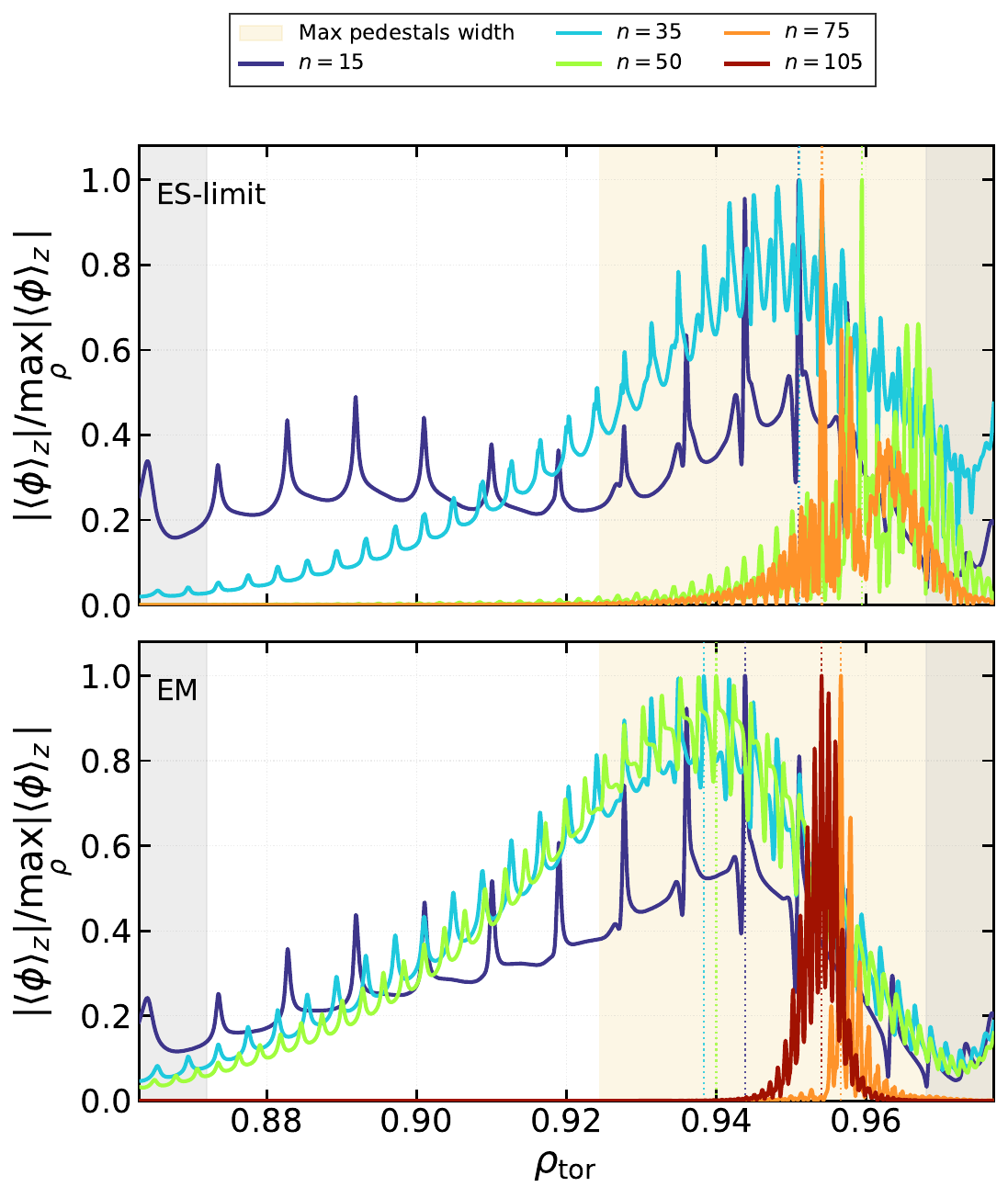}
\makebox[0.49\textwidth][c]{\small (a) Type-I ELM}\hfill
\makebox[0.49\textwidth][c]{\small (b) BSE}
\caption{Normalized radial structures of the Jacobian-weighted, parallel-averaged $\hat{\phi})_1$ for selected toroidal mode numbers in (a) the type-I ELM discharge \#97395 and (b) the BSE discharge \#94442. The upper and lower panels compare the $\mathrm{ES}_l$ and EM simulations, respectively. Each mode is normalized independently. Dotted vertical lines mark the positions of the mode maxima, the yellow regions denote the union of the fitted pedestal intervals, and the grey regions denote the numerical radial buffers.}
\label{fig:linear_phi_structures}
\end{figure}

The effect of EM physics on the global radial structures is shown in figure~\ref{fig:linear_phi_structures} for selected toroidal mode numbers across the linear spectra. For each mode, the normalized radial structure of $\phi_1$ is defined as
\begin{equation}
\phi_n(\rhot) = \frac{\left|\langle\phi_1(\rhot,z;n)\rangle_z\right|} {\displaystyle\max_{\rhot}\left|\langle\phi_1(\rhot,z;n)\rangle_z\right|},
\label{eq:normalized_phi_structure}
\end{equation}
where $\langle\cdots\rangle_z$ denotes the Jacobian-weighted parallel average (see appendix~\ref{app:linear_fingerprints} for definition). We note that the narrow spikes in $\phi_n$ are located at the mode-rational surfaces, where $q(\rhot)=m/n$, and result from the mode structure near surfaces with vanishing parallel wavenumber. The radial structure is particularly relevant in view of the nonlinear coupling among modes localized at different radial positions, which can affect the radial diffusivity profiles and the resulting pedestal structure. In addition, the radial positions of the mode maxima are reported in table~\ref{tab:linear_phi_maxima} and can be compared with the pedestal structures shown in figure~\ref{fig:input_profiles_basic} and table~\ref{tab:pedestal_structure}.

\begin{table*}[t]
\caption{Radial positions $\rhot^{\phi,\max}$ of the maxima of the normalized linear modes structures shown in figure~\ref{fig:linear_phi_structures}.}
\label{tab:linear_phi_maxima}
\centering
\begin{tabular}{c|cc|cc}
\hline
& \multicolumn{2}{c|}{\#97395} & \multicolumn{2}{c}{\#94442} \\
$n$ & $\mathrm{ES}_l$ & EM & $\mathrm{ES}_l$ & EM \\
\hline
15  & 0.932 & 0.932 & 0.951 & 0.944 \\
35  & 0.967 & 0.952 & 0.951 & 0.938 \\
50  & 0.957 & 0.951 & 0.959 & 0.940 \\
75  & 0.967 & 0.952 & 0.954 & 0.957 \\
105 & 0.967 & 0.960 & N.A.  & 0.954 \\
\hline
\end{tabular}
\end{table*}

For the type-I ELM discharge, the largest-scale mode, $n=15$, has a broad radial structure in both the $\mathrm{ES}_l$ and EM simulations, extending over a large fraction of the simulated radial domain and therefore coupling the outer-core region to the pedestal. For $n=35$--$50$, corresponding to the KBM-compatible branch destabilized by EM effects, the EM modes peak around $\rhot^{\mathrm{ped}}$ and extend across most of the pedestal. This differs from the corresponding $\mathrm{ES}_l$ modes, which develop multi-peaked radial structures and do not exhibit the same clear localization across the pedestal. The difference between the $\mathrm{ES}_l$ and EM structures also persists for $n=75$--$105$: in the $\mathrm{ES}_l$ case, these modes tend to peak towards the pedestal bottom, whereas in the EM case they are more radially localized and their maxima are shifted towards the region of stronger drive. For the BSE discharge, the largest-scale mode, $n=15$, also develops a broad radial structure in both the $\mathrm{ES}_l$ and EM simulations and does not peak clearly within the fitted pedestal region. For $n=35$--$50$, the electromagnetic destabilization is accompanied by an inward shift of the mode-structure maxima towards the pedestal-top region, from $\rhot=0.951$--$0.959$ in the $\mathrm{ES}_l$ simulations to $\rhot=0.938$--$0.940$ in the EM simulations. At higher toroidal mode numbers, $n=75$--$105$, corresponding to the intermediate-scale ETG branch, electromagnetic effects modify the radial extent of the structures, which become strongly localized around the steepest-gradient region. The global linear analysis therefore provides evidence that EM effects associated with finite$-\beta$ are relevant to the stability and radial organization of modes over the complete range from the largest to the smaller retained scales in both discharges. 

In summary, the resulting linear picture is as follows: in the type-I ELMy pedestal, the low- and intermediate-$n$ linear spectrum is dominated by a KBM-compatible electromagnetic branch extending across the full pedestal and peaking close to the maximum drive, while the higher-$n$ TEM and intermediate-scale ETG branches are more localized around the strongest-gradient region; in the BSE edge, the low- and intermediate-$n$ linear spectrum is dominated by hybrid ITG--TEM modes modified by finite$-\beta$ effects, whose structures extend from the outer core to the pedestal and peak around the pedestal top, while the higher-$n$ intermediate-scale ETG modes are strongly localized around the maximum-gradient region.

The effect of these instabilities on the nonlinear transport is addressed in the next section. Because the focus of the present work is the edge region, and because structures whose radial extent becomes comparable to the complete simulation domain can be artificially affected by the homogeneous Dirichlet B.C., the minimum nonzero toroidal mode number retained in the nonlinear simulations is $n=5$. This choice minimizes the number of modes that may be artificially influenced by the radial boundaries while retaining the largest-scale structures that can be represented consistently in the present edge domain. Simulations covering a larger radial domain from the core to the edge would be required to capture still lower-$n$ modes consistently and are outside the scope of this work.


\section{Nonlinear turbulent transport: electromagnetic effects and equilibrium $E\times B$ flow}
\label{sec:nonlinear}

In this section, the turbulent transport resulting from the instabilities identified in the linear analysis is investigated by means of nonlinear global simulations. The analysis focuses on the impact of electromagnetic effects and the equilibrium $E\times B$ flow, which are known to be key elements in edge-turbulence dynamics~\cite{scott_drift_2005,scott_nonlinear_2010,burrell_role_2020}.

\subsection{Impact of electromagnetic effects and equilibrium flow on effective diffusivities}

\begin{figure}[t]
\centering
\includegraphics[width=0.43\textwidth]{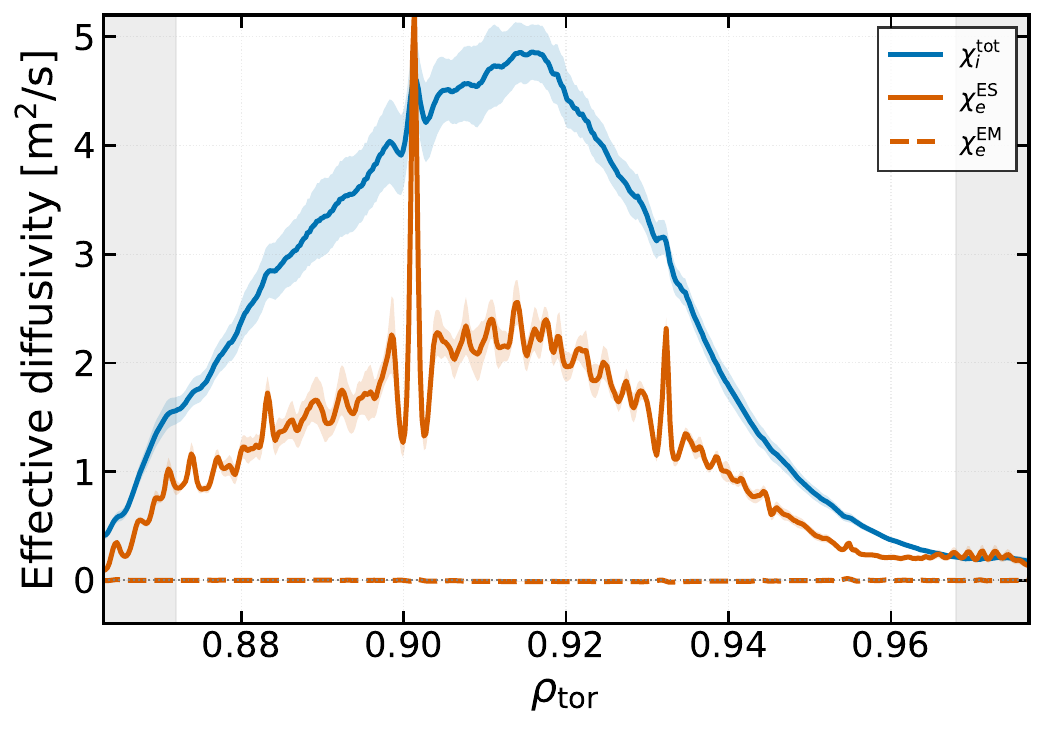}\qquad
\includegraphics[width=0.43\textwidth]{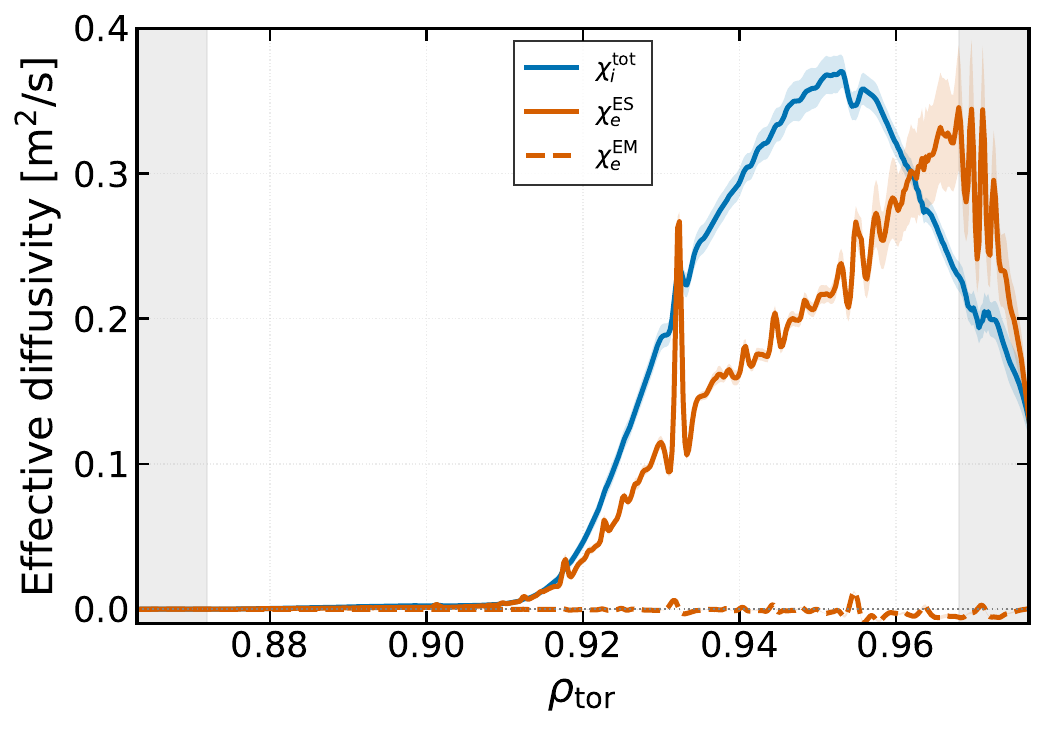}
\makebox[0.43\textwidth][c]{\small (a) $\mathrm{ES}_l$, without $E\times B$}\qquad
\makebox[0.43\textwidth][c]{\small (b) $\mathrm{ES}_l$, with $E\times B$}\\
\includegraphics[width=0.43\textwidth]{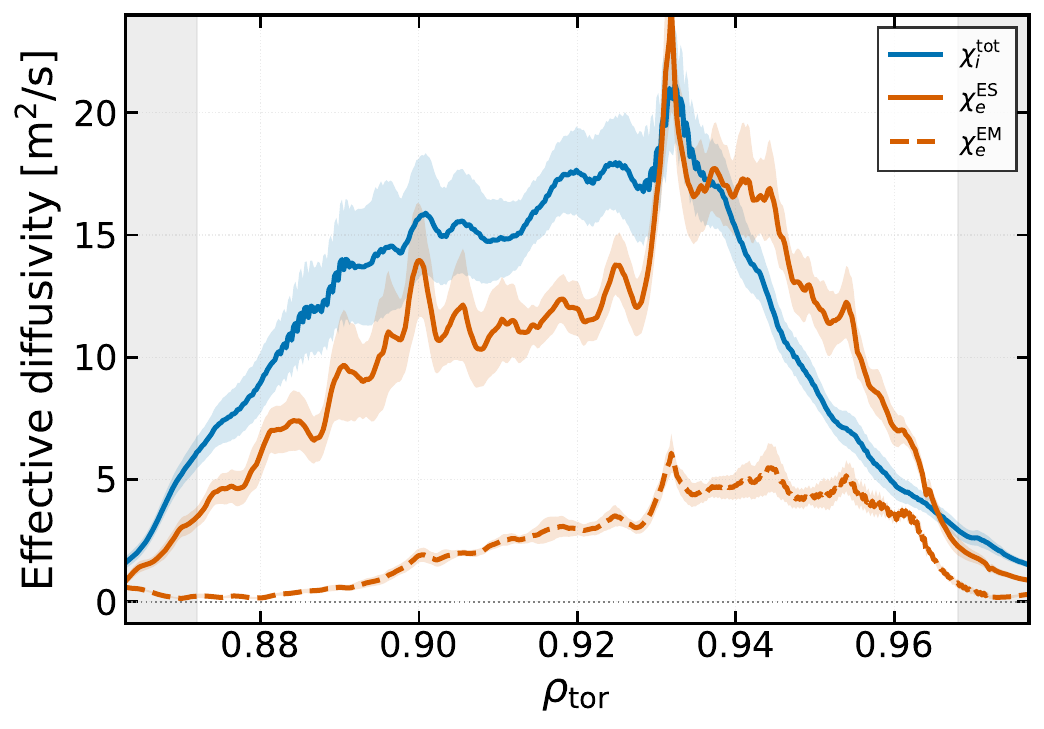}\qquad
\includegraphics[width=0.43\textwidth]{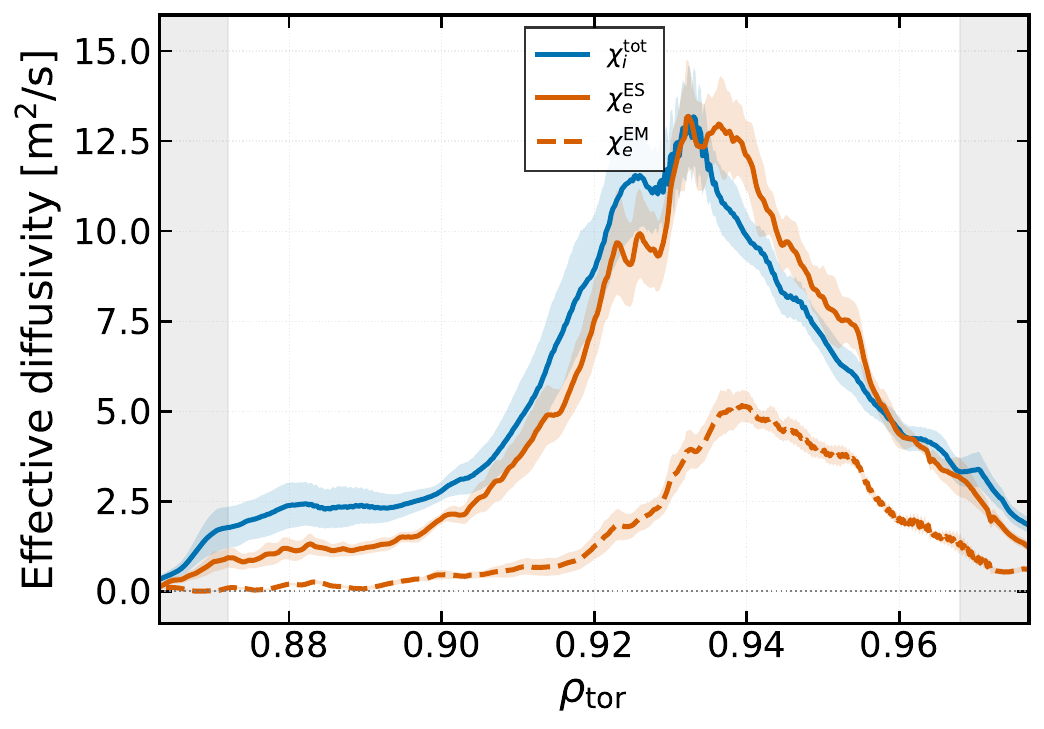}
\makebox[0.43\textwidth][c]{\small (c) EM, without $E\times B$}\qquad
\makebox[0.43\textwidth][c]{\small (d) EM, with $E\times B$}
\caption{Time-averaged radial effective transport coefficients for the type-I ELM pedestal. The four panels compare the $\mathrm{ES}_l$ and EM simulations without and with the nominal equilibrium $E\times B$ flow. The quantities shown are the total ion heat diffusivity $\chi_i^{\mathrm{tot}}$, and the electrostatic and electromagnetic contributions to the electron heat diffusivity, $\chi_e^{\mathrm{ES}}$ and $\chi_e^{\mathrm{EM}}$. Shaded bands denote the statistical uncertainty, while grey regions denote the numerical radial buffers.}
\label{fig:nonlinear_transport_typeI}
\end{figure}

\begin{figure}[t]
\centering
\includegraphics[width=0.43\textwidth]{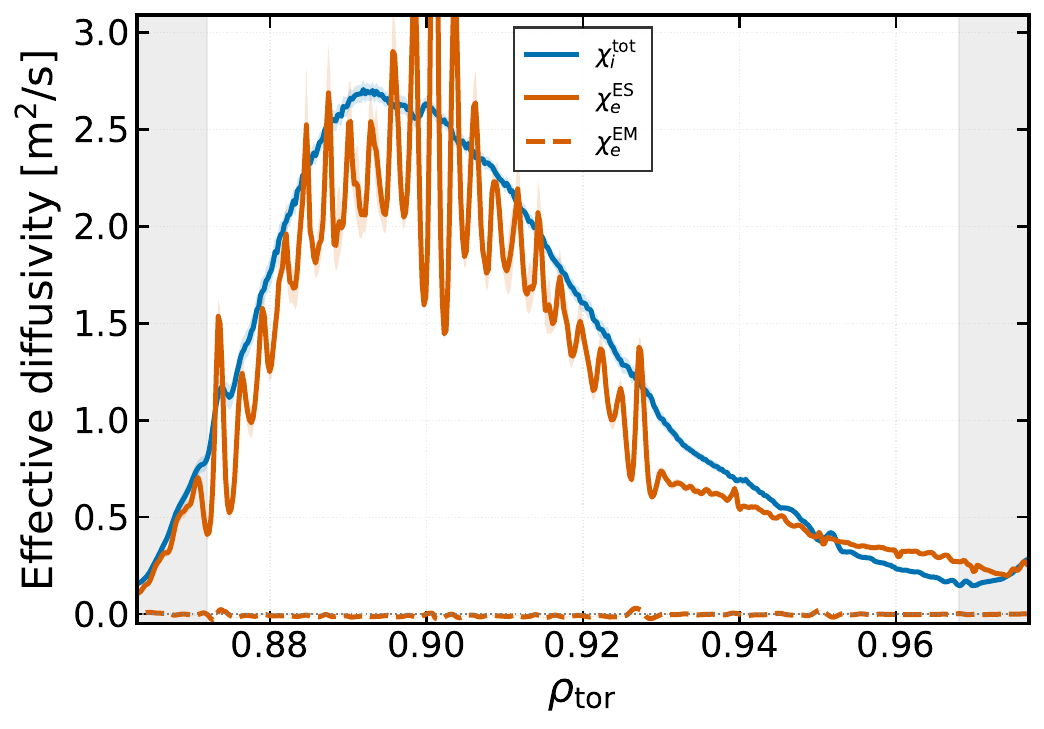}\qquad
\includegraphics[width=0.43\textwidth]{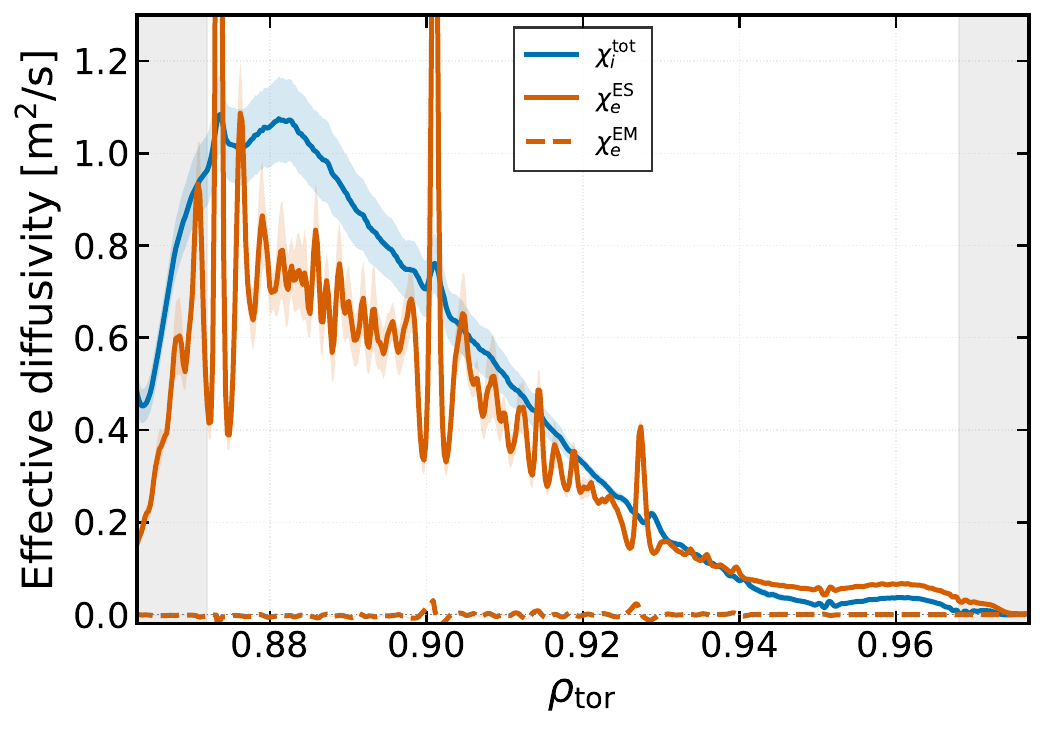}
\makebox[0.43\textwidth][c]{\small (a) $\mathrm{ES}_l$, without $E\times B$}\qquad
\makebox[0.43\textwidth][c]{\small (b) $\mathrm{ES}_l$, with $0.5\cdot E\times B$}\\
\includegraphics[width=0.43\textwidth]{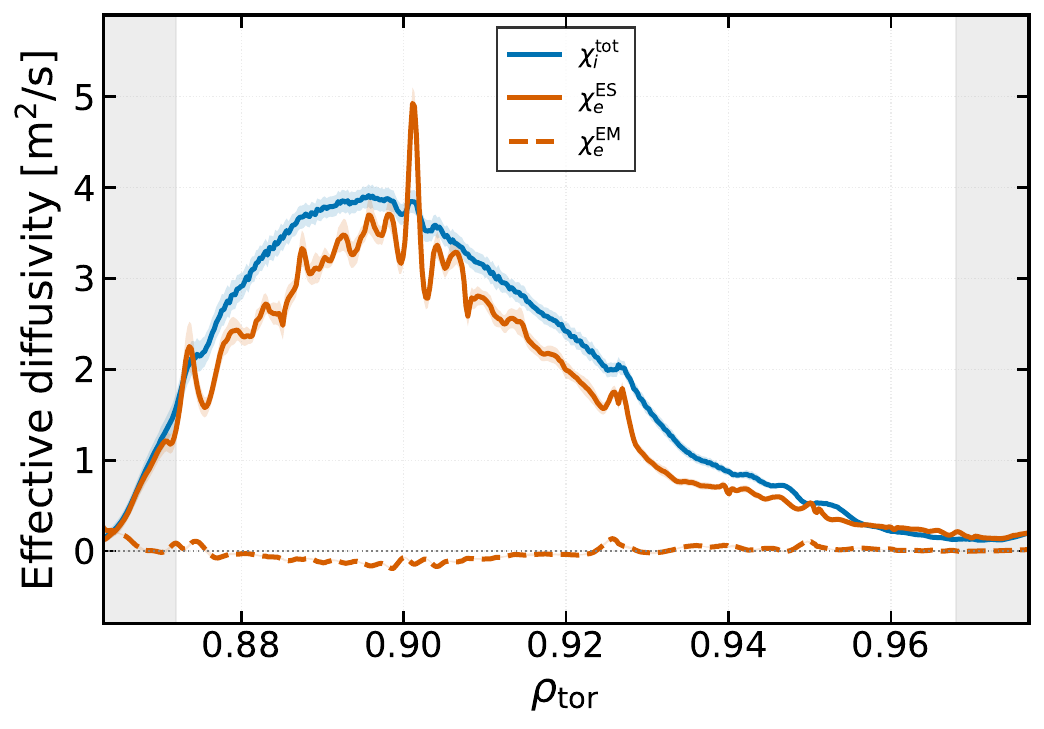}\qquad
\includegraphics[width=0.43\textwidth]{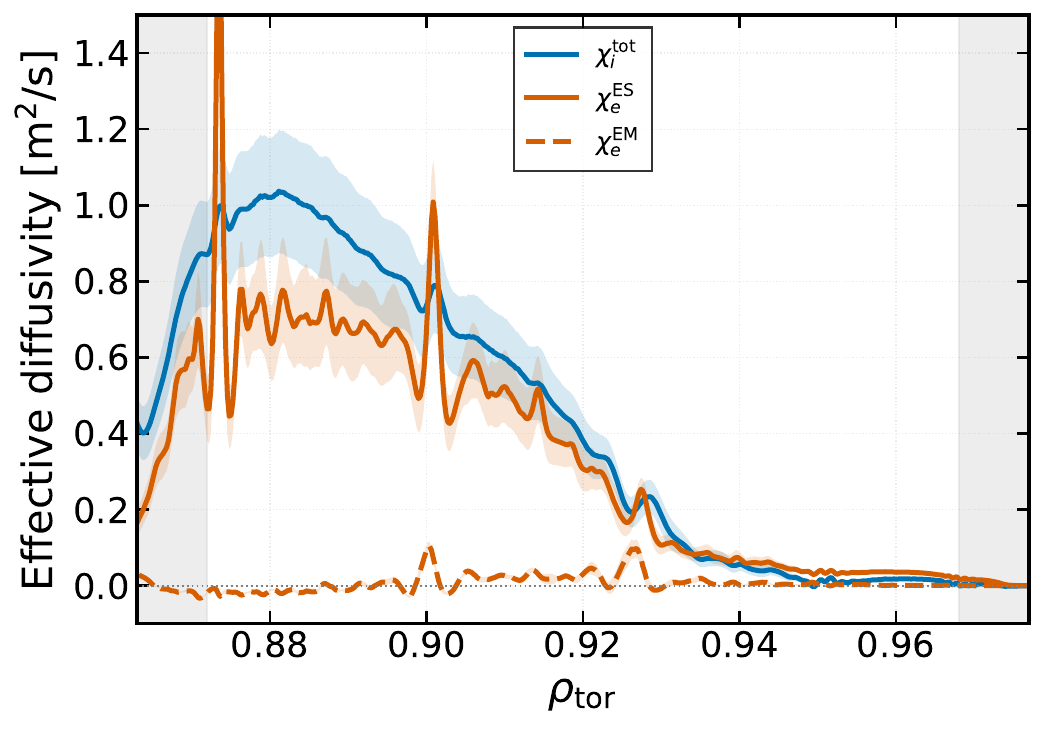}
\makebox[0.43\textwidth][c]{\small (c) EM, without $E\times B$}\qquad
\makebox[0.43\textwidth][c]{\small (d) EM, with $0.5\cdot E\times B$}\\
\includegraphics[width=0.43\textwidth]{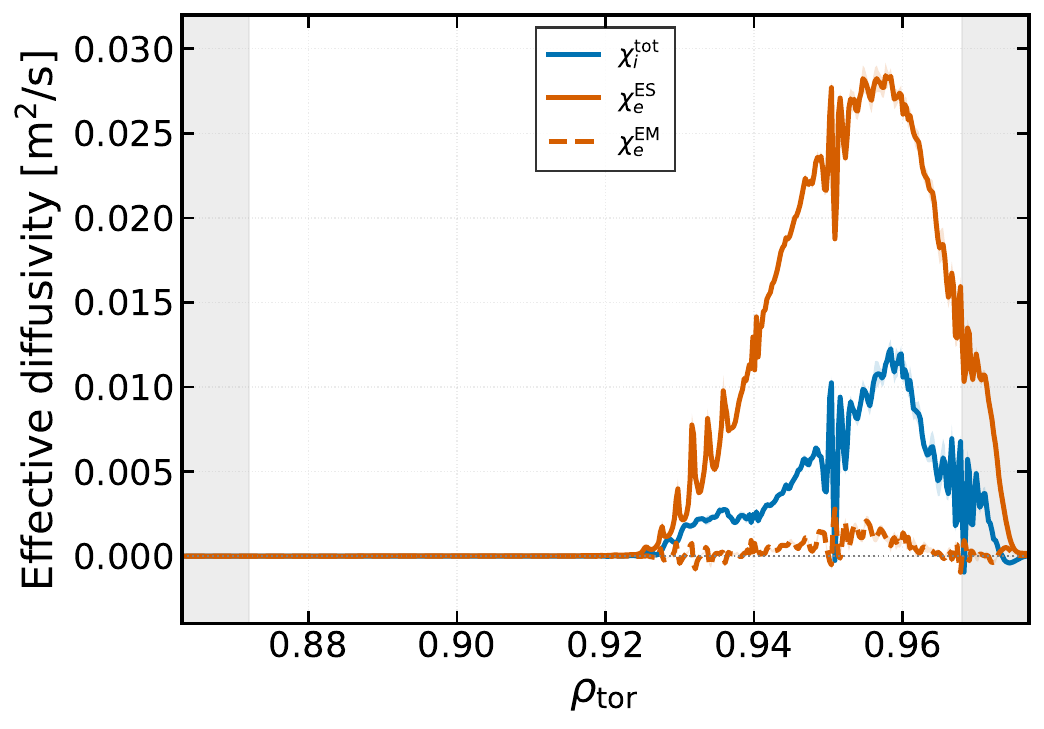}\qquad
\includegraphics[width=0.43\textwidth]{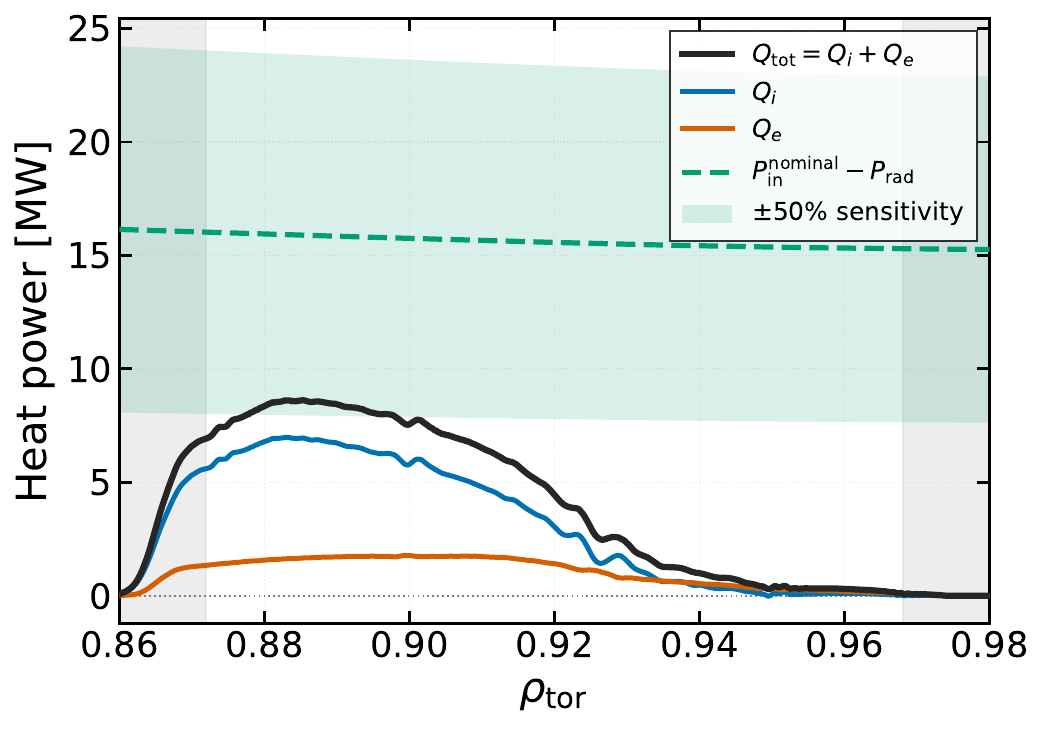}
\makebox[0.43\textwidth][c]{\small (e) EM, with $0.8\cdot E\times B$}\qquad
\makebox[0.43\textwidth][c]{\small (f) EM, with $0.5\cdot E\times B$}
\caption{Time-averaged radial transport for the BSE edge. Panels (a)--(e) show the effective transport coefficients. The simulations without equilibrium $E\times B$ flow and with $f_E=0.5$ are compared; panel (e) shows the EM sensitivity case with $f_E=0.8$. Panel (f) shows the ion, electron, and total turbulent heat powers for the EM simulation with $f_E=0.5$. The heat power estimate $P_{\mathrm{in}}-P_{\mathrm{rad}}(<\rhot)$ is shown and the shaded green region denotes a $\pm 50\%$ sensitivity range. Grey regions denote the numerical radial buffers.}
\label{fig:nonlinear_transport_smallELM}
\end{figure}

\begin{table*}[t]
\caption{Nonlinear radially averaged diffusivity ratios. For simulations with equilibrium $E\times B$ flow, the ratios use $f_E=1$ for the type-I ELM discharge and $f_E=0.5$ for the BSE discharge.}
\label{tab:transport_response}
\centering
\begin{tabular}{llcccc}
\hline
Discharge & Diff. &
$S_{\mathrm{EM}/\mathrm{ES}_l}^{\mathrm{w/o}~E\times B}$ &
$S_{\mathrm{EM}/\mathrm{ES}_l}^{\mathrm{w/}~E\times B}$ &
$S_{(E\times B)/(\mathrm{w/o}~E\times B)}^{\mathrm{ES}_l}$ &
$S_{(E\times B)/(\mathrm{w/o}~E\times B)}^{\mathrm{EM}}$ \\
\hline
Type-I ELM
& $\chi_i$ & $4.64$ & $43.98$& $0.0497$ & $0.471$ \\
& $\chi_e$ & $10.25$ & $76.62$ & $0.0715$ & $0.535$ \\
\hline
BSE
& $\chi_i$ & $1.470$ & $1.009$ & $0.307$ & $0.210$ \\
& $\chi_e$ & $1.399$ & $0.934$ & $0.310$ & $0.207$ \\
\hline
\end{tabular}
\end{table*}

The radial profiles of the time-averaged total effective deuterium heat diffusivity $\chi_i^{\mathrm{tot}}$, the electrostatic and electromagnetic contributions to the electron heat diffusivity, $\chi_e^{\mathrm{ES}}$ and $\chi_e^{\mathrm{EM}}$, are shown in figures~\ref{fig:nonlinear_transport_typeI} and~\ref{fig:nonlinear_transport_smallELM} for the type-I ELM and BSE discharges, respectively. The definitions of the effective transport coefficients are given in appendix~\ref{app:linear_fingerprints}. For completeness, the contributions of the individual toroidal mode numbers to the ion heat transport, $Q_i(\rhot,n)$, are reported in appendix~\ref{app:Qi_maps}. To summarize the comparison quantitatively, the radially averaged diffusivity ratios are defined as
\begin{equation}
S_{\mathrm{EM}/\mathrm{ES}_l}^{j}(\chi_s) = \frac{\langle \chi \rangle_{\mathrm{EM},j}} {\langle\chi\rangle_{\mathrm{ES}_l,j}}, \qquad  j\in\{\mathrm{w/o}~E\times B,\mathrm{w/}~E\times B\},
\label{eq:S_EM}
\end{equation}
and
\begin{equation}
S_{(E\times B)/(\mathrm{w/o}~E\times B)}^{m}(\chi_s) = \frac{\langle\chi\rangle_{\mathrm{w/}~E\times B,m}} {\langle\chi\rangle_{\mathrm{w/o}~E\times B,m}}, \qquad m\in\{\mathrm{ES}_l,\mathrm{EM}\}.
\label{eq:S_Er}
\end{equation}
The resulting values are reported in table~\ref{tab:transport_response}.

Focusing first on the type-I ELM discharge, in the $\mathrm{ES}_l$ simulation, the turbulent transport driven by the linearly identified ITG, TEM, and intermediate-scale ETG branches gives $\chi_i/\chi_e>1$ over most of the radial domain. This turbulence regime is strongly regulated by the equilibrium flow. In the region of weaker instability drive and relatively larger shearing rate, approximately $0.87\lesssim\rhot\lesssim0.92$, the turbulent transport is almost completely suppressed, while the residual transport becomes localized closer to the strong-gradient region, as shown in figure~\ref{fig:nonlinear_transport_typeI}(b).

Introducing EM effects changes the radial distribution of the diffusivities and produces a strong increase in all transport channels, although by different factors. As reported in table~\ref{tab:transport_response}, without equilibrium flow, $S_{\mathrm{EM}/\mathrm{ES}_l}^{\mathrm{w/o}~E\times B}\simeq4.64$ for $\chi_i$, while it reaches approximately $10.25$ for $\chi_e$. This difference is partly associated with the activation of the electromagnetic electron heat-transport channel $\chi_e^{\mathrm{EM}}$ by the KBM-driven turbulence, and the resulting heat-channel ratio becomes $\chi_i/\chi_e\simeq1$ over a large fraction of the edge region. Nevertheless, the explicitly electromagnetic contribution does not account for the complete transport increase: finite$-\beta$ effects also strongly modify the electrostatic contribution to the heat flux within the EM simulation.

The resulting EM turbulence regime is only partially regulated by the nominal equilibrium flow. Averaged over the radial domain, $\chi_i$ and $\chi_e$ are reduced to approximately $47\%$ and $54\%$, respectively, of the corresponding values without equilibrium flow. This stabilization is not radially uniform. In particular, the transport is only weakly reduced over approximately $0.92\lesssim\rhot\lesssim0.96$, corresponding to the region of maximum pressure-gradient drive, minimum $|\omega_{E\times B}^{\mathrm{HB}}|$, and radial localization of the linearly identified KBM modes. The electromagnetic transport driven by this branch is therefore comparatively resilient to the equilibrium flow in the strongly driven pedestal region. The very large values $S_{\mathrm{EM}/\mathrm{ES}_l}^{\mathrm{w/}~E\times B}\simeq43.98$ for $\chi_i$ and $76.62$ for $\chi_e$ result from the different response of the ES$_l$ and EM simulations to the equilibrium flow. Indeed, in the $\mathrm{ES}_l$ case, the equilibrium flow strongly reduces $\chi_i$ and $\chi_e$, whereas the EM turbulence regime is less sensitive to it.

For the BSE discharge, the $\mathrm{ES}_l$ turbulent transport driven by the linearly identified hybrid ITG--TEM branch produces $\chi_i/\chi_e\simeq1$ over most of the radial domain. This turbulence regime is strongly suppressed by the reconstructed equilibrium flow. Sensitivity tests were therefore performed by varying the amplitude of the $E_r$ profile within the associated uncertainty. The principal comparison uses $f_E=0.5$, for which finite turbulent transport is retained. As shown in figure~\ref{fig:nonlinear_transport_smallELM}(b), introducing this reduced-amplitude equilibrium flow strongly modifies the radial diffusivity profiles and reduces both $\chi_i$ and $\chi_e$ by approximately the same factor. In particular, $S_{(E\times B)/(\mathrm{w/o}~E\times B)}^{\mathrm{ES}_l}\simeq0.307$ for $\chi_i$ and $0.310$ for $\chi_e$, corresponding to an average reduction by approximately a factor of three in both heat channels.

Introducing EM effects in the BSE case produces a milder modification of the radial diffusivity profiles than in the type-I ELM case. Without equilibrium flow, $\chi_i$ and $\chi_e$ increase by approximately $47\%$ and $40\%$, respectively, while the explicitly electromagnetic contribution $\chi_e^{\mathrm{EM}}$ remains negligible over most of the radial domain. The resulting turbulence is therefore predominantly electrostatic, but modified by finite$-\beta$ effects. In particular, finite$-\beta$ effects modify the magnitude and radial distribution of the electrostatic flux without reaching the KBM-driven transport regime found in the type-I ELM pedestal.

When the reduced equilibrium flow, $f_E=0.5$, is included in the BSE EM simulation, both heat-transport channels are reduced by approximately a factor of five, with $S_{(E\times B)/(\mathrm{w/o}~E\times B)}^{\mathrm{EM}}\simeq0.210$ for $\chi_i$ and $0.207$ for $\chi_e$. At fixed equilibrium flow, the diffusivity ratios are then close to $1$, with $S_{\mathrm{EM}/\mathrm{ES}_l}^{\mathrm{w/}~E\times B}\simeq1.009$ for $\chi_i$ and $0.934$ for $\chi_e$. Thus, although finite$-\beta$ effects increase transport without equilibrium flow by approximately $40$--$47\%$, the combination of finite$-\beta$ effects and equilibrium flow yields similar radially averaged heat transport in the $\mathrm{ES}_l$ and EM simulations at $f_E=0.5$.

At the larger amplitude, $f_E=0.8$, the BSE turbulent transport is almost completely suppressed, as shown in figure~\ref{fig:nonlinear_transport_smallELM}(e). The radial--spectral decomposition in appendix~\ref{app:Qi_maps} shows that the low-$n$ hybrid ITG--TEM contribution is strongly stabilized and that only a small residual contribution from modes with $n\gtrsim50$ survives. This residual transport peaks over approximately $0.93\lesssim\rhot\lesssim0.97$, corresponding to the radial region in which the intermediate-scale ETG branch was found in the linear analysis. The comparison between $f_E=0.5$ and $f_E=0.8$ therefore demonstrates the strong sensitivity of the two-species BSE turbulent transport to the amplitude of the equilibrium flow.

Finally, the total turbulent heat power $Q_{\mathrm{tot}}=Q_i+Q_e$ for the BSE EM simulation with $f_E=0.5$ is compared with the experimental heat power, estimated as $P_{\mathrm{in}}^{\mathrm{nominal}}-P_{\mathrm{rad}}$, where $P_{\mathrm{in}}^{\mathrm{nominal}}$ is the nominal input power and $P_{\mathrm{rad}}$ is the radiated power. Although a strict power-balance comparison is not attempted, the simulated heat power reaches approximately $8$--$\qty{9}{\mega\watt}$ in the main transport-carrying region and is therefore of the same order as the reference range.

The main nonlinear results can be summarized as follows. The steep pre-type-I ELM pedestal is dominated by an electromagnetic turbulence regime with $\chi_i/\chi_e\simeq1$, for which the equilibrium flow is comparatively ineffective in the most strongly driven region. By contrast, the BSE edge is regulated by a predominantly electrostatic turbulence regime driven by hybrid ITG--TEM modes. Finite$-\beta$ effects modify its radial diffusivity profiles, and the equilibrium flow strongly reduces its transport. In both regimes, the combination of EM effects and the radially varying equilibrium flow is essential for determining not only the magnitude of transport, but also its radial distribution. In contrast to previous analyses of JET pedestals in which MTMs were found to be a dominant electron heat-transport mechanism~\cite{hatch_gyrokinetic_2017,kotschenreuther_gyrokinetic_2019}, no dominant MTM branch is identified in the present two discharges.

\subsection{Radial turbulence organization and zonal-flow response}

To shed light on the differences in the nonlinear transport response, the radial organization of the resulting turbulence regimes is characterized. The complete ion heat-transport distributions $Q_i(\rhot,n)$ are reported in appendix~\ref{app:Qi_maps}. These maps show that EM effects and equilibrium flow modify not only the overall transport amplitude, but also the dominant toroidal mode numbers, radial localization, and radial extent of the transport. In the type-I ELM case, the EM simulations retain broad transport over the pedestal and strong contributions from low and intermediate $n$, whereas in the BSE case the equilibrium flow reduces the low-$n$ contribution and reorganizes the residual transport towards more localized radial regions.

\begin{figure}[t]
\centering
\includegraphics[width=0.45\textwidth]{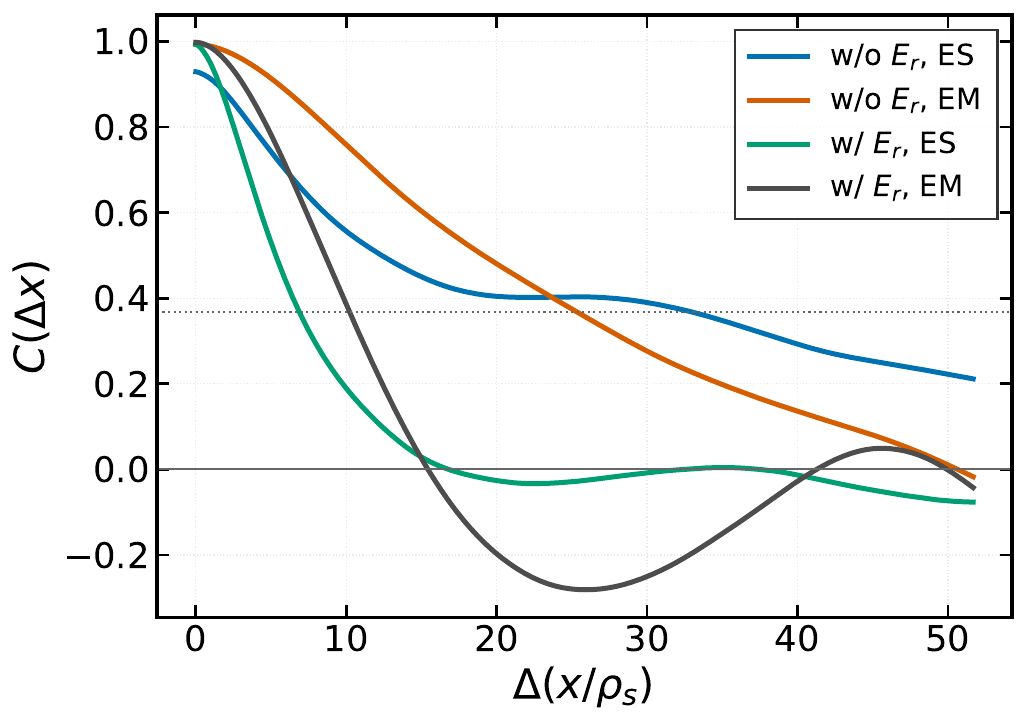}\qquad
\includegraphics[width=0.45\textwidth]{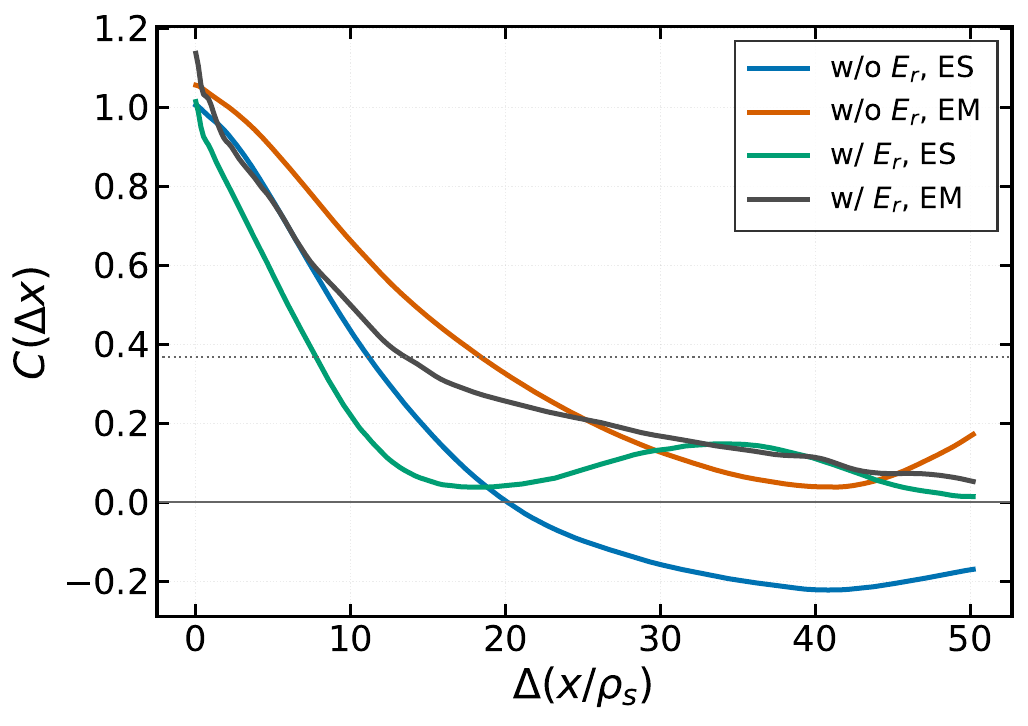}
\makebox[0.45\textwidth][c]{\small (a) Type-I ELM}\qquad
\makebox[0.45\textwidth][c]{\small (b) BSE}
\caption{Radial autocorrelation of the nonlinear ion heat-flux fluctuations for (a) the type-I ELM and (b) the BSE. The $\mathrm{ES}_l$ and EM simulations without and with the equilibrium $E\times B$ flow are compared. The horizontal dashed line corresponds to $C=1/e$.}
\label{fig:radial_autocorrelation}
\end{figure}

The radial organization is quantified using the autocorrelation of the ion heat-flux fluctuations $Q_i(x,t)$ as a proxy for the turbulent structures. The signal is standardized independently at each radius according to
\begin{equation}
q(t,x) = \frac{Q_i(t,x)-\langle Q_i(t,x)\rangle_t}{\sigma_t[Q_i(t,x)]},
\label{eq:standardized_heat_flux}
\end{equation}
and the radial correlation is defined as
\begin{equation}
C(\Delta x_k) = \left\langle q(t,x_i)q(t,x_{i+k}) \right\rangle_{t,i}, \qquad \Delta x_k=k\Delta x.
\label{eq:radial_correlation}
\end{equation}
Starting from this quantity, the integral radial correlation length is defined as
\begin{equation}
\lambda_{\mathrm{int}} = \int_0^{\Delta x_0} C(\Delta x)\,\mathrm{d}\Delta x,
\label{eq:lambda_integral}
\end{equation}
where $\Delta x_0$ is the first zero crossing of $C(\Delta x)$, or the largest retained radial lag if no zero crossing occurs.

The corresponding results are shown in figure~\ref{fig:radial_autocorrelation} and summarized in table~\ref{tab:turbulence_organization}. For the type-I ELM discharge, the $\mathrm{ES}_l$ and EM simulations without equilibrium flow develop turbulent structures with large radial coherence across the domain, with $\lambda_{\mathrm{int}}=22.43\rho_s$ and $21.65\rho_s$, respectively. Introducing the nominal equilibrium flow produces a strong reduction of the radial coherence in both field models, yielding $\lambda_{\mathrm{int}}=6.16\rho_s$ in the $\mathrm{ES}_l$ simulation, and $8.52\rho_s$ in the EM simulation, consistently with the shearing apart of radially extended turbulent structures. In the EM simulation with equilibrium flow, despite the strongly reduced time-averaged radial coherence, the resulting effective diffusivities remain distributed over a relatively broad radial domain, as shown in figure~\ref{fig:nonlinear_transport_typeI}(d). This differs from the $\mathrm{ES}_l$ case, for which the reduction in radial coherence is also reflected in the much stronger radial localization of the residual diffusivities shown in figure~\ref{fig:nonlinear_transport_typeI}(b).

A different trend is observed for the BSE discharge. In the absence of equilibrium flow, the $\mathrm{ES}_l$ and EM simulations develop turbulent structures with significantly different radial coherence: $\lambda_{\mathrm{int}}$ increases from $9.41\rho_s$ in the $\mathrm{ES}_l$ case to $17.31\rho_s$ in the EM case, i.e. by almost a factor of $2$. Introducing the equilibrium flow modifies the radial correlation less abruptly than in the type-I ELM turbulence. In the $\mathrm{ES}_l$ simulation, $\lambda_{\mathrm{int}}$ remains almost unchanged, decreasing only from $9.41\rho_s$ to $9.35\rho_s$, whereas in the EM simulation it decreases from $17.31\rho_s$ to $15.31\rho_s$. Therefore, in the BSE case the main separation of the radial-correlation lengths is associated with EM effects rather than with the equilibrium flow. 

These results quantitatively demonstrate a non-trivial modification of the turbulence organization by the interplay between EM effects and the equilibrium $E\times B$ flow, with a response that depends on the edge turbulence regime.

\begin{figure}[t]
\centering
\includegraphics[width=0.45\textwidth]{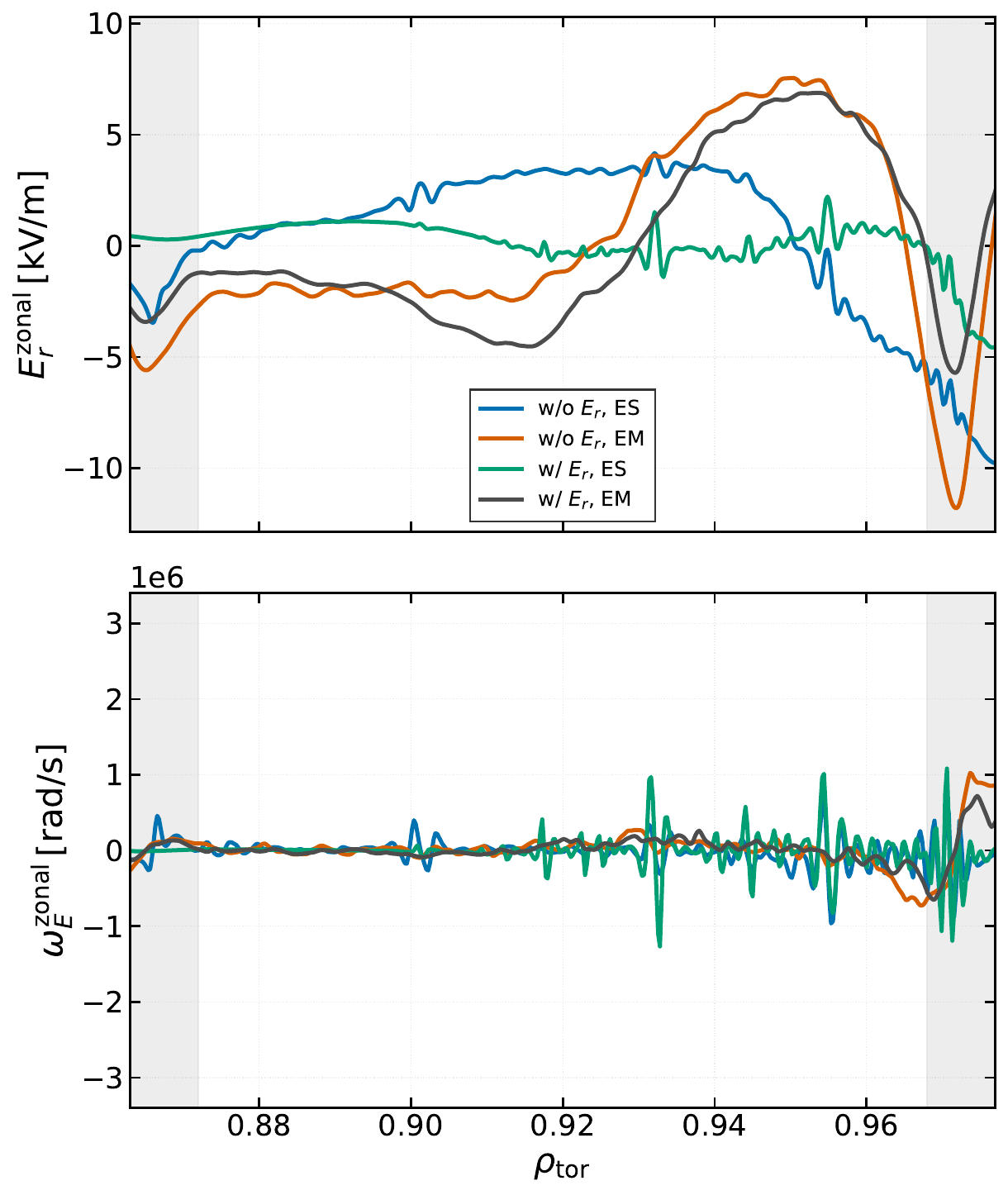}\qquad
\includegraphics[width=0.45\textwidth]{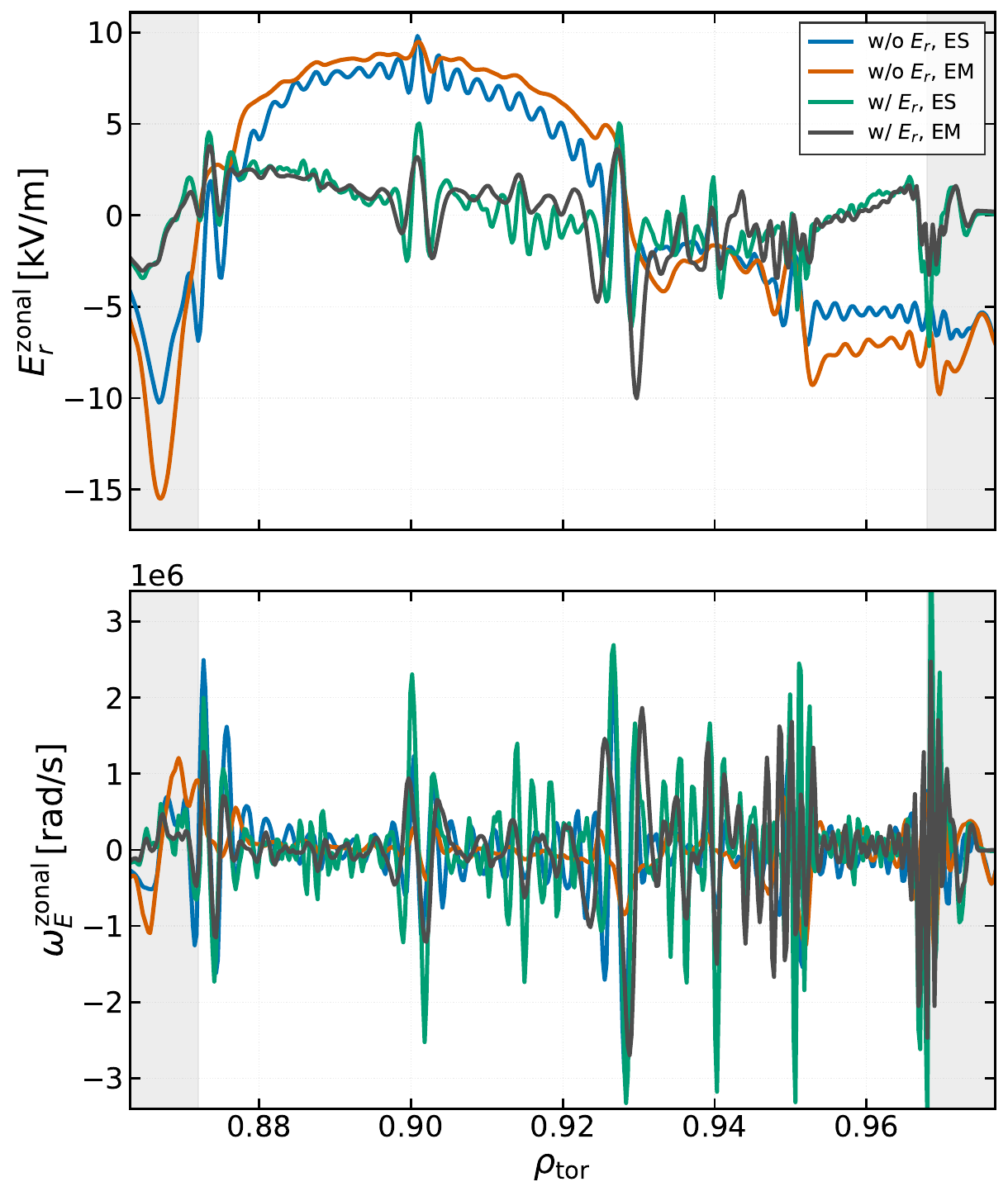}
\makebox[0.45\textwidth][c]{\small (a) Type-I ELM}\qquad
\makebox[0.45\textwidth][c]{\small (b) BSE}
\caption{Turbulence-generated zonal radial electric field and zonal $E\times B$ shearing rate for (a) the type-I ELM and (b) the BSE. The $\mathrm{ES}_l$ and EM simulations without and with the equilibrium flow are compared. The zonal quantities contain only the turbulence-generated axisymmetric contribution and are distinct from the equilibrium field. Grey regions denote the numerical radial buffers.}
\label{fig:zonal_Er_shear}
\end{figure}

Finally, to investigate the modification of turbulence self-regulation by zonal flows~\cite{diamond_zonal_2005}, and their role in the saturation of the two turbulence regimes, the zonal radial electric field and zonal shearing rate are shown in figure~\ref{fig:zonal_Er_shear}. They are defined as
\begin{equation}
E_r^{\mathrm{ZF}} = -\frac{\partial\phi_0}{\partial x}, \qquad v_E^{\mathrm{ZF}} = -\frac{E_r^{\mathrm{ZF}}}{C_{xy}}, \qquad \omega_E^{\mathrm{ZF}} = -\frac{\partial v_E^{\mathrm{ZF}}}{\partial x},
\label{eq:zonal_shearing}
\end{equation}
where $\phi_0(t,x)=\langle\phi(t,x,k_y=0,z)\rangle_z$ is the flux-surface-averaged zonal potential, $\langle\cdots\rangle_z$ is the Jacobian-weighted parallel average (see appendix~\ref{app:linear_fingerprints}), and $C_{xy}$ is a GENE geometrical factor (see appendix~\ref{app:Er}). In addition, table~\ref{tab:turbulence_organization} reports reduced metrics characterizing the zonal-shear amplitude and temporal autocorrelation, which are relevant to its effectiveness in regulating turbulence \cite{kim_random_2004,singh_how_2026}. Defining
\begin{equation}
\overline{\omega}_E(x) = \left\langle\omega_E^{\mathrm{ZF}}(t,x)\right\rangle_t,
\qquad
\omega'_E(t,x) = \omega_E^{\mathrm{ZF}}(t,x)-\overline{\omega}_E(x),
\end{equation}
the local temporal autocorrelation is
\begin{equation}
C_E(x,\tau) = \frac{\displaystyle\sum_t \omega'_E(t,x)\omega'_E(t+\tau,x)} {\displaystyle\sum_t[\omega'_E(t,x)]^2}.
\end{equation}
The global RMS zonal shearing rate, coherence time, and shearing Kubo number
reported in table~\ref{tab:turbulence_organization} are defined as
\begin{align}
\omega_{\mathrm{ZF,rms}} &= \left[ \left\langle \left\langle[\omega_E^{\mathrm{ZF}}(t,x)]^2\right\rangle_t \right\rangle_x \right]^{1/2}, \\
\overline{C}_E(\tau) &= \left\langle C_E(x,\tau)\right\rangle_x, \\
\tau_C &= \min\{\tau>0:\overline{C}_E(\tau)=e^{-1}\}, \\
K_{\mathrm{ZF}} &= \omega_{\mathrm{ZF,rms}}\tau_C .
\label{eq:zonal_persistence_local}
\end{align}
Here, the brackets $\langle\cdots\rangle_x$ denote a buffer-excluded radial average. The reduced value $K_{\mathrm{ZF}}$ combines the total RMS zonal-shear amplitude with the coherence time of its fluctuating component and is used as a comparative persistence diagnostic.

\begin{table*}[t]
\caption{Diagnostics for nonlinear turbulence organization. $\lambda_{\mathrm{int}}$ is the integral radial correlation length of the ion heat-flux fluctuations in units of $\rho_s$, $\omega_{\mathrm{ZF,rms}}^{\mathrm{SI}}$ is the global RMS zonal shearing rate, and shearing Kubo number $K_{\mathrm{ZF}}=\omega_{\mathrm{ZF,rms}}\tau_C$ is the dimensionless zonal-shear persistence parameter.}
\label{tab:turbulence_organization}
\centering
\begin{tabular}{l|rl|ccc}
\hline
Edge profile & Model & $E\times B$ flow & $\lambda_{\mathrm{int}}$ & $\omega_{\mathrm{ZF,rms}}^{\mathrm{SI}}$ & $K_{\mathrm{ZF}}$ \\
\hline
Type-I ELM & $\mathrm{ES}_l$ & without & 22.43 & $3.00\times10^{5}$ & 2.63 \\
Type-I ELM & EM & without & 21.65 & $2.91\times10^{5}$ & 0.62 \\
Type-I ELM & $\mathrm{ES}_l$ & with & 6.16 & $4.28\times10^{5}$ & 6.93 \\
Type-I ELM & EM & with & 8.52 & $2.63\times10^{5}$ & 0.65 \\
\hline
BSE & $\mathrm{ES}_l$ & without & 9.41 & $9.61\times10^{5}$ & 7.10 \\
BSE & EM & without & 17.31 & $6.10\times10^{5}$ & 2.66 \\
BSE & $\mathrm{ES}_l$ & with & 9.35 & $1.44\times10^{6}$ & 15.65 \\
BSE & EM & with & 15.31 & $1.26\times10^{6}$ & 29.21 \\
\hline
\end{tabular}
\end{table*}

For the type-I ELM discharge without equilibrium flow, the $\mathrm{ES}_l$ and EM simulations develop significantly different radial structures of $E_r^{\mathrm{ZF}}$. In the EM case, the zonal field reaches large amplitudes in the radial region corresponding to the strongest KBM, although the RMS zonal shearing rates are similar: $\omega_{\mathrm{ZF,rms}}^{\mathrm{SI}}=3.00\times10^5~\mathrm{rad\,s^{-1}}$ in the $\mathrm{ES}_l$ simulation and $2.91\times10^5~\mathrm{rad\,s^{-1}}$ in the EM simulation. The main difference concerns the temporal coherence: $K_{\mathrm{ZF}}$ decreases from $2.63$ in the $\mathrm{ES}_l$ case to $0.62$ in the EM case. This reduced persistence accompanies the strong increase in the effective diffusivities produced by electromagnetic effects.

Introducing the equilibrium flow into the type-I ELM $\mathrm{ES}_l$ simulation reduces the amplitude of $E_r^{\mathrm{ZF}}$, consistently with the reduction of the turbulence that drives the zonal potential. Nevertheless, the corresponding RMS zonal shearing rate increases from $3.00\times10^5$ to $4.28\times10^5~\mathrm{rad\,s^{-1}}$, while $K_{\mathrm{ZF}}$ increases from $2.63$ to $6.93$. This response is consistent with a positive feedback between the equilibrium flow and the turbulence-generated zonal shear in the $\mathrm{ES}_l$ regime. By contrast, in the EM turbulence regime driven by the KBM branch, the equilibrium flow has only a weak effect: $\omega_{\mathrm{ZF,rms}}^{\mathrm{SI}}$ changes from $2.91\times10^5$ to $2.63\times10^5~\mathrm{rad\,s^{-1}}$, and $K_{\mathrm{ZF}}$ remains close to $0.6$. The zonal shear therefore remains substantially less persistent than in the corresponding $\mathrm{ES}_l$ simulation.

In the BSE simulations, the zonal structures extend across the whole radial domain, in contrast to the stronger localization towards the steep-gradient region observed in the type-I ELM case. Without equilibrium flow, the $\mathrm{ES}_l$ and EM simulations develop zonal flows with similar large-scale radial shapes but different shearing-rate patterns. Electromagnetic effects increase the radial size of the shearing structures while reducing their amplitude: $\omega_{\mathrm{ZF,rms}}^{\mathrm{SI}}$ decreases from $9.61\times10^5$ to $6.10\times10^5~\mathrm{rad\,s^{-1}}$, corresponding to a reduction of approximately $37\%$. At the same time, the temporal coherence decreases by approximately a factor of $1.7$, and $K_{\mathrm{ZF}}$ decreases by more than a factor of $2$. These results indicate less persistent zonal-flow regulation in the EM simulation than in the corresponding $\mathrm{ES}_l$.

Introducing the equilibrium flow into the BSE $\mathrm{ES}_l$ simulation again reduces the amplitude of the zonal $E_r$ structures while increasing the associated shearing rate from $9.61\times10^5$ to $1.44\times10^6~\mathrm{rad\,s^{-1}}$. The persistence measure increases from $K_{\mathrm{ZF}}=7.10$ to $15.65$, again indicating a positive feedback between the equilibrium flow and turbulence self-regulation in the $\mathrm{ES}_l$ regime. A different response is obtained in the BSE EM simulation. Relative to the corresponding case without equilibrium flow, the RMS zonal shearing rate approximately doubles, while its temporal coherence also increases. Consequently, $K_{\mathrm{ZF}}$ rises by approximately one order of magnitude. The resulting picture therefore suggests that the interplay between finite$-\beta$ effects and the equilibrium $E\times B$ flow strongly modifies the nonlinear self-regulation mechanisms in the BSE turbulence regime.

A direct attribution of these changes to a specific zonal-flow generation or damping mechanism is not possible from the present diagnostics alone. In particular, demonstrating electromagnetic depletion or enhancement of zonal flows would require a direct balance of Reynolds and Maxwell stresses, together with the neoclassical, parallel-streaming, and trapping contributions to the zonal-flow evolution equation. Such an analysis is left for future work.

\begin{figure}[t]
\centering
\includegraphics[width=0.94\textwidth]{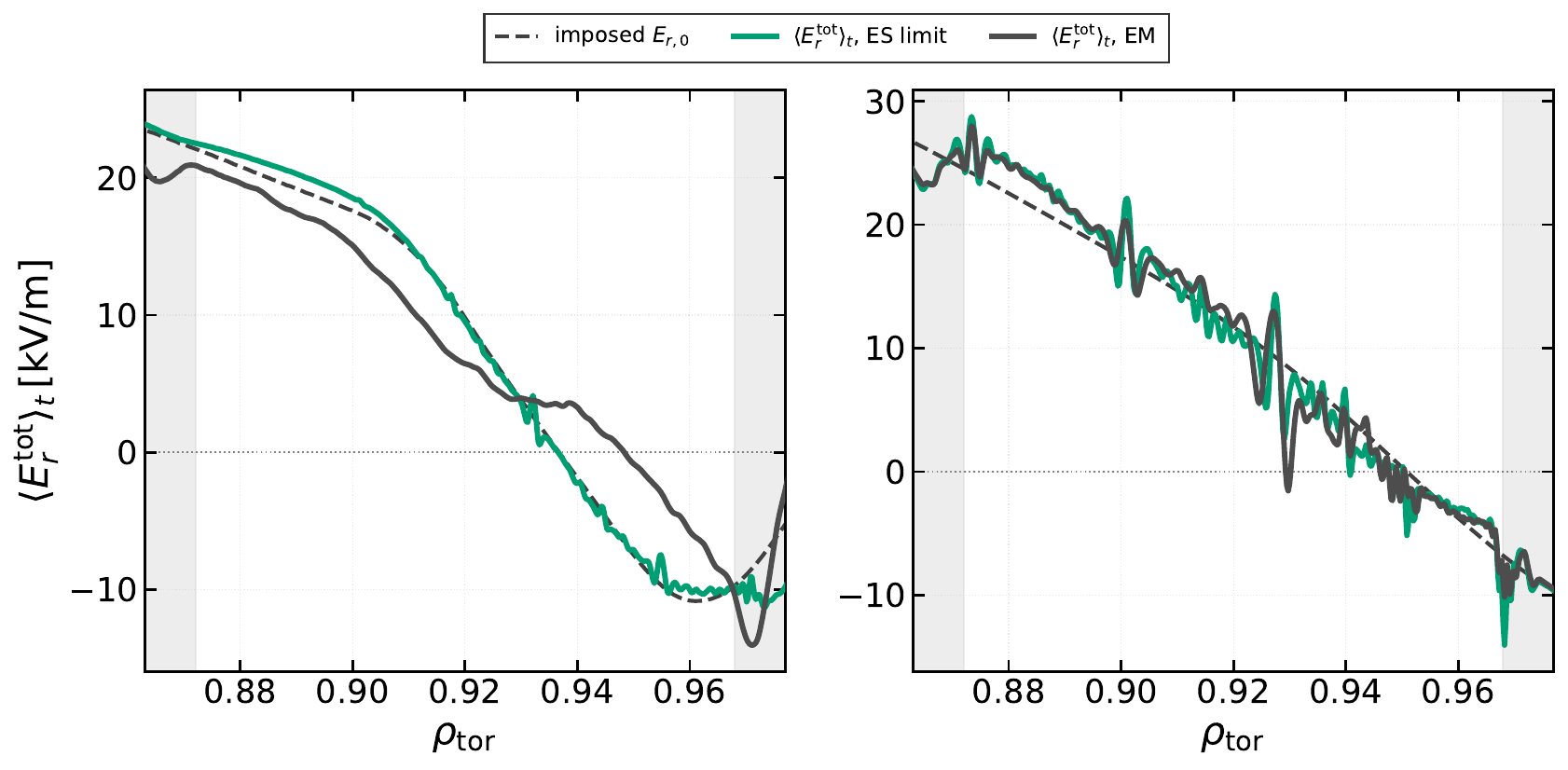}\par
\makebox[0.55\textwidth][c]{\small (a) Type-I ELM}\hfill
\makebox[0.45\textwidth][c]{\small (b) BSE}
\caption{Time-averaged total axisymmetric radial electric field for (a) the type-I ELM and (b) the BSE. Solid curves show the sum of the equilibrium $E_r$ and the time-averaged turbulence-generated zonal field for the $\mathrm{ES}_l$ and EM simulations. Dashed curves show the equilibrium $E_r^{\mathrm{ext}}=f_EE_r^{\mathrm{FB}}$. Grey regions denote the numerical radial buffers.}
\label{fig:total_Er}
\end{figure}

Finally, the total axisymmetric radial electric field resulting from the simulations with equilibrium flow is shown in figure~\ref{fig:total_Er}. Because the equilibrium field is stationary, its time average is
\begin{equation}
\left\langle E_r^{\mathrm{tot}}(\rhot,t) \right\rangle_t = E_r^{\mathrm{ext}}(\rhot) + \left\langle E_r^{\mathrm{ZF}}(\rhot,t) \right\rangle_t.
\label{eq:total_Er}
\end{equation}

For the type-I ELM discharge, $\langle E_r^{\mathrm{tot}}\rangle_t$ is only weakly modified relative to the equilibrium field in the $\mathrm{ES}_l$ simulation. In the EM simulation, the stronger turbulence-generated zonal field reduces the amplitude of the $E_r$ over part of the domain and therefore locally opposes it. The resulting total field is also radially shifted and does not develop the same clear well within the considered domain. For the BSE discharge, the zonal contribution produces a weaker modification of the large-scale equilibrium $E_r$ profile but introduces additional staircase-like radial structures that differ between the $\mathrm{ES}_l$ and EM simulations. Therefore, the time-averaged zonal field locally reinforces or opposes the equilibrium $E_r$ and introduces additional radial structure into the total axisymmetric $E_r$.


\section{Discussion and conclusions}
\label{sec:conclusion}

The turbulent transport mechanisms in the edge of JET plasmas with the Be wall and W divertor, featuring different ELM dynamics and pedestal structures, have been investigated by means of global linear and nonlinear gyrokinetic simulations. Through a direct comparison, we have identified the different roles played by finite$-\beta$ effects and the equilibrium $E\times B$ flow, which may contribute to the experimentally observed differences between the two regimes.

At low-$n$, turbulence in the type-I ELM regime is driven by a KBM branch destabilized at the nominal plasma $\beta$, whereas the same spectral range in the BSE case is dominated by hybrid ITG--TEM modes destabilized and modified by finite$-\beta$ effects. These results are consistent with the established role of pressure-gradient-driven ballooning transport in conventional type-I ELM pedestals~\cite{snyder_development_2009,urano_pedestal_2014} and identify a different transport mechanism in the BSE edge. The resulting linear picture is also consistent with previous local simulations of pedestal-top turbulence, which showed that the BSE profiles remain below the KBM threshold because of their lower pressure and $\beta$~\cite{dicorato_gyrokinetic_2022,dicorato_turbulent_2024}. The difference persists at intermediate ion--electron scales, $70<n<120$, where TEM and intermediate-scale ETG branches are found in the type-I ELM case, whereas modes compatible with ETGs and exhibiting mixed parity appear in the BSE case. In contrast to previous gyrokinetic studies of JET pedestals~\cite{hatch_gyrokinetic_2017,kotschenreuther_gyrokinetic_2019}, MTMs are not found to be dominant. In both discharges, EM effects modify the growth rates and frequencies, transport fingerprints, and radial mode structures, supporting their relevance beyond the destabilization of the KBM branch.

Consistently with the linear picture, EM effects produce different nonlinear turbulence regimes in the two discharges. In the type-I ELM case, a strong transport enhancement associated with KBM-driven turbulence is obtained, with a finite magnetic-flutter contribution to the electron heat transport. In the BSE regime, EM effects instead produce a moderate transport increase without activating another dominant transport channel. The relevance of EM effects is consistent with previous studies of edge turbulence~\cite{scott_drift_2005,scott_nonlinear_2010}, since both discharges reach $\widehat{\beta}_{T_e}=\beta_e(qR/L_{T_e})^2>1$ in the steep-gradient region. The effectiveness of the equilibrium $E\times B$ flow as a turbulence-suppression mechanism~\cite{hahm_flow_1995,burrell_role_2020} is found to depend on the EM character of the turbulence, while generally modifying the radial diffusivity profiles. In the weakly electromagnetic BSE regime, similarly to the $\mathrm{ES}_l$ case, a sufficiently large equilibrium $E\times B$ flow strongly suppresses the ion-scale turbulence, leaving only a weak residual contribution associated mainly with the intermediate-scale ETG branch. In the type-I ELM case, by contrast, a clear separation appears between the EM and $\mathrm{ES}_l$ simulations: the KBM-driven turbulence is only weakly stabilized and maintains strong transport across the pedestal, with $\chi_i\simeq\chi_e$. This change in the transport response when the KBM branch is accessed supports pressure-gradient-driven electromagnetic turbulence as a candidate mechanism for regulating the type-I ELM pedestal.

Concerning the saturation physics, the analysis of the interplay between zonal-flow self-regulation and the equilibrium $E\times B$ flow reveals complex dynamics~\cite{diamond_zonal_2005,kim_random_2004}. In the type-I ELM regime, the zonal-flow response is more sensitive to the KBM-compatible turbulence than to the equilibrium flow, highlighting the relevance of turbulence self-regulation to saturation. A more complex picture emerges for the BSE regime, where the simultaneous presence of EM effects and equilibrium $E\times B$ flow produces a spatio-temporal reorganization of the zonal flows that increases their combined amplitude and persistence. Further analysis of the physical origin of these differences, including Reynolds- and Maxwell-stress and nonlinear free-energy-transfer diagnostics, is required. Compared with the previous local analysis at the BSE pedestal top~\cite{dicorato_turbulent_2024}, a different balance between EM and equilibrium-flow stabilization is found. This raises questions concerning the applicability of the local approximation in the steep-gradient region, although the difference cannot be attributed uniquely to radial non-locality because the local and global simulations also differ in their equilibrium flow implementation, species content, sources, and boundary conditions.

The absolute turbulent heat power obtained for the BSE EM simulation with $f_E=0.5$ is of the same order as the estimated available power. Although this does not constitute a quantitative power-balance validation, it supports the physical relevance of the simulated transport levels. For the type-I ELM discharge, the profiles represent the high-gradient state immediately before an ELM crash; the strong simulated transport is therefore interpreted as the response of a highly unstable pre-ELM pedestal. The simulation results are thus physically relevant for interpreting the observed pedestal structures. In the type-I ELM regime, the dominant KBM-driven turbulence is not readily stabilized by the equilibrium $E\times B$ flow and produces $\chi_i\simeq\chi_e$ across the fully developed pedestal. It may therefore contribute to limiting both the $T_i$ and $T_e$ pedestal profiles before the type-I ELM is triggered. At the same time, the pre-ELM pedestal approaches the peeling--ballooning stability boundary, providing the conventional mechanism for the type-I ELM crash~\cite{snyder_edge_2002,snyder_pedestal_2009}. In the BSE regime, a different picture emerges: the retained ion-scale turbulence remains below the KBM threshold and is efficiently stabilized by the equilibrium $E\times B$ flow. Since the retained ion and electron
heat-transport channels are reduced by similar relative factors, their response alone does not explain $p_i/p_e>1$. Additional electron heat transport from the unresolved electron-scale ETG turbulence~\cite{dicorato_turbulent_2024} could limit $p_e$ and would therefore be consistent with sustaining $p_i/p_e>1$. This condition is favored by the simultaneous reduction in density, which lowers $\beta$, and the increase in $T_i$ without a correspondingly strong normalized ion-temperature gradient. The strong sensitivity to the equilibrium $E\times B$ flow also implies that, if the BSE pedestal were regulated only by the two-species ion-scale mechanisms retained here, the ion-scale transport would not provide a sufficient limit to continued profile growth. This raises the question of which additional mechanisms saturate the pedestal evolution and whether they are connected to the small-ELM dynamics, whose onset is not explained by the ideal peeling--ballooning and local ideal-ballooning stability analyses of the nominal BSE profiles~\cite{de_la_luna_exploring_2024}. One possible contribution arises from kinetic impurities, which are not negligible in the BSE plasma and were previously found to modify the local linear spectrum~\cite{dicorato_gyrokinetic_2022,dicorato_turbulent_2024}. A second contribution may arise from complete electron-scale coverage and ion--electron cross-scale coupling, which could modify the saturation physics and provide additional transport channels that survive the nominal equilibrium $E\times B$ flow. ETG turbulence has indeed been found to produce significant electron heat transport in JET-ILW pedestals and in the previous local analysis of the BSE pedestal top~\cite{chapman-oplopoiou_role_2022,dicorato_turbulent_2024}. Global multispecies and multiscale simulations are therefore required to address these open questions.

In conclusion, the present global analysis identifies two distinct turbulent transport regimes in the JET edge. The pre-type-I ELM pedestal develops KBM-driven electromagnetic turbulence that remains comparatively resilient to the equilibrium $E\times B$ flow, whereas the BSE edge remains below the KBM threshold and is dominated by finite$-\beta$-modified drift-wave turbulence that is strongly regulated by the equilibrium flow. These results show that finite$-\beta$ effects remain relevant below KBM onset and that the effectiveness of equilibrium-flow stabilization depends on the underlying turbulence regime. These elements should be retained in reduced models of edge transport, together with the radial variation and localization of the turbulent response.


\ack{The first author would like to thank C. Angioni, G. Dif-Pradalier, X. Garbet, G. Lo-Cascio, and participants to TTF US--EU 2026 for discussions. Simulations have been performed on Pitagora EUROfusion HPC hosted at Cineca. This work has been carried out within the framework of the EUROfusion Consortium, funded by the European Union via the Euratom Research and Training Programme (Grant Agreement No 101052200 - EUROfusion). Views and opinions expressed are however those of the author(s) only and do not necessarily reflect those of the European Union or the European Commission. Neither the European Union nor the European Commission can be held responsible for them. The first author also acknowledges the use of the Chat-AI service provided by the Gesellschaft f\"ur wissenschaftliche Datenverarbeitung mbH G\"ottingen (GWDG), through which OpenAI's ChatGPT model was used for coding and text draft assistance.}



\data{The data supporting the findings of this study are available from the
corresponding author upon reasonable request.}


\appendix

\section{Pedestal fits and profile definitions}
\label{app:profiles}

The pedestal profiles are characterized using the five-parameter modified-hyperbolic-tangent function
\begin{equation}
\mathrm{mtanh}(\rhot)
=
\frac{h_1-h_0}{2}
\left[
\frac{(1+s x)e^x-e^{-x}}{e^x+e^{-x}}+1
\right]
+h_0,
\label{eq:mtanh}
\end{equation}
where
\begin{equation}
x=\frac{p-\rhot}{w/2}.
\label{eq:mtanh_x}
\end{equation}
Here $h_1$ is the pedestal height, $h_0$ is the outer-profile offset, $p$ is the modified-hyperbolic-tangent centre parameter, $w$ is the fitted pedestal width, and $s$ is related to the inner-profile slope. Since the fits are performed using $\rhot$ as the radial coordinate, the widths are expressed in $\Delta\rhot$. The interval $[p-w/2,p+w/2]$ is used to indicate the radial region associated with the fitted pedestal width. The pedestal position is defined as the location of the maximum absolute gradient of the fitted profile
\begin{equation}
\rhot^{\mathrm{ped}}
=
\underset{\rhot}{\operatorname{arg\,max}}
\left|
\frac{\mathrm{d}(\mathrm{mtanh})}{\mathrm{d}\rhot}
\right|.
\label{eq:pedestal_position}
\end{equation}
The pedestal-top position $\rhot^{\mathrm{top}}$ is defined as the radial position corresponding to the pedestal height $h_1$ obtained from the fit, following the definition adopted for the JET pedestal database~\cite{frassinetti_pedestal_2021}. The fitted values of $h_1$, $w$, $\rhot^{\mathrm{ped}}$, and $\rhot^{\mathrm{top}}$ are reported in table~\ref{tab:pedestal_structure}.

\section{GENE code extension and numerical setup}
\label{app:Er}

\subsection{Reconstruction of the equilibrium radial electric field}

The equilibrium radial electric field is reconstructed from the radial projection of the force-balance equation for the main-ion species,
\begin{equation}
n_{0i}q_i
\left(
\boldsymbol{E}_0
+
\boldsymbol{V}_i\times\boldsymbol{B}_0
\right)
-
\nabla p_{0i}
=
0.
\label{eq:force_balance}
\end{equation}
The straight-field-line coordinate system is $(\Psi,\theta,\varphi)$, where $\Psi$ is the poloidal magnetic flux, $\theta$ is the poloidal angle, and $\varphi$ is the toroidal angle. The field-aligned GENE coordinates are defined as
\begin{equation}
x=C_x(\Psi)+x_0,
\qquad
y=C_y\left(\sigma_{I_p}q\theta-\varphi\right),
\qquad
z=\theta,
\label{eq:gene_coordinates}
\end{equation}
and the equilibrium magnetic field can be written as
\begin{equation}
\boldsymbol{B}_0 = \nabla\Psi\times\nabla \left( \sigma_{I_p}q\theta-\varphi \right) = \mathcal C(x)\nabla x\times\nabla y,
\label{eq:gene_magnetic_field}
\end{equation}
where
\begin{equation}
\mathcal C(x)=C_{xy}(x) = \frac{1}{C_y} \left( \frac{\mathrm{d}C_x}{\mathrm{d}\Psi} \right)^{-1},
\qquad
C_{xy}^{-1} = C_y\frac{\mathrm{d}C_x}{\mathrm{d}\Psi}.
\label{eq:Cxy_definition}
\end{equation}

The equilibrium electric field and the main-ion pressure gradient are
\begin{equation}
\boldsymbol{E}_0
=
-\nabla\Phi_0(\Psi),
\qquad
\nabla p_{0i}
=
\frac{\partial p_{0i}}{\partial\Psi}\nabla\Psi.
\label{eq:E_pressure_expansion}
\end{equation}
Expanding the magnetic field and main-ion velocity in their poloidal and toroidal components gives
\begin{equation}
\boldsymbol{B}_0
=
\sigma_{I_p}|B^\theta|\boldsymbol{e}_\theta
+
\sigma_{B_t}|B^\varphi|\boldsymbol{e}_\varphi,
\qquad
\boldsymbol{V}_i
=
\sigma_{V_p}|V_i^\theta|\boldsymbol{e}_\theta
+
\sigma_{\Omega}|V_i^\varphi|\boldsymbol{e}_\varphi.
\label{eq:B_V_expansion}
\end{equation}
The resulting radial force-balance equation is
\begin{equation}
-\frac{\partial\Phi_0}{\partial\Psi}
=
\frac{1}{n_{0i}q_i}
\frac{\partial p_{0i}}{\partial\Psi}
-
\left[
\sigma_{V_p}\sigma_{B_t}|V_i^\theta|
\frac{|\boldsymbol{B}\cdot\nabla\varphi|}
{|\boldsymbol{B}\cdot\nabla\theta|}
-
\sigma_{\Omega}\sigma_{I_p}|V_i^\varphi|
\right].
\label{eq:radial_force_balance_psi}
\end{equation}
Transforming the radial derivative from $\Psi$ to the GENE coordinate $x$ yields
\begin{equation}
E_r
=
-\partial_x\Phi_0(x)
=
\frac{\partial_xp_{0i}(x)}{n_{0i}q_i}
-
\left(
\frac{\mathrm{d}x}{\mathrm{d}\Psi}
\right)^{-1}
\left[
\sigma_{V_p}\sigma_{B_t}|V_i^\theta|
\frac{|\boldsymbol{B}\cdot\nabla\varphi|}
{|\boldsymbol{B}\cdot\nabla\theta|}
-
\sigma_{\Omega}\sigma_{I_p}|V_i^\varphi|
\right].
\label{eq:radial_force_balance_gene}
\end{equation}
The contravariant velocity components $V_i^\theta=\boldsymbol{V}_i\cdot\nabla\theta$ and $V_i^\varphi=\boldsymbol{V}_i\cdot\nabla\varphi$ correspond to the poloidal and toroidal angular frequencies, respectively. The sign convention is such that positive toroidal rotation corresponds to clockwise rotation when viewed from above, while positive poloidal rotation corresponds to counter-clockwise rotation in the poloidal plane.

In the implementation used here, the complete force-balance field is decomposed as
\begin{equation}
E_r^{\mathrm{FB}}
=
E_r^{\mathrm{diam}}
+
E_r^{\Omega_\varphi}
+
E_r^{v_\theta},
\label{eq:Er_complete}
\end{equation}
where the three contributions in dimensional units are
\begin{align}
E_r^{\mathrm{diam}}
={}&
-
\frac{\omega_{n_i}+\omega_{T_i}}{C_{xy}}
\frac{T_iT_{i,\mathrm{prof}}}{Z_i}
\frac{10^3T_{\mathrm{ref}}}{L_{\mathrm{ref}}},
\label{eq:Er_gene_diam}
\\
E_r^{\Omega_\varphi}
={}&
C_yB_{\mathrm{ref}}L_{\mathrm{ref}}\Omega_{\varphi i},
\label{eq:Er_gene_toroidal}
\\
E_r^{v_\theta}
={}&
-\sigma_{B_\varphi}
C_yB_{\mathrm{ref}}L_{\mathrm{ref}}
\frac{|F|}{\fsavg{R^2}}
U_{\theta i}.
\label{eq:Er_gene_poloidal}
\end{align}
Here, $\omega_{n_i}$ and $\omega_{T_i}$ are the logarithmic gradients, $T_i$ and $T_{i,\mathrm{prof}}$ are the reference and radially varying main-ion temperature factors in GENE, $Z_i$ is the main-ion charge, $C_y$ and $C_{xy}$ are the GENE coordinate factors, $F=RB_\varphi$ is the diamagnetic function, $\fsavg{R^2}$ is the flux-surface-averaged squared major radius, $\Omega_{\varphi i}$ is the toroidal angular rotation frequency, and $U_{\theta i}$ is the neoclassical poloidal-flow quantity calculated with NCLASS~\cite{houlberg_bootstrap_1997}. Equations~\eqref{eq:Er_gene_diam}--\eqref{eq:Er_gene_poloidal} are the expressions used to calculate the three contributions shown in figure~\ref{fig:equilibrium_Er_shear}.

The radial electric field is normalized in GENE according to
\begin{equation}
\widehat E_r^{\mathrm{FB}}
=
E_r^{\mathrm{FB}}
\frac{L_{\mathrm{ref}}}{10^3T_{\mathrm{ref}}}
=
-\partial_{\widehat{x}}\widehat{\Phi}_0.
\label{eq:Er_gene_normalization}
\end{equation}
The field imposed in a given simulation is
\begin{equation}
E_r^{\mathrm{ext}}(\rhot)
=
f_EE_r^{\mathrm{FB}}(\rhot),
\label{eq:Er_ext_appendix}
\end{equation}
where the values of $f_E$ are listed in table~\ref{tab:simulation_matrix}.

\subsection{Explicit inclusion of equilibrium $E\times B$ advection}
The equilibrium electrostatic potential $\Phi_0(x)$ is included in the gyrocentre equations as an additional term in the fluctuating generalized potential 
\begin{equation}
\overline{\xi}_1 = \overline{\phi}_1 - \frac{v_\parallel}{c}\overline{A}_{1\parallel}.
\label{eq:generalized_potential}
\end{equation}
The full gyrocentre distribution function is then separated into a stationary local Maxwellian equilibrium $F_{0\sigma}$ and a time-dependent perturbed contribution $F_{1\sigma}$,
\begin{equation}
F_\sigma = F_{0\sigma} + F_{1\sigma},
\qquad 
\frac{F_{1\sigma}}{F_{0\sigma}} \sim \epsilon_\delta \ll1.
\label{eq:delta_f_splitting}
\end{equation}
After applying this $\delta f$ splitting, additional terms in the gyrokinetic Vlasov equation describing the advection by the equilibrium $E\times B$ velocity are thus obtained, in dimensional GENE coordinates,
\begin{equation}
\frac{c}{\mathcal C(x)} \frac{B_0}{B_\parallel^*} \partial_x\Phi_0(x) \left[ \partial_yF_{1\sigma} - \frac{q_\sigma}{cm_\sigma} \partial_y\overline{A}_{1\parallel} \partial_{v_\parallel}F_{0\sigma} \right].
\label{eq:equilibrium_ExB_term_dimensional}
\end{equation}
Using the standard GENE normalization, these terms become
\begin{equation}
\frac{1}{\widehat{\mathcal C}} \frac{\widehat B_0}{\widehat B_\parallel^*} \partial_{\widehat{x}}\widehat{\Phi}_0 \left[ \partial_{\widehat{y}}\widehat F_{1\sigma} - \frac{\widehat q_\sigma} {\widehat m_\sigma\widehat v_{T\sigma}(x_0)} \partial_{\widehat{y}}\widehat{\overline A}_{1\parallel} \partial_{\widehat v_\parallel}\widehat F_{0\sigma} \right].
\label{eq:equilibrium_ExB_term}
\end{equation}
The first term in brackets represents the binormal advection of the perturbed distribution function by the equilibrium $E\times B$ velocity. The second is the associated electromagnetic contribution involving $\overline{A}_{1\parallel}$. The complete radial profile of $\partial_x\Phi_0$ is retained in the global simulations. In the present implementation, the gyroaverage of the equilibrium potential is neglected, so that $\Phi_0$, rather than $\overline{\Phi}_0$, enters the equilibrium-flow terms. The extension to a gyroaveraged equilibrium $E\times B$ flow and its impact will be addressed in future work.

\subsection{Numerical setup}
\label{app:numerics}

The global nonlinear simulations cover the radial domain $0.86\leq\rhot\leq0.98$, with buffer regions occupying $10\%$ of the domain at each boundary. The retained nonzero toroidal mode number range is $5\leq n\leq235$. The production-run phase-space resolutions and Krook-source coefficients are summarized in table~\ref{tab:numerical_parameters}. The additional BSE EM sensitivity run with $f_E=0.8$ uses the same numerical resolution and Krook-source coefficients as the BSE EM run with $f_E=0.5$.

\begin{table*}[t]
\caption{Production run numerical parameters of the global nonlinear simulations. All simulations use $0.86\leq\rhot\leq0.98$, $10\%$ lower and upper radial buffers, and the nonzero toroidal mode number range $5\leq n\leq235$. Here $n_{k_y}$ is the number of retained non-negative binormal Fourier modes, and $c_{K,T}$ and $c_{K,n}$ are the heat- and particle-source coefficients, respectively.}
\label{tab:numerical_parameters}
\centering
\begin{tabular}{lllcccccc}
\hline
Edge profile & Model & Flow &
$n_x$ & $n_{k_y}$ & $n_z$ &
$n_{v_\parallel}$ & $n_\mu$ &
$(c_{K,T},c_{K,n})$ \\
\hline
Type-I ELM & $\mathrm{ES}_l$ & without & 512 & 48 & 64 & 48 & 24 & $(0.10,0.05)$ \\
Type-I ELM & $\mathrm{ES}_l$ & with    & 512 & 48 & 64 & 48 & 24 & $(0.10,0.05)$ \\
Type-I ELM & EM             & without & 512 & 48 & 64 & 48 & 24 & $(0.10,0.05)$ \\
Type-I ELM & EM             & with    & 512 & 48 & 64 & 48 & 24 & $(0.10,0.05)$ \\
\hline
BSE & $\mathrm{ES}_l$ & without & 512 & 48 & 64 & 48 & 24 & $(0.05,0.05)$ \\
BSE & $\mathrm{ES}_l$ & with    & 512 & 48 & 64 & 48 & 24 & $(0.05,0.05)$ \\
BSE & EM             & without & 512 & 48 & 64 & 48 & 24 & $(0.05,0.05)$ \\
BSE & EM             & with    & 512 & 48 & 64 & 48 & 24 & $(0.05,0.05)$ \\
\hline
\end{tabular}
\end{table*}

In GENE normalized units, the radial box sizes are $l_x=86.58\rho_s$ and $84.08\rho_s$ for the type-I ELM and BSE cases, respectively, giving $\Delta x=l_x/n_x\simeq0.169\rho_s$ and $0.164\rho_s$.

The block-structured velocity grid uses 8 blocks radially. Within each block, the reference quantities defining $(v_\parallel,\mu)$ are rescaled so that an equivalent local flux-tube velocity-space domain and resolution are retained despite the radial variation of temperature. The GENE parameter \texttt{vsp\_std}, which controls the velocity-domain extent, was varied over $3.5$, and $5.0$; the production simulations use $\texttt{vsp\_std}=5.0$ to ensure sufficiently large velocity-space domains. The resulting resolution $(n_{v_\parallel},n_\mu)=(48,24)$ is comparable to a well-resolved flux-tube velocity grid in each radial block. Additional sensitivity tests varied the minimum retained toroidal mode number between $n_{\min}=5$ and $10$, the Krook-source strengths, and the radial-buffer widths. For the type-I ELM simulations, the source coefficients were initially $(c_{K,T},c_{K,n})=(0.05,0.05)$; the heat-source coefficient was increased to the production value $(0.10,0.05)$ to limit excessive relaxation of the input profiles. These tests were used to select the production parameters reported in table~\ref{tab:numerical_parameters}.

\section{Diagnostics: effective diffusivities, transport fingerprints, and mode parity}
\label{app:linear_fingerprints}

\subsection{Parallel average}
For a field $\mathcal F(\rhot,z)$, the Jacobian-weighted parallel average is defined as
\begin{equation}
\langle\mathcal F\rangle_z = \frac{\displaystyle\sum_z\mathcal J(\rhot,z)\mathcal F(\rhot,z)} {\displaystyle\sum_z\mathcal J(\rhot,z)},
\label{eq:parallel_average}
\end{equation}
where $\mathcal J$ is the Jacobian of the GENE field-aligned coordinate system.

\subsection{Effective diffusivities}
The nonlinear turbulent particle and heat transport is characterized using effective diffusivity coefficients. Defining the logarithmic density and temperature gradients as
\begin{equation}
\omega_{n,s} = -\frac{\mathrm{d}\ln n_s}{\mathrm{d}r},
\qquad
\omega_{T,s} = -\frac{\mathrm{d}\ln T_s}{\mathrm{d}r},
\label{eq:log_gradients}
\end{equation}
the effective particle and heat diffusivities are
\begin{equation}
D_s^c(t,r) = \frac{\Gamma_s^c(t,r)}{n_s(t,r)\omega_{n,s}(t,r)\fsavg{g^{rr}}},
\qquad
\chi_s^c(t,r) = \frac{Q_s^c(t,r)}{n_s(t,r)T_s(t,r)\omega_{T,s}(t,r)\fsavg{g^{rr}}},
\label{eq:nonlinear_diffusivities}
\end{equation}
where $g^{rr}=|\nabla r|^2$ and $c\in\{\mathrm{ES},\mathrm{EM}\}$ denotes the electrostatic and explicitly electromagnetic flux contributions. At a fixed denominator, the total coefficients are
\begin{equation}
D_s^{\mathrm{tot}} = D_s^{\mathrm{ES}} + D_s^{\mathrm{EM}},
\qquad
\chi_s^{\mathrm{tot}} = \chi_s^{\mathrm{ES}} + \chi_s^{\mathrm{EM}}.
\label{eq:total_diffusivities}
\end{equation}
These quantities are effective coefficients and do not provide a decomposition into diffusive and convective transport. In particular, $D_e^{\mathrm{tot}}$ does not distinguish particle diffusion from a convective pinch.

For a radially varying quantity $A(r)$, the buffer-excluded radial average is defined as
\begin{equation}
\langle A\rangle_r = \frac{1}{r_2-r_1}\int_{r_1}^{r_2} A(r)\,\mathrm{d}r,
\label{eq:radial_average}
\end{equation}
where $r_1$ and $r_2$ delimit the physical domain after removing the radial buffers.

\subsection{Linear transport fingerprints}
The transport fingerprints extend the mode-identification method introduced in reference~\cite{kotschenreuther_gyrokinetic_2019} to the radially global simulations. For each toroidal mode number $n$, the effective coefficients are calculated from the corresponding total linear particle and heat fluxes,
\begin{equation}
D_s(r,n) = \frac{\Gamma_s^{\mathrm{tot}}(r,n)}{n_s(r)\omega_{n,s}(r)},
\qquad
\chi_s(r,n) = \frac{Q_s^{\mathrm{tot}}(r,n)}{n_s(r)T_s(r)\omega_{T,s}(r)}.
\label{eq:linear_diffusivities}
\end{equation}
The reduced transport fingerprints are formed from separately averaged coefficients,
\begin{equation}
R_\chi(n) = \frac{\langle\chi_i(r,n)\rangle_r}{\langle\chi_e(r,n)\rangle_r},
\qquad
R_D(n) = \frac{\langle D_e(r,n)\rangle_r}{\langle\chi_e(r,n)\rangle_r}.
\label{eq:reduced_fingerprints}
\end{equation}

\begin{table}[t]
\caption{Properties used in the identification of microinstabilities. The real-frequency sign is positive or negative for propagation in the ion- or electron-diamagnetic direction, respectively. The transport fingerprint ratios are adapted from ref.~\cite{kotschenreuther_gyrokinetic_2019}.}
\label{tab:mode_fingerprints}
\centering
\begin{tabular}{llllll}
\hline
Mode & Scale & Main drive & $\mathrm{sgn}(\omega)$ & $\chi_i/\chi_e$ & $D_e/\chi_e$ \\
\hline
KBM & $k_y\rho_i\sim0.1$ & $L_p^{-1},\beta_e$ & $>0$ & $\sim1$ & $\sim2/3$ \\
MTM & $k_y\rho_i\sim0.1$ & $L_{T_e}^{-1},\beta_e$ & $<0$ & $\sim0$ & $\sim0$ \\
ITG & $0.1<k_y\rho_i\lesssim1$ & $L_{T_i}^{-1}$ & $>0$ & $\gtrsim1$ & $-0.2$--$1$ \\
TEM & $k_y\rho_i\gtrsim0.1$ & $L_{T_e}^{-1},L_{n_e}^{-1}$ & $<0$ & $\mathcal O(1)$ & $-0.2$--$1$ \\
ETG & $k_y\rho_i\gtrsim1$ & $L_{T_e}^{-1}$ & $<0$ & $\sim0$ & $\sim0$ \\
\hline
\end{tabular}
\end{table}

The fingerprints listed in table~\ref{tab:mode_fingerprints} are used as qualitative mode-characterization criteria. In the steep-gradient edge, the unstable modes can combine properties of different standard branches; the classification as hybrid ITG--TEM or KBM therefore also uses the mode frequency, finite$-\beta$ response, parity, and radial localization.

\begin{figure}[t]
\centering
\includegraphics[width=0.47\textwidth]{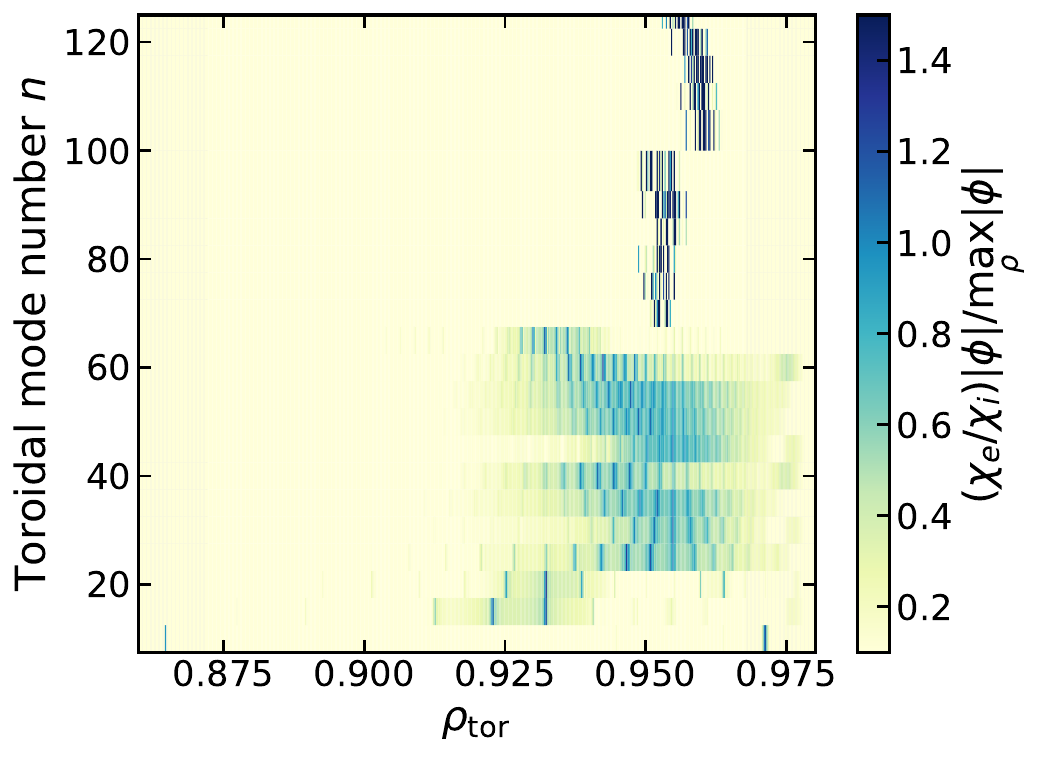}\qquad
\includegraphics[width=0.47\textwidth]{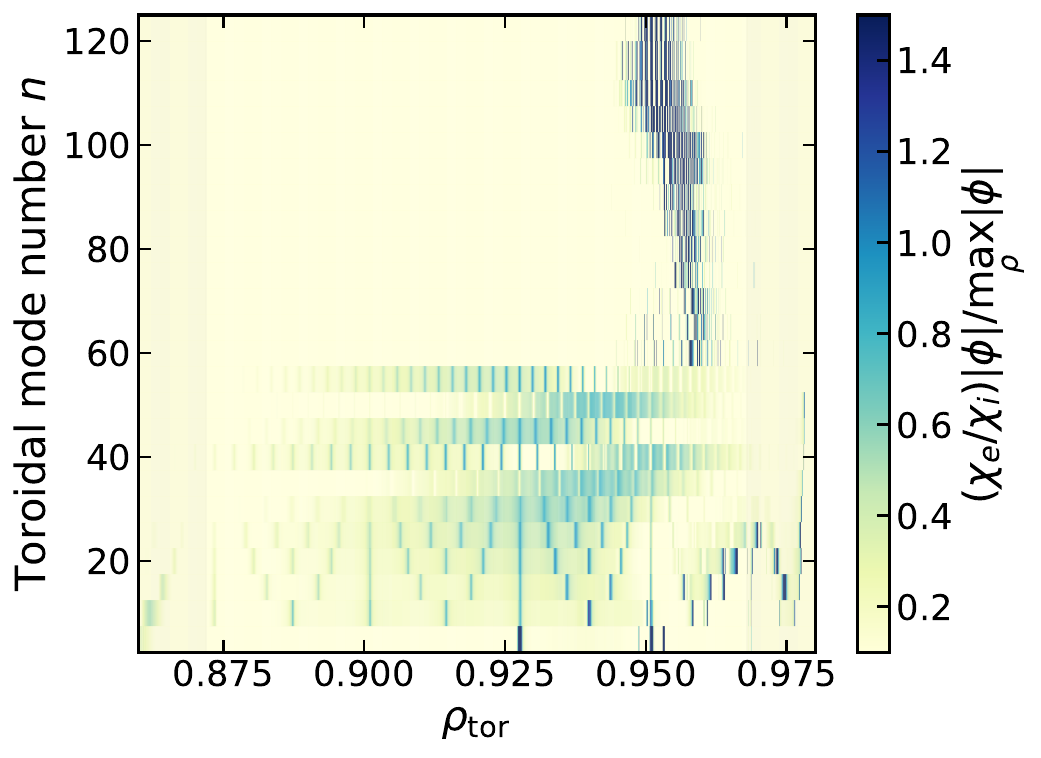}
\makebox[0.47\textwidth][c]{\small (a) Type-I ELM}\qquad
\makebox[0.47\textwidth][c]{\small (b) BSE}
\caption{Radially resolved linear electron-to-ion heat-transport fingerprints in the $(\rhot,n)$ plane for (a) the type-I ELM discharge and (b) the BSE discharge. The pointwise ratio is weighted by $\hat{\phi}_1$ eigenfunction envelope.}
\label{fig:linear_chie_over_chii_appendix}
\end{figure}

\begin{figure}[t]
\centering
\includegraphics[width=0.47\textwidth]{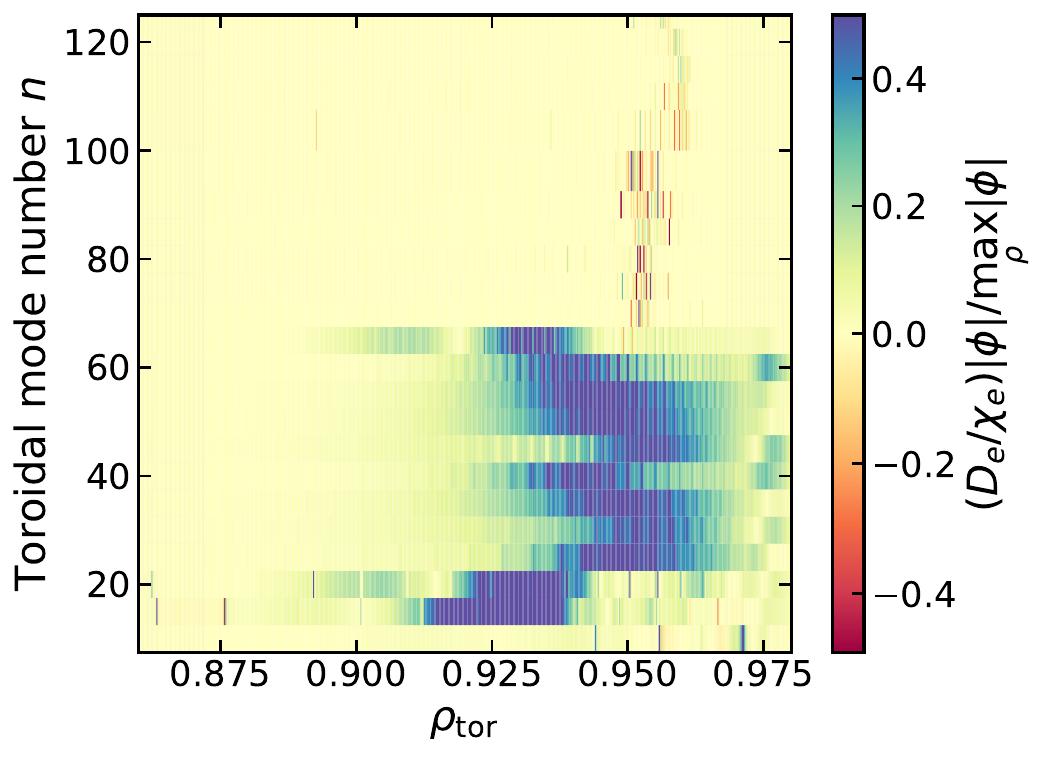}\qquad
\includegraphics[width=0.47\textwidth]{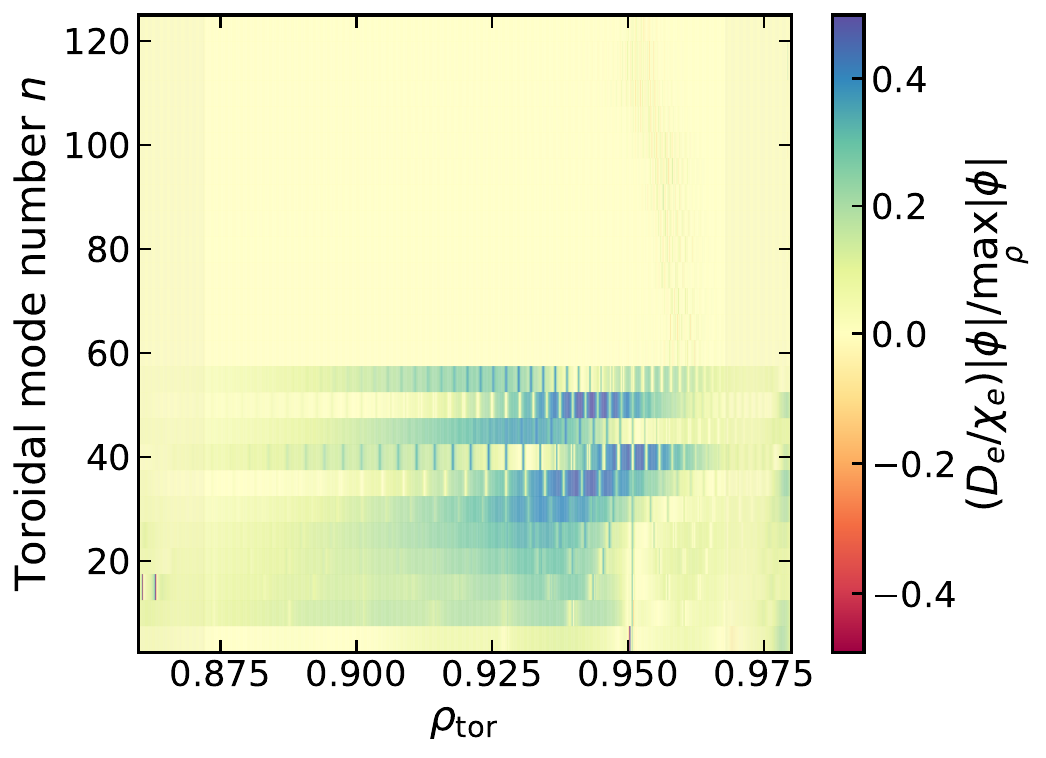}
\makebox[0.47\textwidth][c]{\small (a) Type-I ELM}\qquad
\makebox[0.47\textwidth][c]{\small (b) BSE}
\caption{Radially resolved particle-to-electron-heat transport fingerprints in the $(\rhot,n)$ plane for (a) the type-I ELM discharge and (b) the BSE discharge. The ratio is weighted by $\hat{\phi}_1$ eigenfunction envelope. The effective particle diffusivity does not distinguish diffusive from convective transport.}
\label{fig:linear_De_over_chie_appendix}
\end{figure}

\subsection{Mode parity}
For a complex field $\mathcal{F}(\rhot,z;n)$, the parity diagnostic is defined as
\begin{equation}
\mathcal P_{\mathcal{F}}(\rhot,n) = \frac{\left|\displaystyle\int \mathcal{F}(\rhot,z;n)\,\mathrm{d}z\right|} {\displaystyle\int|\mathcal{F}(\rhot,z;n)|\,\mathrm{d}z}.
\label{eq:parity_local}
\end{equation}
Thus, $\mathcal P_{\mathcal{F}}\rightarrow1$ indicates an even structure, while $\mathcal P_{\mathcal{F}}\rightarrow0$ indicates an odd structure. The reduced radial parity is
\begin{equation}
\left\langle\mathcal P_{\mathcal{F}}\right\rangle_{\rhot} (n)= \frac{ \displaystyle\int W_{\mathcal{F}}(\rhot,n)\mathcal P_{\mathcal{F}}(\rhot,n)\,\mathrm{d}\rhot }{ \displaystyle\int W_{\mathcal{F}}(\rhot,n)\,\mathrm{d}\rhot },
\label{eq:parity_reduced}
\end{equation}
where
\begin{equation}
W_{\mathcal{F}}(\rhot,n) = \left[ \int|{\mathcal{F}}(\rhot,z;n)|^2\,\mathrm{d}z\right]^{1/2}. 
\label{eq:parity_weight}
\end{equation}
The weighting by the radial field envelope prevents regions with negligible mode amplitude from dominating the reduced parity. For a conventional ballooning-parity mode, $\hat{\phi}_1$ is approximately even and $\hat{A}_{1\parallel}$ approximately odd, whereas the opposite ordering is characteristic of tearing parity. In steep-gradient regions, however, higher-order and hybrid eigenstates can exhibit mixed parity~\cite{pueschel_microinstabilities_2019}; parity is therefore used together with the others mode-identification diagnostics.

\section{Radial--spectral decomposition of nonlinear ion heat transport}
\label{app:Qi_maps}

The nonlinear ion heat transport is decomposed in the $(\rhot,n)$ plane to identify changes in the dominant toroidal scales, radial localization, and radial extent of the transport produced by electromagnetic effects and the equilibrium flow.

\begin{figure}[t]
\centering
\includegraphics[width=0.25\textwidth]{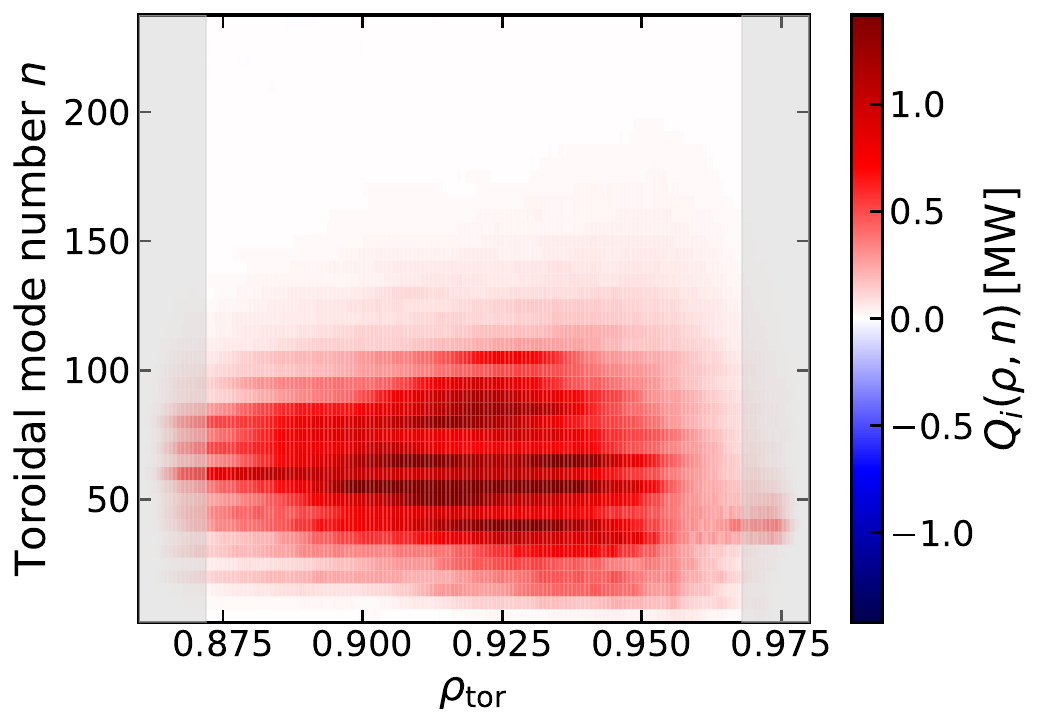}\hfill
\includegraphics[width=0.25\textwidth]{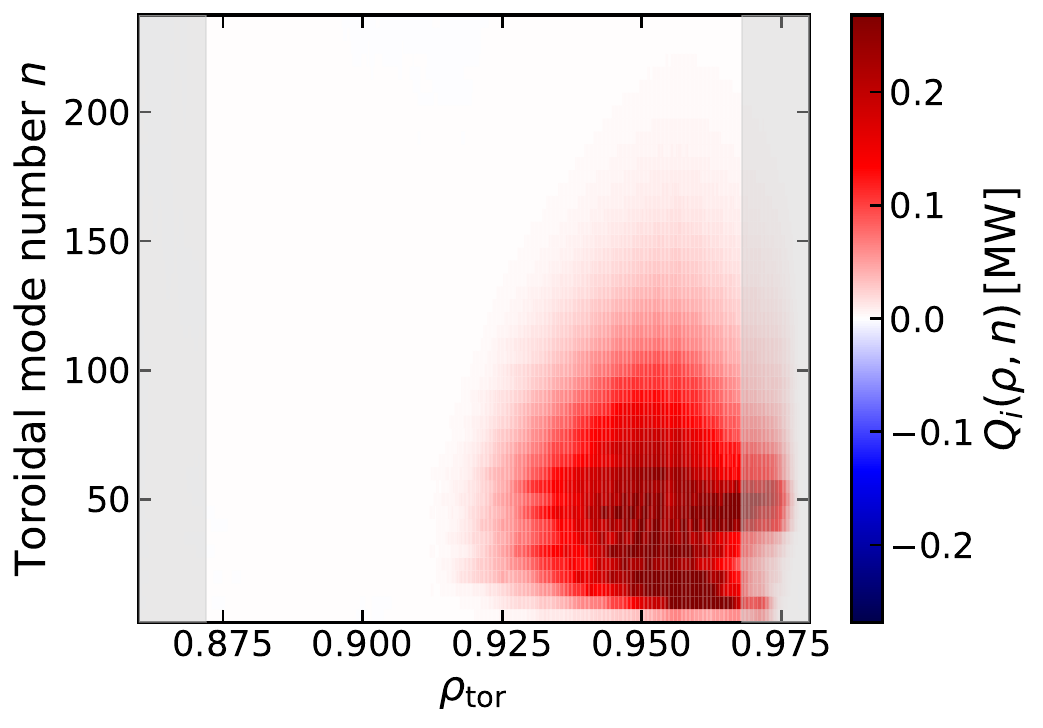}\hfill
\includegraphics[width=0.25\textwidth]{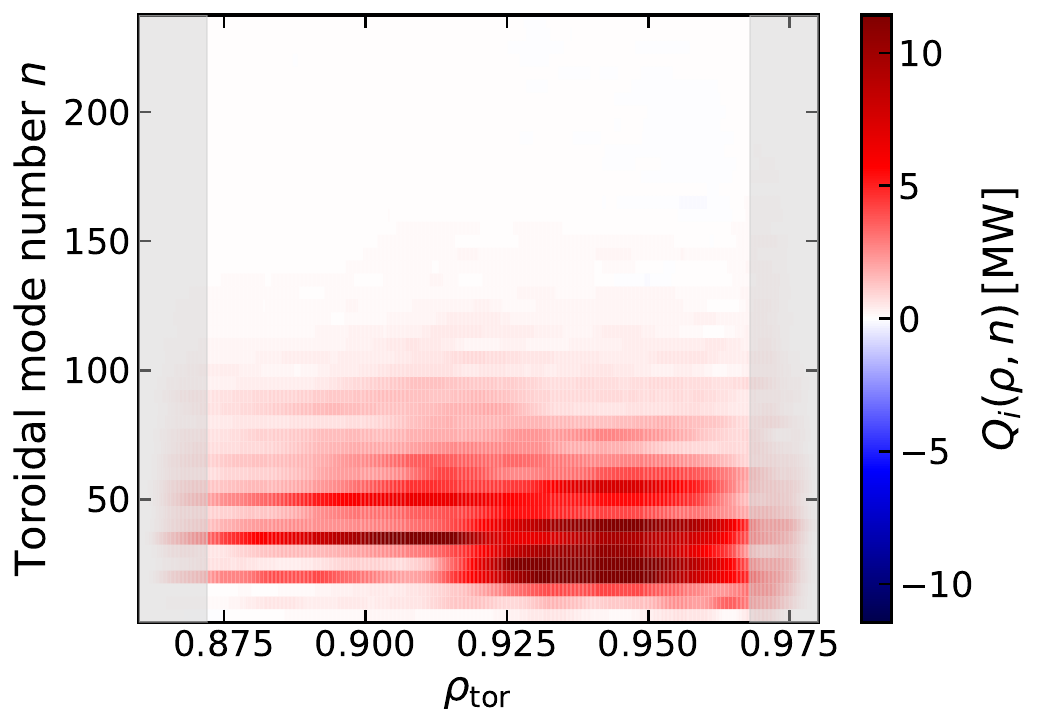}\hfill
\includegraphics[width=0.25\textwidth]{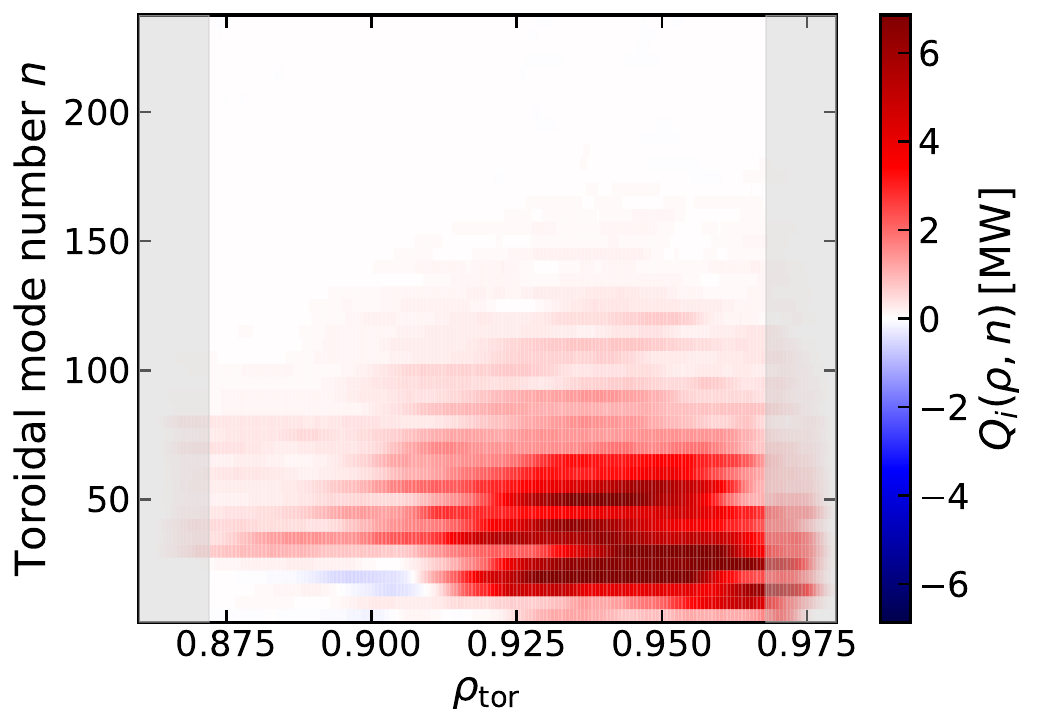}\\
\makebox[0.25\textwidth][c]{\small (a) $\mathrm{ES}_l$, without $E\times B$}\hfill
\makebox[0.25\textwidth][c]{\small (b) $\mathrm{ES}_l$, with $E\times B$}\hfill
\makebox[0.25\textwidth][c]{\small (c) EM, without $E\times B$}\hfill
\makebox[0.25\textwidth][c]{\small (d) EM, with $E\times B$}
\caption{Time-averaged nonlinear ion heat transport $Q_i(\rhot,n)$ for the type-I ELM pedestal. The panels compare the $\mathrm{ES}_l$ and EM simulations without and with the nominal equilibrium $E\times B$ flow. Grey regions denote the numerical radial buffers. Each panel uses an independent colour scale.}
\label{fig:typeI_Qi_xn_appendix}
\end{figure}

Without equilibrium $E\times B$ flow, the type-I ELM $\mathrm{ES}_l$ transport is distributed over approximately $20\lesssim n\lesssim120$ and extends radially from the inner edge region to the pedestal. Introducing the nominal equilibrium flow strongly reduces the transport amplitude and shifts the dominant contribution towards approximately $20\lesssim n\lesssim80$ and $\rhot\gtrsim0.92$. Electromagnetic effects produce a strong increase in the transport and broaden its radial distribution. In the EM simulation without flow, the dominant transport is carried by approximately $20\lesssim n\lesssim100$ over a broad radial region, with the strongest contribution around $n\simeq35$--$60$ and $\rhot\simeq0.92$--$0.96$. When the equilibrium flow is included, substantial transport remains over the same pedestal region, consistently with the resilience of the KBM-compatible electromagnetic turbulence identified from the effective diffusivities.

\begin{figure}[t]
\centering
\includegraphics[width=0.25\textwidth]{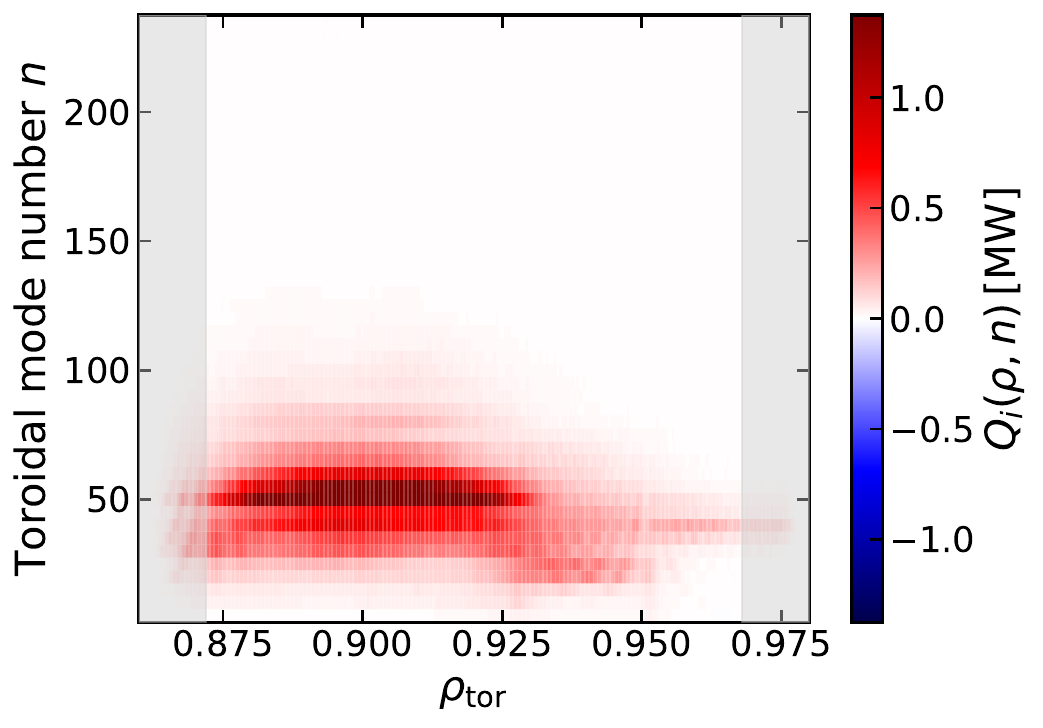}\hfill
\includegraphics[width=0.25\textwidth]{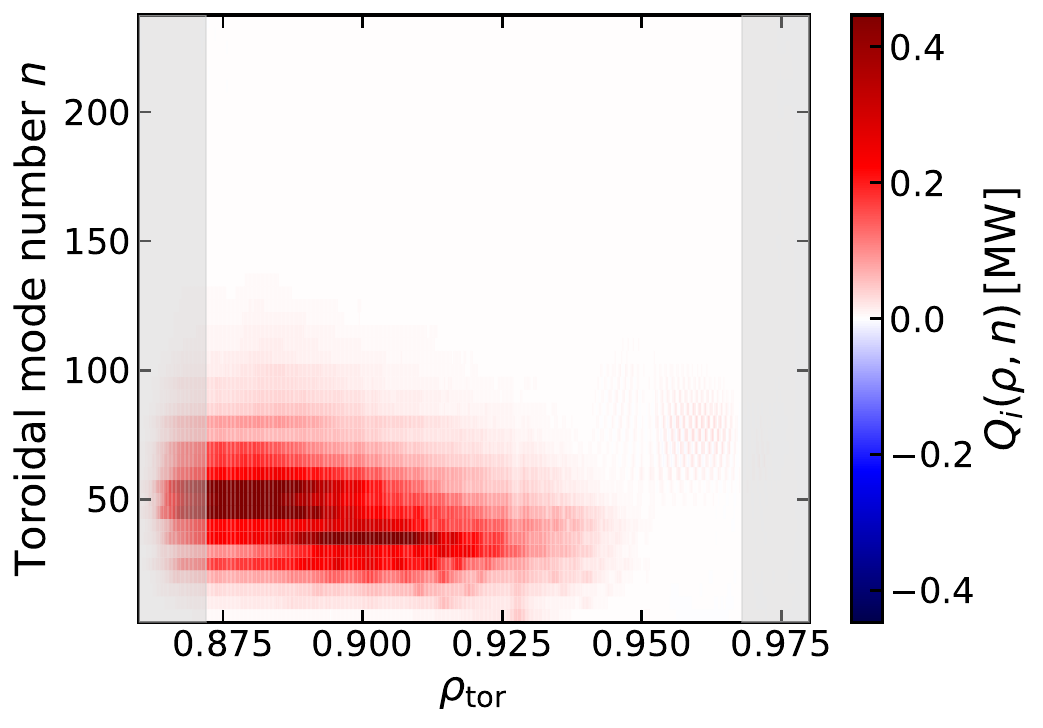}\hfill
\includegraphics[width=0.25\textwidth]{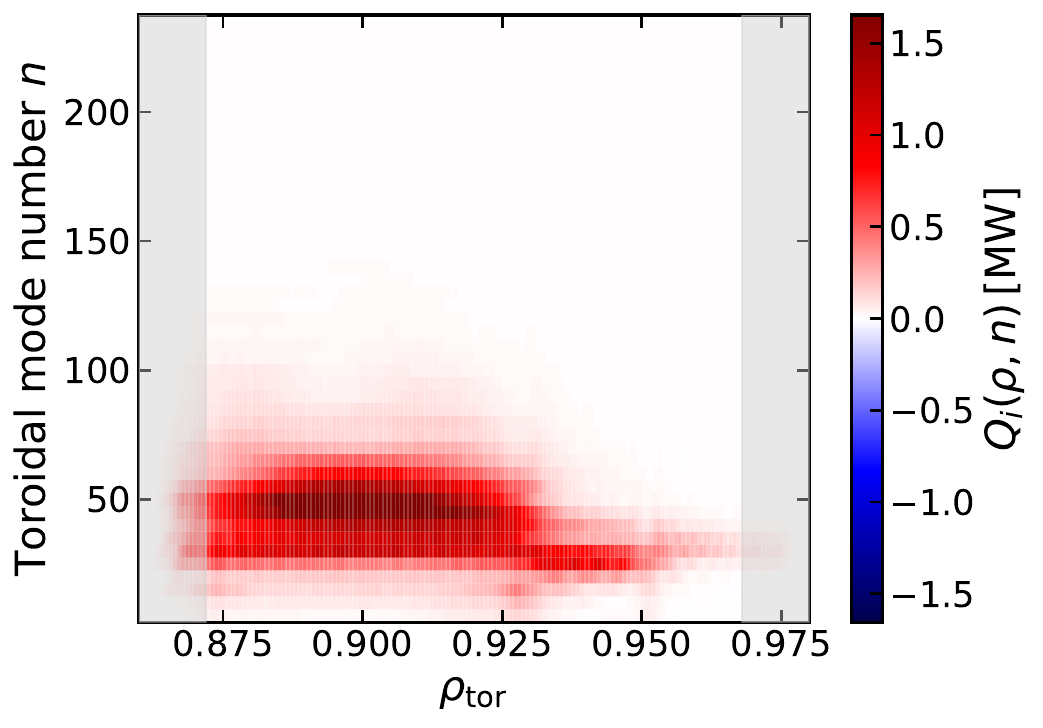}\hfill
\includegraphics[width=0.25\textwidth]{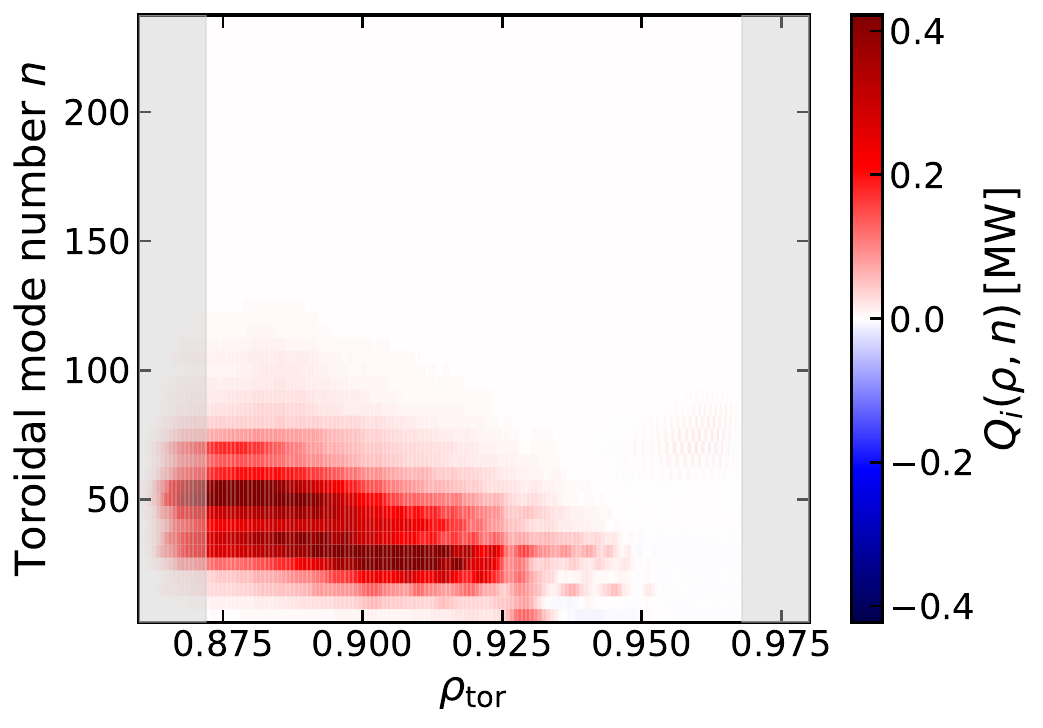}\\
\makebox[0.25\textwidth][c]{\small (a) $\mathrm{ES}_l$, without $E\times B$}\hfill
\makebox[0.25\textwidth][c]{\small (b) $\mathrm{ES}_l$, with $0.5\cdot E\times B$}\hfill
\makebox[0.25\textwidth][c]{\small (c) EM, without $E\times B$}\hfill
\makebox[0.25\textwidth][c]{\small (d) EM, with $0.5\cdot E\times B$}\\
\includegraphics[width=0.25\textwidth]{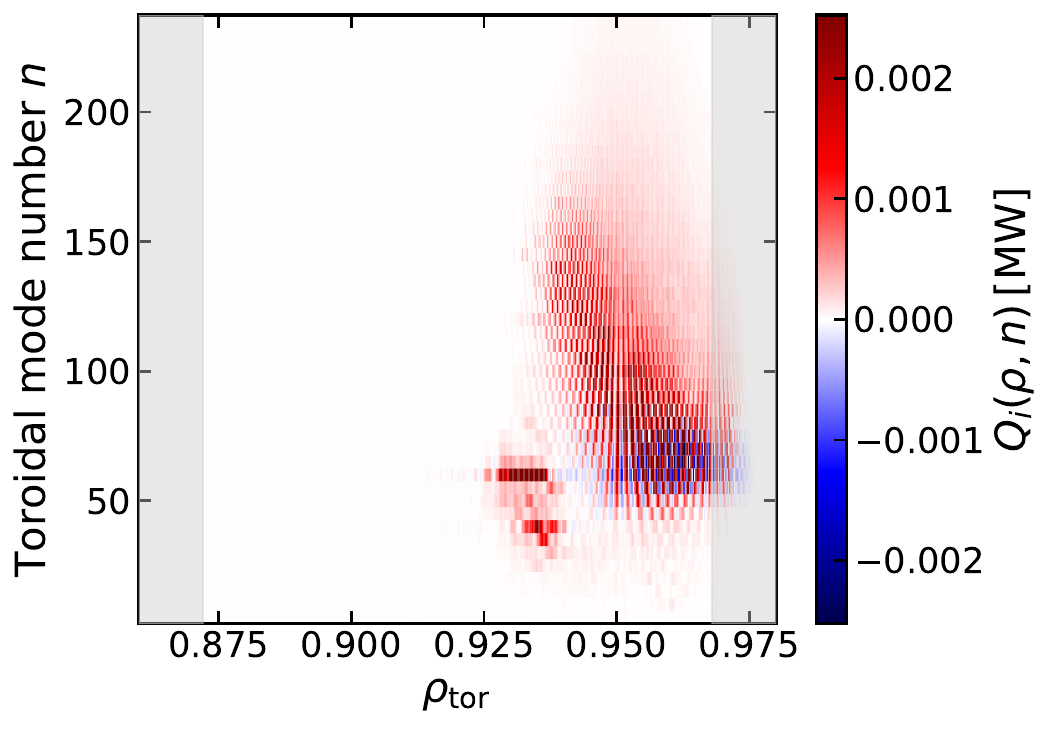}\\
\makebox[0.25\textwidth][c]{\small (e) EM, with $0.8\cdot E\times B$}
\caption{Time-averaged nonlinear ion heat transport $Q_i(\rhot,n)$ for the BSE edge. The panels compare the $\mathrm{ES}_l$ and EM simulations without equilibrium flow, with $f_E=0.5$, and, in panel (e), with $f_E=0.8$. Grey regions denote the numerical radial buffers. Each panel uses an independent colour scale.}
\label{fig:smallELM_Qi_xn_appendix}
\end{figure}

In the BSE simulations without equilibrium $E\times B$ flow, the dominant ion heat transport is carried by approximately $20\lesssim n\lesssim80$ and extends over a broad radial region from the outer core to the pedestal. Electromagnetic effects moderately increase the transport and modify its radial and spectral distribution without producing the low-$n$ ballooning-dominated state found in the type-I ELM pedestal. Introducing the reduced equilibrium flow, $f_E=0.5$, strongly decreases the low- and intermediate-$n$ contributions in both the $\mathrm{ES}_l$ and EM simulations and radially reorganizes the residual transport. At $f_E=0.8$, the large-scale hybrid ITG--TEM contribution is almost completely suppressed, and only weak transport associated mainly with $n\gtrsim50$ remains around $\rhot\simeq0.93$--$0.97$, consistently with the radial localization of the intermediate-scale ETG branch found in the linear analysis.


\bibliographystyle{unsrt}
\bibliography{references}

@article{confinement_chapter_1999,
	title = {Chapter 2: Plasma confinement and transport},
	volume = {39},
	issn = {0029-5515},
	url = {https://dx.doi.org/10.1088/0029-5515/39/12/302},
	doi = {10.1088/0029-5515/39/12/302},
	shorttitle = {Chapter 2},
	pages = {2175},
	number = {12},
	journal = {Nuclear Fusion},
	shortjournal = {Nucl. Fusion},
	author = {Confinement, ITER Physics Expert Group on and {Transport} and Modelling, ITER Physics Expert Group on Confinement and {Database} and Editors, ITER Physics Basis},
	urlyear = {2024-05-02},
	year = {1999-12},
	langid = {english},
}

@article{mcmillan_system_2010,
	title = {System Size Effects on Gyrokinetic Turbulence},
	volume = {105},
	issn = {0031-9007, 1079-7114},
	url = {https://link.aps.org/doi/10.1103/PhysRevLett.105.155001},
	doi = {10.1103/PhysRevLett.105.155001},
	pages = {155001},
	number = {15},
	journal = {Physical Review Letters},
	shortjournal = {Phys. Rev. Lett.},
	author = {{McMillan}, B. F. and Lapillonne, X. and Brunner, S. and Villard, L. and Jolliet, S. and Bottino, A. and Görler, T. and Jenko, F.},
	urlyear = {2024-03-21},
	year = {2010-10-04},
	langid = {english},
}

@article{burrell_role_2020,
	title = {Role of sheared \textit{E × B} flow in self-organized, improved confinement states in magnetized plasmas},
	volume = {27},
	issn = {1070-664X, 1089-7674},
	url = {https://pubs.aip.org/pop/article/27/6/060501/153617/Role-of-sheared-E-B-flow-in-self-organized},
	doi = {10.1063/1.5142734},
	pages = {060501},
	number = {6},
	journal = {Physics of Plasmas},
	author = {Burrell, K. H.},
	urlyear = {2024-03-20},
	year = {2020-06-01},
	langid = {english},
}

@article{scott_nonlinear_2010,
	title = {Nonlinear Dynamics in the Tokamak Edge},
	volume = {50},
	issn = {1521-3986},
	url = {https://onlinelibrary.wiley.com/doi/abs/10.1002/ctpp.201010039},
	doi = {10.1002/ctpp.201010039},
	pages = {228--241},
	number = {3},
	journal = {Contributions to Plasma Physics},
	author = {Scott, B. D. and Kendl, A. and Ribeiro, T.},
	urlyear = {2024-03-19},
	year = {2010},
	langid = {english},
	note = {\_eprint: https://onlinelibrary.wiley.com/doi/pdf/10.1002/ctpp.201010039},
}

@article{scott_drift_2005,
	title = {Drift wave versus interchange turbulence in tokamak geometry: Linear versus nonlinear mode structure},
	volume = {12},
	issn = {1070-664X, 1089-7674},
	url = {https://pubs.aip.org/pop/article/12/6/062314/762231/Drift-wave-versus-interchange-turbulence-in},
	doi = {10.1063/1.1917866},
	shorttitle = {Drift wave versus interchange turbulence in tokamak geometry},
	pages = {062314},
	number = {6},
	journal = {Physics of Plasmas},
	author = {Scott, Bruce D.},
	urlyear = {2024-03-19},
	year = {2005-06-01},
	langid = {english},
}

@article{pueschel_microinstabilities_2019,
	title = {On microinstabilities and turbulence in steep-gradient regions of fusion devices},
	volume = {61},
	issn = {0741-3335},
	url = {https://dx.doi.org/10.1088/1361-6587/aaf8c1},
	doi = {10.1088/1361-6587/aaf8c1},
	pages = {034002},
	number = {3},
	journal = {Plasma Physics and Controlled Fusion},
	shortjournal = {Plasma Phys. Control. Fusion},
	publisher = {{IOP} Publishing},
	author = {Pueschel, M. J. and Hatch, D. R. and Ernst, D. R. and Guttenfelder, W. and Terry, P. W. and Citrin, J. and Connor, J. W.},
	urlyear = {2024-03-13},
	year = {2019-02},
	langid = {english},
}

@article{snyder_first-principles_2011,
	title = {A first-principles predictive model of the pedestal height and width: development, testing and {ITER} optimization with the {EPED} model},
	volume = {51},
	issn = {0029-5515, 1741-4326},
	url = {https://iopscience.iop.org/article/10.1088/0029-5515/51/10/103016},
	doi = {10.1088/0029-5515/51/10/103016},
	shorttitle = {A first-principles predictive model of the pedestal height and width},
	pages = {103016},
	number = {10},
	journal = {Nuclear Fusion},
	shortjournal = {Nucl. Fusion},
	author = {Snyder, P.B. and Groebner, R.J. and Hughes, J.W. and Osborne, T.H. and Beurskens, M. and Leonard, A.W. and Wilson, H.R. and Xu, X.Q.},
	urlyear = {2024-03-06},
	year = {2011-10-01},
	langid = {english},
}

@article{de_la_luna_exploring_2024,
	title = {Exploring the physics of a high-performance H-mode scenario with small {ELMs} at low collisionality in {JET} with Be/W wall},
	volume = {64},
	issn = {0029-5515, 1741-4326},
	url = {https://iopscience.iop.org/article/10.1088/1741-4326/ad5fa0},
	doi = {10.1088/1741-4326/ad5fa0},
	pages = {096014},
	number = {9},
	journal = {Nuclear Fusion},
	shortjournal = {Nucl. Fusion},
	author = {De La Luna, E. and Garcia, J. and Sertoli, M. and Lomas, P. and Mazzi, S. and Štancar, Ž. and Dunne, M. and Aiba, N. and Silburn, S. and Faitsch, M. and Szepesi, G. and Auriemma, F. and Balboa, I. and Frassinetti, L. and Garzotti, L. and Menmuir, S. and Refy, D. and Rimini, F. and Solano, E.R. and Sozzi, C. and Vecsei, M. and {JET Contributors}},
	urlyear = {2024-08-12},
	year = {2024-09-01},
	langid = {english},
}

@article{jenko_electron_2000,
	title = {Electron temperature gradient driven turbulence},
	volume = {7},
	issn = {1070-664X},
	url = {https://doi.org/10.1063/1.874014},
	doi = {10.1063/1.874014},
	pages = {1904--1910},
	number = {5},
	journal = {Physics of Plasmas},
	shortjournal = {Physics of Plasmas},
	author = {Jenko, F. and Dorland, W. and Kotschenreuther, M. and Rogers, B. N.},
	urlyear = {2024-06-05},
	year = {2000-05-01},
}

@article{frassinetti_pedestal_2021,
	title = {Pedestal structure, stability and scalings in {JET}-{ILW}: the {EUROfusion} {JET}-{ILW} pedestal database},
	volume = {61},
	issn = {0029-5515, 1741-4326},
	url = {https://iopscience.iop.org/article/10.1088/1741-4326/abb79e},
	doi = {10.1088/1741-4326/abb79e},
	shorttitle = {Pedestal structure, stability and scalings in {JET}-{ILW}},
	pages = {016001},
	number = {1},
	journal = {Nuclear Fusion},
	shortjournal = {Nucl. Fusion},
	author = {Frassinetti, L. and Saarelma, S. and Verdoolaege, G. and Groth, M. and Hillesheim, J.C. and Bilkova, P. and Bohm, P. and Dunne, M. and Fridström, R. and Giovannozzi, E. and Imbeaux, F. and Labit, B. and De La Luna, E. and Maggi, C. and Owsiak, M. and Scannell, R. and {JET contributors}},
	urlyear = {2024-01-10},
	year = {2021-01-01},
	langid = {english},
}

@article{kotschenreuther_gyrokinetic_2019,
	title = {Gyrokinetic analysis and simulation of pedestals to identify the culprits for energy losses using ‘fingerprints’},
	volume = {59},
	issn = {0029-5515},
	url = {https://dx.doi.org/10.1088/1741-4326/ab1fa2},
	doi = {10.1088/1741-4326/ab1fa2},
	pages = {096001},
	number = {9},
	journal = {Nuclear Fusion},
	shortjournal = {Nucl. Fusion},
	publisher = {{IOP} Publishing},
	author = {Kotschenreuther, M. and Liu, X. and Hatch, D. R. and Mahajan, S. and Zheng, L. and Diallo, A. and Groebner, R. and {TEAM}, the {DIII}-D. and Hillesheim, J. C. and Maggi, C. F. and Giroud, C. and Koechl, F. and Parail, V. and Saarelma, S. and Solano, E. R. and Chankin, A. and {{JET Contributors}}},
	urlyear = {2023-11-28},
	year = {2019-07},
	langid = {english},
}

@article{chapman-oplopoiou_role_2022,
	title = {The role of {ETG} modes in {JET}–{ILW} pedestals with varying levels of power and fuelling},
	volume = {62},
	issn = {0029-5515},
	url = {https://dx.doi.org/10.1088/1741-4326/ac7476},
	doi = {10.1088/1741-4326/ac7476},
	pages = {086028},
	number = {8},
	journal = {Nuclear Fusion},
	shortjournal = {Nucl. Fusion},
	publisher = {{IOP} Publishing},
	author = {Chapman-Oplopoiou, B. and Hatch, D. R. and Field, A. R. and Frassinetti, L. and Hillesheim, J. C. and Horvath, L. and Maggi, C. F. and Parisi, J. F. and Roach, C. M. and Saarelma, S. and Walker, J. and {{JET Contributors}}},
	urlyear = {2023-11-11},
	year = {2022-06},
	langid = {english},
}

@article{hatch_gyrokinetic_2017,
	title = {A gyrokinetic perspective on the {JET}-{ILW} pedestal},
	volume = {57},
	issn = {0029-5515},
	url = {https://dx.doi.org/10.1088/1741-4326/aa51e1},
	doi = {10.1088/1741-4326/aa51e1},
	pages = {036020},
	number = {3},
	journal = {Nuclear Fusion},
	shortjournal = {Nucl. Fusion},
	publisher = {{IOP} Publishing},
	author = {Hatch, D. R. and Kotschenreuther, M. and Mahajan, S. and Valanju, P. and Liu, X.},
	urlyear = {2023-11-11},
	year = {2017-01},
	langid = {english},
}

@article{diamond_zonal_2005,
	title = {Zonal flows in plasma—a review},
	volume = {47},
	issn = {0741-3335, 1361-6587},
	url = {https://iopscience.iop.org/article/10.1088/0741-3335/47/5/R01},
	doi = {10.1088/0741-3335/47/5/R01},
	pages = {R35--R161},
	number = {5},
	journal = {Plasma Physics and Controlled Fusion},
	shortjournal = {Plasma Phys. Control. Fusion},
	author = {Diamond, P H and Itoh, S-I and Itoh, K and Hahm, T S},
	urlyear = {2023-11-10},
	year = {2005-05-01},
	langid = {english},
}

@article{urano_pedestal_2014,
	title = {Pedestal structure in H-mode plasmas},
	volume = {54},
	issn = {0029-5515},
	url = {https://dx.doi.org/10.1088/0029-5515/54/11/116001},
	doi = {10.1088/0029-5515/54/11/116001},
	pages = {116001},
	number = {11},
	journal = {Nuclear Fusion},
	shortjournal = {Nucl. Fusion},
	publisher = {{IOP} Publishing},
	author = {Urano, Hajime},
	urlyear = {2023-11-10},
	year = {2014-11},
	langid = {english},
}

@article{beer_field-aligned_1995,
	title = {Field-aligned coordinates for nonlinear simulations of tokamak turbulence},
	volume = {2},
	issn = {1070-664X, 1089-7674},
	url = {https://pubs.aip.org/pop/article/2/7/2687/261544/Field-aligned-coordinates-for-nonlinear},
	doi = {10.1063/1.871232},
	pages = {2687--2700},
	number = {7},
	journal = {Physics of Plasmas},
	author = {Beer, M. A. and Cowley, S. C. and Hammett, G. W.},
	urlyear = {2023-10-26},
	year = {1995-07-01},
	langid = {english},
}

@article{dicorato_gyrokinetic_2022,
	title = {Gyrokinetic Stability Analysis of {JET} Pedestal Top Plasmas with Small-{ELMs}},
	volume = {2397},
	issn = {1742-6596},
	url = {https://dx.doi.org/10.1088/1742-6596/2397/1/012007},
	doi = {10.1088/1742-6596/2397/1/012007},
	pages = {012007},
	number = {1},
	journal = {Journal of Physics: Conference Series},
	shortjournal = {J. Phys.: Conf. Ser.},
	publisher = {{IOP} Publishing},
	author = {Dicorato, M. and Muraglia, M. and Camenen, Y. and Garcia, J. and Garbet, X. and Frigione, D. and Garzotti, L. and Lomas, P. and Rimini, F. and Sozzi, C. and Eester, D. Van},
	urlyear = {2023-09-29},
	year = {2022-12},
	langid = {english},
}

@article{garcia_new_2022,
	title = {New H-mode regimes with small {ELMs} and high thermal confinement in the Joint European Torus},
	volume = {29},
	issn = {1070-664X, 1089-7674},
	url = {https://pubs.aip.org/pop/article/29/3/032505/2847797/New-H-mode-regimes-with-small-ELMs-and-high},
	doi = {10.1063/5.0072236},
	pages = {032505},
	number = {3},
	journal = {Physics of Plasmas},
	author = {Garcia, J. and De La Luna, E. and Sertoli, M. and Casson, F. J. and Mazzi, S. and Štancar, Ž. and Szepesi, G. and Frigione, D. and Garzotti, L. and Rimini, F. and Van Eester, D. and Lomas, P. and Sozzi, C. and Aiba, N. and Dicorato, M. and Mariani, A. and Coelho, R. and Frassinetti, L. and Huijsmans, G. T. A. and Liu, F. and {JET Contributors}},
	urlyear = {2023-10-17},
	year = {2022-03-01},
	langid = {english},
}

@article{groebner_elements_2023,
	title = {Elements of H-mode pedestal structure},
	volume = {65},
	issn = {0741-3335},
	url = {https://dx.doi.org/10.1088/1361-6587/acd478},
	doi = {10.1088/1361-6587/acd478},
	pages = {073001},
	number = {7},
	journal = {Plasma Physics and Controlled Fusion},
	shortjournal = {Plasma Phys. Control. Fusion},
	publisher = {{IOP} Publishing},
	author = {Groebner, R. J. and Saarelma, S.},
	urlyear = {2023-09-21},
	year = {2023-05},
	langid = {english},
}

@article{brizard_foundations_2007,
	title = {Foundations of nonlinear gyrokinetic theory},
	volume = {79},
	url = {https://link.aps.org/doi/10.1103/RevModPhys.79.421},
	doi = {10.1103/RevModPhys.79.421},
	pages = {421--468},
	number = {2},
	journal = {Reviews of Modern Physics},
	shortjournal = {Rev. Mod. Phys.},
	publisher = {American Physical Society},
	author = {Brizard, A. J. and Hahm, T. S.},
	urlyear = {2023-08-21},
	year = {2007-04-02},
}

@article{garbet_gyrokinetic_2010,
	title = {Gyrokinetic simulations of turbulent transport},
	volume = {50},
	issn = {0029-5515},
	url = {https://dx.doi.org/10.1088/0029-5515/50/4/043002},
	doi = {10.1088/0029-5515/50/4/043002},
	pages = {043002},
	number = {4},
	journal = {Nuclear Fusion},
	shortjournal = {Nucl. Fusion},
	author = {Garbet, X. and Idomura, Y. and Villard, L. and Watanabe, T. H.},
	urlyear = {2023-08-21},
	year = {2010-03},
	langid = {english},
}

@article{snyder_edge_2002,
	title = {Edge localized modes and the pedestal: A model based on coupled peeling–ballooning modes},
	volume = {9},
	issn = {1070-664X, 1089-7674},
	url = {https://pubs.aip.org/pop/article/9/5/2037/1070288/Edge-localized-modes-and-the-pedestal-A-model},
	doi = {10.1063/1.1449463},
	shorttitle = {Edge localized modes and the pedestal},
	pages = {2037--2043},
	number = {5},
	journal = {Physics of Plasmas},
	author = {Snyder, P. B. and Wilson, H. R. and Ferron, J. R. and Lao, L. L. and Leonard, A. W. and Osborne, T. H. and Turnbull, A. D. and Mossessian, D. and Murakami, M. and Xu, X. Q.},
	urlyear = {2024-11-13},
	year = {2002-05-01},
	langid = {english},
}

@article{dicorato_turbulent_2024,
	title = {Turbulent transport mechanisms and their impact on the pedestal top of {JET} plasmas with small-{ELMs}},
	volume = {66},
	issn = {0741-3335},
	url = {https://dx.doi.org/10.1088/1361-6587/ad867b},
	doi = {10.1088/1361-6587/ad867b},
	pages = {125002},
	number = {12},
	journal = {Plasma Physics and Controlled Fusion},
	shortjournal = {Plasma Phys. Control. Fusion},
	publisher = {{IOP} Publishing},
	author = {Dicorato, M. and Muraglia, M. and Camenen, Y. and Garcia, J. and Garbet, X. and Hatch, D. R. and Merlo, G. and Luna, E. de la and Štancar, Ž and Garzotti, L. and Zotta, V. K. and Rimini, F. and Frigione, D. and {{JET Contributors}} and Team, {EUROfusion} Tokamak Exploitation},
	urlyear = {2024-10-23},
	year = {2024-10},
	langid = {english},
}

@book{scott_turbulence_2021,
	title = {Turbulence and Instabilities in Magnetised Plasmas, Volume 1: Fluid drift turbulence},
	isbn = {978-0-7503-2504-2},
	url = {https://iopscience.iop.org/book/mono/978-0-7503-2504-2},
	shorttitle = {Turbulence and Instabilities in Magnetised Plasmas, Volume 1},
	publisher = {{IOP} Publishing},
	author = {Scott, Bruce},
	urlyear = {2024-09-01},
	year = {2021-03-01},
	langid = {english},
}

@article{snyder_development_2009,
	title = {Development and validation of a predictive model for the pedestal height},
	volume = {16},
	issn = {1070-664X, 1089-7674},
	url = {https://pubs.aip.org/pop/article/16/5/056118/677838/Development-and-validation-of-a-predictive-model},
	doi = {10.1063/1.3122146},
	pages = {056118},
	number = {5},
	journal = {Physics of Plasmas},
	author = {Snyder, P. B. and Groebner, R. J. and Leonard, A. W. and Osborne, T. H. and Wilson, H. R.},
	urlyear = {2024-08-27},
	year = {2009-05-01},
	langid = {english},
}

@article{snyder_pedestal_2009,
	title = {Pedestal stability comparison and {ITER} pedestal prediction},
	volume = {49},
	issn = {0029-5515},
	url = {https://dx.doi.org/10.1088/0029-5515/49/8/085035},
	doi = {10.1088/0029-5515/49/8/085035},
	pages = {085035},
	number = {8},
	journal = {Nuclear Fusion},
	shortjournal = {Nucl. Fusion},
	author = {Snyder, P. B. and Aiba, N. and Beurskens, M. and Groebner, R. J. and Horton, L. D. and Hubbard, A. E. and Hughes, J. W. and Huysmans, G. T. A. and Kamada, Y. and Kirk, A. and Konz, C. and Leonard, A. W. and Lönnroth, J. and Maggi, C. F. and Maingi, R. and Osborne, T. H. and Oyama, N. and Pankin, A. and Saarelma, S. and Saibene, G. and Terry, J. L. and Urano, H. and Wilson, H. R.},
	urlyear = {2024-08-27},
	year = {2009-07},
	langid = {english},
}

@thesis{leppin_turbulence_nodate,
	title = {Turbulence simulations of the high confinement mode pedestal in tokamak fusion experiments},
	type = {PhD Thesis},
	author = {Leppin, Leonhard Andrew},
	langid = {english},
}

@article{gorler_global_2011,
	title = {The global version of the gyrokinetic turbulence code {GENE}},
	volume = {230},
	issn = {0021-9991},
	url = {https://www.sciencedirect.com/science/article/pii/S0021999111003457},
	doi = {10.1016/j.jcp.2011.05.034},
	pages = {7053--7071},
	number = {18},
	journal = {Journal of Computational Physics},
	shortjournal = {Journal of Computational Physics},
	author = {Görler, T. and Lapillonne, X. and Brunner, S. and Dannert, T. and Jenko, F. and Merz, F. and Told, D.},
	urlyear = {2025-02-13},
	year = {2011-08-01},
}

@book{scott_turbulence_2021-1,
	title = {Turbulence and Instabilities in Magnetised Plasmas, Volume 2: Gyrokinetic theory and gyrofluid turbulence},
	isbn = {978-0-7503-3855-4},
	url = {https://iopscience.iop.org/book/mono/978-0-7503-3855-4},
	shorttitle = {Turbulence and Instabilities in Magnetised Plasmas, Volume 2},
	publisher = {{IOP} Publishing},
	author = {Scott, Bruce},
	urlyear = {2025-04-24},
	year = {2021-11-01},
	langid = {english},
}

@article{fenstermacher_progress_2025,
	title = {Progress in pedestal and edge physics: Chapter 3 of the special issue: on the path to tokamak burning plasma operation},
	volume = {65},
	issn = {0029-5515},
	url = {https://dx.doi.org/10.1088/1741-4326/adb1f3},
	doi = {10.1088/1741-4326/adb1f3},
	shorttitle = {Progress in pedestal and edge physics},
	pages = {053001},
	number = {5},
	journal = {Nuclear Fusion},
	shortjournal = {Nucl. Fusion},
	publisher = {{IOP} Publishing},
	author = {Fenstermacher, M.E. and Baylor, L.R. and de la Luna, E. and Dunne, M.G. and Huijsmans, G.T.A. and Kirk, A. and Laggner, F.M. and Osborne, T.H. and Paz-Soldan, C. and Saarelma, S. and Snyder, P.B. and Viezzer, E. and Becoulet, M. and Burrell, K.H. and Cathey, A. and Chen, X. and Hoelzl, M. and Hughes, J.W. and Maingi, R. and Nelson, A.O. and Urano, H. and Wolfrum, E. and Xu, X.Q. and Diallo, A. and Frassinetti, L. and Futatani, S. and Gil, L. and Groebner, R. and Happel, T. and Kim, S.H. and King, J. and Labit, B. and Lang, P.T. and Liu, Y.Q. and Liu, Z.X. and Lunsford, R. and Park, G.Y. and Sheikh, U. and Suttrop, W. and Vanovac, B. and Wilcox, R.S. and Wingen, A. and Zhang, T.},
	urlyear = {2025-05-14},
	year = {2025-04},
	langid = {english},
}

@article{houlberg_bootstrap_1997,
	title = {Bootstrap current and neoclassical transport in tokamaks of arbitrary collisionality and aspect ratio},
	volume = {4},
	issn = {1070-664X, 1089-7674},
	url = {https://pubs.aip.org/pop/article/4/9/3230/457506/Bootstrap-current-and-neoclassical-transport-in},
	doi = {10.1063/1.872465},
	pages = {3230--3242},
	number = {9},
	journal = {Physics of Plasmas},
	author = {Houlberg, W. A. and Shaing, K. C. and Hirshman, S. P. and Zarnstorff, M. C.},
	urlyear = {2025-09-08},
	year = {1997-09-01},
	langid = {english},
}

@article{ida_non-local_2022,
	title = {Non-local transport nature revealed by the research in transient phenomena of toroidal plasma},
	volume = {6},
	issn = {2367-3192},
	url = {https://doi.org/10.1007/s41614-022-00064-6},
	doi = {10.1007/s41614-022-00064-6},
	pages = {2},
	number = {1},
	journal = {Reviews of Modern Plasma Physics},
	shortjournal = {Rev. Mod. Plasma Phys.},
	author = {Ida, Katsumi},
	urlyear = {2026-01-09},
	year = {2022-01-27},
	langid = {english},
}

@article{li_multi-scale_2025,
	title = {Multi-scale Interaction Mechanism for Edge-Localized-Mode Suppression in the Tokamak Edge},
	volume = {16},
	issn = {2041-1723},
	url = {https://www.nature.com/articles/s41467-025-66313-7},
	doi = {10.1038/s41467-025-66313-7},
	pages = {11526},
	number = {1},
	journal = {Nature Communications},
	shortjournal = {Nat Commun},
	author = {Li, Zeyu and Diamond, P. H. and Chen, Xi and Khabanov, F. and Xu, Xueqiao and Hong, R. J. and Chan, V. S. and Muscatello, C. M. and Zeng, L. and Yu, G. Y. and Rhodes, T. and {McKee}, G. R. and Yan, Zheng and Austin, M. E.},
	urlyear = {2026-05-11},
	year = {2025-11-21},
	langid = {english},
}

@article{jarema_block-structured_2016,
	title = {Block-structured grids for Eulerian gyrokinetic simulations},
	volume = {198},
	issn = {0010-4655},
	url = {https://www.sciencedirect.com/science/article/pii/S0010465515003458},
	doi = {10.1016/j.cpc.2015.09.007},
	pages = {105--117},
	journal = {Computer Physics Communications},
	shortjournal = {Computer Physics Communications},
	author = {Jarema, D. and Bungartz, H. J. and Görler, T. and Jenko, F. and Neckel, T. and Told, D.},
	urlyear = {2026-08-31},
	year = {2016-01-01},
}

@article{hahm_flow_1995,
	title = {Flow shear induced fluctuation suppression in finite aspect ratio shaped tokamak plasma},
	volume = {2},
	issn = {1070-664X},
	url = {https://doi.org/10.1063/1.871313},
	doi = {10.1063/1.871313},
	pages = {1648--1651},
	number = {5},
	journal = {Physics of Plasmas},
	shortjournal = {Phys. Plasmas},
	author = {Hahm, T. S. and Burrell, K. H.},
	urlyear = {2026-09-08},
	year = {1995-05-01},
}

@article{geng_physics_2020,
	title = {The physics of a small-scale tearing mode in collisionless slab plasmas},
	volume = {62},
	issn = {0741-3335},
	url = {https://doi.org/10.1088/1361-6587/ab97f1},
	doi = {10.1088/1361-6587/ab97f1},
	pages = {085009},
	number = {8},
	journal = {Plasma Physics and Controlled Fusion},
	shortjournal = {Plasma Phys. Control. Fusion},
	publisher = {{IOP} Publishing},
	author = {Geng, Chen and Dickinson, David and Wilson, Howard},
	urlyear = {2026-09-11},
	year = {2020-07},
	langid = {english},
}

@misc{singh_how_2026,
	title = {How nonlinear spectral back transfer limits the temporal coherency of zonal modes?},
	url = {http://arxiv.org/abs/2604.03421},
	doi = {10.48550/arXiv.2604.03421},
	number = {{arXiv}:2604.03421},
	publisher = {{arXiv}},
	author = {Singh, Rameswar and Diamond, P. H.},
	urlyear = {2026-09-16},
	year = {2026-06-03},
	langid = {english},
	eprinttype = {arxiv},
	eprint = {2604.03421 [physics.plasm-ph]},
}

@article{kim_random_2004,
	title = {Random shearing by zonal flows and transport reduction},
	volume = {11},
	issn = {1070-664X},
	url = {https://doi.org/10.1063/1.1808455},
	doi = {10.1063/1.1808455},
	pages = {L77--L80},
	number = {12},
	journal = {Physics of Plasmas},
	shortjournal = {Phys. Plasmas},
	author = {Kim, Eun-jin and Diamond, P. H.},
	urlyear = {2026-09-16},
	year = {2004-12-01},
}

@thesis{crandall_collisional_2019,
	title = {Collisional and Electromagnetic Physics in Gyrokinetic Models},
	url = {https://escholarship.org/uc/item/1jk6561c},
	institution = {{UCLA}},
	type = {PhD Thesis},
	author = {Crandall, Paul},
	urldate = {2024-03-02},
	date = {2019},
	langid = {english},
}

@article{germaschewski2021,
    author = {Germaschewski, K. and Allen, B. and Dannert, T. and Hrywniak, M. and Donaghy, J. and Merlo, G. and Ethier, S. and D'Azevedo, E. and Jenko, F. and Bhattacharjee, A.},
    title = {Toward exascale whole-device modeling of fusion devices: Porting the GENE gyrokinetic microturbulence code to GPU},
    journal = {Physics of Plasmas},
    volume = {28},
    number = {6},
    pages = {062501},
    year = {2021},
    month = {06},
    issn = {1070-664X},
    doi = {10.1063/5.0046327},
    url = {https://doi.org/10.1063/5.0046327},
}

@article{garzotti_2025,
doi = {10.1088/1361-6587/ade008},
url = {https://doi.org/10.1088/1361-6587/ade008},
year = {2025},
month = {jun},
publisher = {IOP Publishing},
volume = {67},
number = {7},
pages = {075011},
author = {Garzotti, L and Frigione, D and Lomas, P and Rimini, F and Van Eester, D and Aleiferis, S and Alessi, E and Auriemma, F and Carvalho, I S and Carvalho, P and Chomiczewska, A and De La Luna, E and Ferreira, D R and Field, A and Fontana, M and Frassinetti, L and Gabriellini, S and Ghani, Z and Giovannozzi, E and Giroud, C and Gromelski, W and Ivanova-Stanik, I and Kiptily, V and Kirov, K and Lennholm, M and Lowry, C and Maggi, C F and Mailloux, J and Menmuir, S and Morales, R B and Nowak, S and Parail, V and Patel, A and Pau, A and Perez von Thun, C and Piron, L and Pucella, G and Reux, C and Sauter, O and Solano, E R and Sozzi, C and Štancar, Ž and Stuart, C and Sun, H and Telesca, G and Tskhakaya, D and Valovič, M and Wendler, N and Zotta, V K and JET Contributors},
title = {Development of high-current baseline scenario for high deuterium–tritium fusion performance at JET},
journal = {Plasma Physics and Controlled Fusion}
}

@article{garzotti_2026,
doi = {10.1088/1361-6587/ae97b5},
url = {https://doi.org/10.1088/1361-6587/ae97b5},
year = {2026},
month = {aug},
publisher = {IOP Publishing},
volume = {68},
number = {8},
pages = {085036},
author = {Garzotti, L and Frigione, D and Lomas, P and Rimini, F and Van Eester, D and Zotta, V K and Aleiferis, S and Alessi, E and Auriemma, F and Morales, R B and Carvalho, I S and Carvalho, P and Chomiczewska, A and Cicioni, R and De La Luna, E and Ferreira, D R and Field, A and Fontana, M and Frassinetti, L and Gabriellini, S and Ghani, Z and Giovannozzi, E and Giroud, C and Gromelski, W and Grzybicka, N and Ivanova Stanik, I and Kiptily, V and Kirov, K and Lennholm, M and Leoni, C and Lombardo, J and Lowry, C and Maggi, C F and Mailloux, J and Menmuir, S and Nowak, S and Parail, V and Patel, A and Pau, A and Peluso, E and Perez von Thun, C and Piron, L and Pucella, G and Reux, C and Saarelma, S and Sauter, O and Solano, E and Sozzi, C and Štancar, Ž. and Stuart, C and Sun, H and Telesca, G and Tskhakaya, D and Valovič, M and Wendler, N and JET Contributors and the EUROfusion Tokamak Exploitation Team},
title = {Development of a high-current, high fusion performance scenario on JET: physics insights and operational challenges},
journal = {Plasma Physics and Controlled Fusion}
}

\end{document}